\documentclass[nofootinbib,prd,aps,groupedaddress,longbibliography,twocolumn
]{revtex4-1}

\usepackage{bbm}
\usepackage[dvipsnames]{xcolor}
\usepackage{amssymb,latexsym}

\usepackage{mathtools}
\usepackage[normalem]{ulem}

\newcommand{\greenn}[1]{{\color{ForestGreen} #1}}

\definecolor{applegreen}{rgb}{0.55, 0.71, 0.0}
\definecolor{armygreen}{rgb}{0.29, 0.33, 0.13}

\definecolor{caribbeangreen}{rgb}{0.0, 0.8, 0.6}

\newcommand{\KSu}{{\varphi}}
\newcommand{\KSv}{{\psi}}
\newcommand{\ezero}{=|_{\charge=0} \ }

\newcommand{\BLphi}{{\tilde{\varphi}}}
\newcommand{\tildet}{{\tilde{t}}}

\newcommand{\BLr}{{\tilde{r}}}

\newcommand{\KStheta}{{\ell}}
\newcommand{\BLtheta}{{\tilde \theta}}
\newcommand{\rhoSquareBL}{{\tilde{\rho}^2}}
\newcommand{\rhoBL}{{\tilde{\rho}}}
\newcommand{\DeltaBL}{{\Delta_{\tilde r}}}
\newcommand{\Deltatheta}{{\Delta_{\tilde\theta}}}
\newcommand{\Deltar}{{\DeltaBL}}

\newcommand{\charge}{{e}}

\newcommand{\mcT}{{\cal T}}
\newcommand{\mcE}{{\cal E}}

\newcommand{\rout}[1]{\xout{\red{#1}}}

\newcommand{\red}[1]{{\color{red} #1}}

\newcommand{\RR}{\hat{r}}

\newcommand{\blue}[1]{{{\color{blue}#1}}}

\DeclareFontFamily{OT1}{rsfs}{}
\DeclareFontShape{OT1}{rsfs}{m}{n}{ <-7> rsfs5 <7-10> rsfs7 <10-> rsfs10}{}
\DeclareMathAlphabet{\mycal}{OT1}{rsfs}{m}{n}

\newcommand{\mcL}{{\cal L}}

{\catcode `\@=11 \global\let\AddToReset=\@addtoreset}
\AddToReset{equation}{section}

{\catcode `\@=11 \global\let\AddToReset=\@addtoreset}
\AddToReset{figure}{section}

\newcounter{mnotecount}[section]

\newcommand{\T}{\mathbb{T}}

\newcommand{\eel}[1]{\label{#1}\end{equation}}
\newcommand{\eeal}[1]{\label{#1}\end{eqnarray}}

\newcommand{\mcK}{{\mycal K}}

\newcommand{\bel}[1]{\begin{equation}\label{#1}}
\newcommand{\bea}{\begin{eqnarray}}
\newcommand{\bean}{\begin{eqnarray}\nonumber}
\newcommand{\beal}[1]{\begin{eqnarray}\label{#1}}
\newcommand{\eea}{\end{eqnarray}}

\newcommand{\be}{\begin{equation}}
\newcommand{\eeq}{\end{equation}}
\newcommand{\ee}{\end{equation}}
\newcommand{\beqa}{\begin{eqnarray}}
\newcommand{\eeqa}{\end{eqnarray}}

\newcommand{\ba}{\begin{array}}
\newcommand{\ea}{\end{array}}

\newcommand{\const}{\mbox{\rm const}} 

\newcommand{\mnote}[1]
{\protect{\stepcounter{mnotecount}}$^{\mbox{\footnotesize
$
\bullet$\themnotecount}}$ \marginpar{
\raggedright\tiny\em
$\!\!\!\!\!\!\,\bullet$\themnotecount: #1} }

\newcommand{\warn}[1]
{\protect{\stepcounter{mnotecount}}$^{\mbox{\footnotesize
$
\bullet$\themnotecount}}$ \marginpar{
\raggedright\tiny\em
$\!\!\!\!\!\!\,\bullet$\themnotecount: {\bf Warning:} #1} }

\newcommand{\C}{\mathbb C}
\newcommand{\R}{\mathbb R}

\newcommand{\Z}{\mathbb Z}

\def\ben{\begin{equation}}
\def\een{\end{equation}}
\def\bena{\begin{eqnarray}}
\def\eena{\end{eqnarray}}

\def\f(#1/#2){\frac{#1}{#2}}
\def\Frac(#1/#2){\left(\frac{#1}{#2}\right)}
\def\chris(#1-#2-#3){{\mit \Gamma}^{#1}{}_{{#2}{#3}} }
\def\tilchris(#1-#2-#3){\tilde{{\mit \Gamma}}^{#1}{}_{{#2}{#3}}}
\def\hatchris(#1-#2-#3){\hat{{\mit \Gamma}}^{#1}{}_{{#2}{#3}}}

\newcommand{\BL}{{\mbox{\tiny BL}}}

\newcommand{\diff}[1]{\text{d}#1}
\newcommand{\bigO}[1]{{{{O}}}\left(#1\right)}

\DeclareFontFamily{OT1}{rsfs}{}
\DeclareFontShape{OT1}{rsfs}{m}{n}{ <-7> rsfs5 <7-10> rsfs7 <10=rsfs10}{}

\renewcommand{\mcK}{{\mathcal K}}
\renewcommand{\rout}[1]{}
\renewcommand{\xout}[1]{}
\renewcommand{\red}[1]{#1}
\renewcommand{\blue}[1]{#1}
\renewcommand{\greenn}[1]{#1}

\begin{document}

\title{
The structure of the singular ring in Kerr-like metrics
\footnote{Preprint UWThPh-2019-37}}

\author{Piotr T.\ Chru\'{s}ciel}
\email[]{piotr.chrusciel@univie.ac.at}
\affiliation{Faculty of Physics, University of Vienna, Vienna 1090, Austria}

\author{Maciej Maliborski}
\email[]{maciej.maliborski@univie.ac.at}
\affiliation{Faculty of Physics, University of Vienna, Vienna 1090, Austria}

\author{Nicol\'as Yunes}
\email[]{nyunes@illinois.edu}
\affiliation{Department of Physics, University of Illinois at Urbana-Champaign, Urbana, Illinois 61801, USA}

\date{\today}

\begin{abstract}
The Kerr geometry is believed to represent the exterior spacetime of astrophysical black holes.
We here re-analyze the geometry of Kerr-like metrics (Kerr, Kerr-Newman, Kerr-de Sitter, and Kerr-anti de Sitter), paying particular attention to the region near the singular set.
We find  that, although the Kretschmann  scalar vanishes at the singular set along a given direction, a certain combination of curvature invariants diverges regardless of the direction of approach.
We also find that the two-dimensional geometry induced by the spacetime metric on the orbits of the isometry group also possesses a singularity regardless of the direction of approach.
Likewise, the two-dimensional geometry in the directions orthogonal to the isometry orbits is $C^{2}$-divergent, but extends continuously at the singular set as a cone with opening angle $4\pi$.
We conclude by showing that tidal forces lead to infinite stresses on neighboring geodesics that approach the singular set, destroying any such observers in finite proper time.
Those geodesics that come in from infinity  and do not hit the singular set but approach it are found to need tremendous energy to get close to the singular set,  experiencing an acceleration transversal to the equatorial plane which grows without bound when the minimal distance of approach goes to zero.
While establishing these results, we also present an alternative description of some other known properties, as well as introducing toroidal coordinates that provide a hands-on description of the double-covering for the geometry near the singular set.

\end{abstract}


\maketitle


\section{Introduction}

There is strong evidence that the Kerr metric is the unique well-behaved, stationary black hole solution of the vacuum Einstein equations, and as such it has deserved a lot of attention in the literature. The foundational references are~\cite{CarterKerr,BONeillK,BoyerLindquist}, and the reader will find many further references in~\cite{VisserKerr}.

 In Kerr-Schild coordinates, the Kerr metric is smooth except at a coordinate set $\R\times S^1$, where $\R$ is the time coordinate and $S^1$ is a circle lying on the equatorial plane. This kind of objects are nowadays called ``strings'' in theoretical physics, or  perhaps ``membranes in spacetime.'' Therefore, the singular coordinate set should really be called  a ``string singularity''  or a ``membrane singularity,'' although historically it has been dubbed ``the ring singularity.'' We will nevertheless stick to the usual ``ring'' terminology.

  While various results concerning the behavior of the metric tensor near this singular ring have been derived for example in~\cite{CarterKerr,ShadiKerr,CarterKerr,IsraelKerr2,IsraelKerr3},  no clear statements about the nature of the singularity exist in the literature. It therefore  seemed of interest to us to study this question and as a byproduct to present an overview of this aspect of the Kerr geometry, with an eye out for what happens when electric charge and a cosmological constant are included. This is the purpose of this work.

The first issue that we address is that of whether the four-dimensional  curvature is singular at the ring. The standard claim is that the Kretschmann scalar of the Kerr metric is unbounded when the ring is approached. As known to experts, this is wrong for some directions of approach, along which the Kretchmann scalar actually vanishes. The same claim and the same problem arise if considering other curvature invariants, such as the Pontryagin scalar. Our first result is the resolution of this issue by exhibiting a curvature invariant that tends to infinity independently of the direction of approach to the ring, for the whole Kerr-Newman (anti)-de Sitter (KN(A)dS) family of metrics with non-zero mass parameter $m$.

 Having this out of the way, the question arises of whether one can develop a more precise and intuitive understanding of the \emph{nature} of this curvature singularity, and of the geometric and physical properties of the metric when the ring is approached. This understanding can be developed more easily by studying meaningful two-dimensional geometries associated with the four-dimensional KN(A)dS spacetime, keeping in mind that a two-dimensional geometry is fully characterized by its Ricci scalar. One can construct such two-dimensional geometries by making use of the stationarity and axisymmetry of the  {KN(A)dS} metrics. Indeed, Killing vectors can be used to show that the region near the singular ring is stationary, it contains causality-violating closed-timelike curves, and a second ergosphere with either the topology of a solid torus (in the spin sub-critical case) or a hollow marble
 (in the spin super-critical case).
The closed-timelike curves appear only close to the ring, but still at macroscopic distances from the ring for stellar-sized objects   and not, as some might be tempted to think, at the Planck length
$(\hbar G/c^{3})^{1/2} \approx   {10^{-33}\; {\rm{cm}}}$, which is typically associated with the scale at which quantum gravitational effects could be expected to modify
the geometry and possibly eliminate or create these curves.

 A useful two-dimensional structure that can be constructed from the Killing vectors is that which describes the geometry of the set of orbits of the isometry group.
To be precise, denote by $G$ the connected component of the identity of the isometry group of the Kerr metric, so that $G$ acts by flowing along linear combinations of the stationary and rotational Killing vectors. Then, one such two-dimensional geometry is defined by the tensor field induced by the spacetime metric $g$ on the orbits of $G$, which we denote by $\chi$. Away from the singular ring, $\chi$ defines a (flat) two-dimensional Lorentzian metric at each orbit of $G$ in spacetime, and the collection of the tensor fields $\chi$ for all orbits  describes how those orbits change when moved transversally.  Our second result is to show that the ring is a $C^2$-singularity for the family of metrics $\chi$, again for the whole KN(A)dS family of metrics. This need not to have been the case just because of the existence of a curvature singularity at the ring. Physically, this result is saying that both the notion of the flow of time (associated with timelike Killing vectors) and the notion of rotation (associated with the rotational Killing vector) break down at the ring.

Another useful two-dimensional geometry is obtained from the \emph{orbit-space metric}, which we denote by $q$, and which is a symmetric, two-dimensional, covariant  tensor field defined away from the set where the orbits of $G$ are null (for a precise definition of $q$, see Eqs.~\eqref{20VIII19.1} and \eqref{20VIII19.2} below). Away from  the singular ring, $q$ defines a two-dimensional Lorentzian metric that can be intuitively thought of as encoding the geometry perpendicular to the isometry flow.  Our third result is the observation that, in the uncharged case and for all values of the cosmological constant, the metric $q$ has a $C^2$\red{-}singularity at the ring, but it extends continuously there as a cone metric with opening angle $4\pi$. In the charged case, we show that this remains true up to a conformal factor which has a direction-dependent limit at the ring. This result is  {unexpected}, given that the $\chi$ geometry and the full four-dimensional geometry are singular at the ring. Physically, one might say that, at a fixed time or a fixed rotation angle, the geometry associated with the {remaining} spatial coordinates (say a radial coordinate and another angular coordinate) is continuous all the way up to the ring, even though the notion of time and rotation break down there.

Last but perhaps not least, we study the tidal forces experienced by geodesics approaching the ring. While it is clear that these forces tend to infinity as geodesics approach the ring, their effect on physical objects could be relatively mild if the resulting integrated stresses, and/or integrated displacements of nearby freely-falling objects were finite.  Our fourth result is to  show that these forces  lead to {both infinite stresses and displacements} on neighboring geodesics  in finite (proper) time.   Finally, we study  a natural family of curves on the equatorial plane that do not hit the ring but that approach it.  We show that the curves  need a tremendous amount of initial energy to come close to the ring, and when they do, they experience a very large acceleration transversal to the equatorial hyperplane, {which would} kick these {curves} out of the equatorial plane.

A useful tool for all of this is a set of toroidal coordinates, which lead to our fifth result: a simple description of the double-covering space for the KN(A)dS geometry. A well-known fact in mathematical relativity regarding the Kerr metric is that certain components in the usual Kerr-Schild coordinates are multivalued when following a closed loop that {circles around} the singular ring~\cite{CarterKerr,Newman:1965tw}. This then prompts the introduction of a double-covering space for the KN(A)dS geometry. We find here that such a double covering can be intuitively and precisely understood with our toroidal coordinate system by realizing that the toroidal angle that winds around the singular ring is actually $4 \pi$ periodic, instead of $2 \pi$ periodic.

While the analysis of the Kerr singularity is of clear geometric interest, the question arises, whether such a study  is of any physical relevance. We emphasise that we do not expect to be able to observe the Kerr singularity either within a white hole or within a ``transient black hole'' whose horizon will evaporate. But the question, whether we will be able to travel near  the singularity and return to tell the story is closely related to the problem of cosmic censorship, which has not been settled yet. Hence, it makes sense to enquire about the observational consequences of the nature of naked singularities, if only to know what we should not expect to observe.

The remainder of this paper details the calculations summarized above. Henceforth, we make use of the conventions of Misner, Thorne and Wheeler~\cite{Misner:1974qy} and we employ geometric units in which $c = 1 = G$.
Section~\ref{s19VII19.1} presents the Kerr metric in Boyer-Lindquist coordinates and indicates the need for a double-covering.
Section~\ref{sec:toroidal} introduces a new set of toroidal coordinates, which are useful to understand the need for a double-covering and to analyze the singular nature of the curvature as the ring is approached.
Section~\ref{sec:near-the-ring} studies the Kerr geometry near the singular ring, through Killing vectors and invariantly-defined  two-dimensional geometric objects.
Section~\ref{s1X19} calculates the tidal forces experienced by a variety of observers as they approach the ring, with Sec.~\ref{s1XI19.11} generalizing the results to accelerated observers.
Section~\ref{sec:electric-CC} generalizes some of the above results to the Carter-Demia\'nski metrics, i.e.\ charged Kerr metrics with a cosmological constant.
%

\section{The Kerr metric}
\label{s19VII19.1}

In this section, we begin by presenting the basics of the Kerr metric in Boyer-Lindquist and Kerr-Schild
coordinates. We then proceed to describe the behavior of the metric across the equatorial plane both
outside and inside the ring singularity.

\subsection{A basic introduction}

In Boyer-Lindquist coordinates $(\tildet ,\tilde{r},\BLtheta,\BLphi)$, the Kerr metric  reads
\begin{eqnarray}
 \nonumber
g
 & =
  &
 - \left(1 - \frac{2 m \tilde{r}}{\rhoSquareBL} \right) \diff{\tildet}\,^{2} - \frac{4 m a \tilde{r}}{\rhoSquareBL} \sin^{2}(\BLtheta) \diff \tildet \,   \diff\BLphi
 \nonumber \\
 & + & \frac{\Sigma_{\BL}}{\rhoSquareBL} \sin^{2}(\BLtheta) \diff\BLphi^{2} + \frac{\rhoSquareBL}{\DeltaBL} \diff{\tilde{r}}^{2} + \rhoSquareBL \diff\BLtheta^{2}
\\
\label{8X19.0}
 & = &
-\frac{\Delta_{\tilde{r}}}{\rhoSquareBL}\left(\diff{\tildet}-a\sin^{2}(\BLtheta)\diff{\BLphi}\right)^{2}
	+ \frac{\rhoSquareBL}{\Delta_{\tilde{r}}}\diff{\tilde{r}}^{2}
	+ \rhoSquareBL\diff{\BLtheta}^{2}
	\nonumber \\
 &
	+ & \frac{\sin^{2}(\BLtheta)}{\rhoSquareBL}\left(a\diff{\tildet} - (\BLr^{2}+a^{2})\diff{\BLphi}\right)^{2}
	\,,
\end{eqnarray}
where we  introduced the usual Boyer-Lindquist functions
\begin{eqnarray}
\label{8X19.1}
\rhoSquareBL &:=& \tilde{r}^{2} + a^{2} \cos^{2}(\BLtheta)\,,
\quad
\DeltaBL := \tilde{r}^{2} -2 m \tilde{r} + a^{2}\,,
\nonumber \\
\Sigma_{\BL} &:=& \left(\tilde{r}^{2} + a^{2}\right)^{2} - a^{2} \DeltaBL \sin^{2}(\BLtheta)\,.
\end{eqnarray}

A few observations are now in order. We denote the radial Boyer-Lindquist coordinate as $\tilde{r}$ to make it clear that this quantity is not equal to $\sqrt{x^{2} + y^{2} + z^{2}}$ in Kerr-Schild coordinates, which play an important role below. The constant $m$ is the (ADM) mass of the black hole spacetime, while $a =S /m$ is the Kerr spin parameter, with $S$ the $z$-component of the (ADM) angular momentum. Recall that the Kerr singularity is located at $\rhoBL = 0$,
which corresponds to a circular string lying on the equatorial plane in Kerr-Schild coordinates, for some obscure reasons referred to in the literature as ``ring''.
Our work focuses on this ring singularity, which means we restrict attention to Kerr spacetimes with $m \ne 0\,$ and $a \ne 0$. Notice that, in particular, neither the sign of $m$ nor a bound such as $a^2<m^2$ matter. However, in order to avoid splitting the analysis into cases, it is convenient to impose $m > 0$, which  will be done henceforth, as can always be achieved by changing the radial Boyer-Lindquist coordinate to its negative (changing time-orientation). Similarly we will assume $a>0$, which can always be achieved by changing $\BLphi$ to its negative (changing space-orientation).

The nature of the Kerr singularity  {can be} revealed by transforming the metric to Kerr-Schild coordinates $(t,x,y,z)$. The transformation between Boyer-Lindquist and Kerr-Schild coordinates is~\cite{Poisson:2009pwt}
\begin{align}
x + i y &= (\tilde{r} + i a) e^{i \left(\BLphi + r^{\#}\right)} \sin(\BLtheta)
 \,,
\nonumber \\
z &= \tilde{r} \cos(\BLtheta)\,,
\quad
t = \tildet  + r^{*} - \tilde{r}\,,
 \label{14VIII19.1}
\end{align}
where we have introduced the tortoise coordinates
\begin{align}
r^{*} := \int \frac{\tilde{r}^{2} + a^{2}}{\DeltaBL} \diff \tilde{r}\,,
\quad
r^{\#} := \int \frac{a}{\DeltaBL} \diff \tilde{r}\,.
 \label{14VIII19.1+}
\end{align}
In Kerr-Schild coordinates, the Kerr metric reads
(cf., e.g., \protect\cite{HE})
\begin{equation}\label{21VI119.14}
g_{\mu\nu} = \eta_{\mu\nu} + H  \KStheta_\mu \KStheta_\nu
 \,,
\end{equation}
where we have defined the factor
\begin{equation}\label{H-eq}
H := \frac{2m{\tilde{r}}^3}{{\tilde{r}}^4+a^2 z^2}\,,
\end{equation}
and where the
covector field associated with the
(ingoing) principal null congruence of the Kerr spacetime is given by
\begin{align}
\KStheta \equiv
 \KStheta_\mu \diff{x^\mu} :=
- &\left(
 \diff{t} +
 \frac 1 {{\tilde{r}}^2+a^2}\left[ {\tilde{r}} (x\diff{x}+y\diff{y})
 \right. \right.
 \nonumber \\
 \label{8V20.01}
 &- \left.\left.
  a (x\diff{y}-y\diff{x})\right] + \frac z {\tilde{r}} \diff{z}
 \right)
 \,,
\end{align}
with ${\tilde{r}}$ defined implicitly as the solution of
the equation
\begin{equation}\label{25VII19.11}
 {\tilde{r}}^4 - {\tilde{r}}^2 ( x^2+y^2+z^2-a^2)-a^2 z^2 = 0
 \,.
\end{equation}
Recall that the covector field $\KStheta$ is null both with respect to the Minkowski metric and to the Kerr metric. This implies
\begin{equation}\label{21VI119.14inv}
 g^{\mu\nu} = \eta^{\mu\nu}- H  \KStheta^\mu \KStheta^\nu
 \,,
\end{equation}
where $\eta^{\mu\nu}$ is the inverse Minkowski metric, and where it does not matter whether the indices on $\KStheta$ have been raised with the Minkowski metric or with the Kerr metric.

A comment on the choice of parameters when presenting some figures later on is now in order. When $a\ne 0$, as assumed here, one can rescale all the coordinates by $a$, $x^\mu \mapsto a x^\mu$, and divide the metric by $a^{-2}$. The metric thus rescaled depends only upon a single parameter $m/a$, which becomes an overall multiplicative factor in the term $H  \KStheta^\mu \KStheta^\nu$. Equivalently, all calculations which do not depend upon an overall constant rescaling of the metric can be carried out by setting $a=1$, keeping in mind that $m$ then is actually $m/a$. We will later find this way of presenting figures more convenient than the usual way in terms of rescalings in $m$. However, when presenting equations throughout the paper, we will leave all factors of $m$ and $a$ explicit.

\subsection{The Kerr metric through the disc}
\label{ss22VII19.1}

In order to make clear the behavior of the metric \emph{near} the {space-}coordinate
disc
\begin{equation}\label{31VII19.1}
  D:=\R\times  \{ x^2 + y^2 < a^2\,, \ z=0\}\subset\R\times \R^3
\,,
\end{equation}
where the first factor represents the $t$--coordinate,
it is best to use the natural differentiable structure arising from the coordinates $(x,y,z)$. For notational convenience,
let us set
\begin{equation}
\label{eq:rho}
 \rho^2:= x^2 +y^2
  \,;
 \end{equation}
{here $\rho$ should} not to be confused with the  \greenn{$\rhoBL$} factor used to write down the Kerr line element in Boyer-Lindquist coordinates  in Eq.~\eqref{8X19.1}. Using this notation, the real valued solutions of {Eq.~\eqref{25VII19.11}}
take the form
\begin{equation}\label{25VII19.2}
\tilde{r} = \pm \frac{\sqrt{\sqrt{\left(a^2-\rho ^2-z^2\right)^2+4 a^2 z^2}-a^2+\rho ^2+z^2}}{\sqrt{2}}
 \,.
\end{equation}
The function $\tilde{r}^2$ clearly extends smoothly through $z=0$ away from $\tilde{r} = 0$,
and it follows that functions such as $\tilde{r}H$ or $H/\tilde{r}$ also extend smoothly there.

As such, for small $z$ we have the expansions
\begin{equation}\label{25VII19.3}
 \tilde{r}
  =
 \pm \left\{
      \begin{array}{ll}
        \frac{a \left| z\right| }{\sqrt{a^2-\rho ^2}}-\frac{ a \rho ^2 \left| z\right| z^2 }{2
   \left(a^2-\rho ^2\right)^{5/2}}+O\left(z^4\right), & \hbox{$\rho < a$;} \\
        \sqrt{\rho ^2-a^2}+\frac{\rho ^2 z^2}{2 \left(\rho ^2-a^2\right)^{3/2}}+O\left(z^3\right), & \hbox{$\rho > a$.}
      \end{array}
    \right.
\end{equation}
This shows that either choice of sign leads to a function that extends smoothly through the hyperplane $z=0$ in the region $\rho>a$.
However, in the region $\rho <a$, given a choice of sign of $\tilde{r}$ for $z>0$, one needs to choose the opposite sign for $z<0$ to obtain a function that extends smoothly through the equatorial hyperplane. One abstract way of achieving this is to take a double cover of $\R\times (\R^3 \setminus \{ \rho=a\,,\ z=0\})$, in which $\tilde{r}$ changes sign each time one crosses the ``disc''  $    \{ t\in \R\,,\ 0\le \rho<a\,,\ z=0\}$.  Another way of capturing this, made explicit below, is to rewrite  the metric functions as  $4\pi$-periodic functions in terms of a natural angular coordinate $\KSv$ around the ring.

Since the ring itself is not part of the spacetime,   any $\Z_{2k}$-covering, $k\ge 1$, or a $\Z$-covering, will also work. Equivalently, the coordinate $\KSv$ could be taken to \rout{range over} \red{be $4 k\pi$-periodic}, with $k\in \Z$, or to range over $\R$.

A further observation stemming from Eq.~\eqref{25VII19.3} is that the function $H$ defined by Eq.~\eqref{H-eq} vanishes on the coordinate disc \xout{$\rho < a$ lying on the equatorial hyperplane $z=0$} \greenn{$\{\rho<a\,,\ z=0\}$}; indeed, the numerator behaves as $|z|^3$ and the denominator as $z^2$. Hence, the metric induced there is simply the three-dimensional Minkowski metric. Perhaps somewhat surprisingly, $H$ does not vanish on the equatorial hyperplane outside of the ring, where instead we have
\begin{equation}\label{27VII19.15}
  H = \pm
   \frac{2 m}{\sqrt{x^{2} + y^{2} - a^{2}}} \quad {\textrm{on}} \quad \{\rho>a, \; z = 0\}\,,
\end{equation}
with the sign of $H$ determined by that of $\tilde{r}$.

Let us here point out that the negative-$\tilde{r}$ region possesses an asymptotically flat region of its own, without Killing horizons, and with the ADM mass of that region equal to  $-m$. But this is irrelevant for the question of interest here, namely the structure of spacetime near the coordinate ring $\{\rho=a, z = 0 \}$.

Before continuing, one should mention the possibility of making an alternative extension, where $\tilde{r}$ remains positive when crossing the  equatorial disc spanned by the ring.
The metric extends then continuously through the disc but some transverse derivatives jump. This leads to a distributional energy-momentum tensor  which has been calculated in~\cite{IsraelKerr2}. The surface matter-density, say $\sigma$, is negative,
\begin{equation}\label{6IX19.1}
  \sigma =- \frac{1}{2}
 \frac{m}{\pi a^2} \left(
 1- \frac{\rho^2}{a^2}\right)^{-3/2}
 \,,
\end{equation}
and therefore we will not consider this option in what follows.
\blue{While the negativity of $\sigma$ perhaps helps to explain the repulsive character of the ring experienced by geodesics~\cite{CarterKerr,BONeillK},
 the resulting spacetime does not provide a physically relevant model.}

It has \red{also} been suggested in~\cite{IsraelKerr2} that \red{this last} \rout{the} positive-$\tilde r$-choice describes a distribution of matter rotating with superluminal speed.  \rout{A proposal} \red{Yet another proposal}, how to associate energy-momentum with the ring has been put forward in \cite{IsraelKerr3}; see also~\cite{BalasinNachbagauer,VickersStrings,GerochTraschen}.

\section{Toroidal coordinates and the Kerr metric}
\label{sec:toroidal}

In this section, we begin by introducing a set of toroidal coordinates
that simplifies the analysis of the Kerr metric near the ring singularity.
This leads us in particular to a rather simpler description of the double-covering construction described in the previous section.
We  conclude this section by studying the curvature singularity
in toroidal coordinates.

\subsection{Coordinates adapted to a torus}

\begin{figure}[thb]
\hspace{-.45cm}\includegraphics[width=\columnwidth,clip=true]{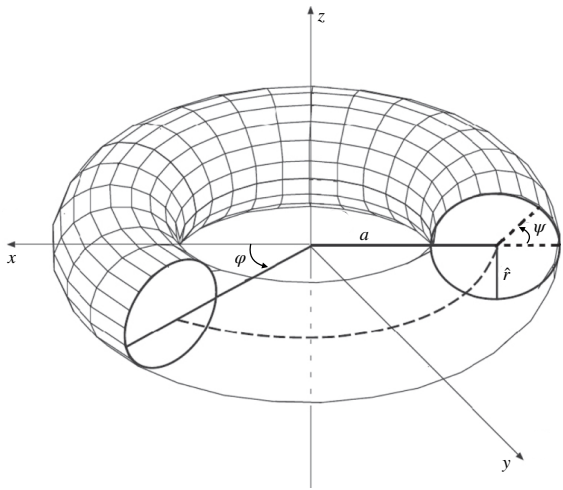}
\caption{Toroidal coordinates. The $\RR$ coordinate represents  the minor radius of the torus, while the major radius is fixed at $a$. The angles $\KSu$ and $\KSv$ are periodic, with $\KSu$ adapted to rotations about the $z$ axis, and $\KSv$ adapted to rotations about the curve that runs the circumference of the ring, shown by the dashed line. Clearly, for these coordinates to make sense, we must restrict $\RR < a$.}
\label{torus-fig}
\end{figure}

The nature of the singularity at the Kerr-Schild coordinate ring  $\{\rho=a,z=0\}$
can be further elucidated by introducing coordinates adapted to it. We thus replace the Kerr-Schild space-coordinates $(x,y,z)$ by new \emph{toroidal} coordinates $(\RR,\KSu,\KSv)$ adapted to a torus with major radius $a$ (i.e.~the distance from the center of the tube to the center of the torus), and minor radius $\RR>0$ (i.e.~the radius of the tube). As shown in Fig.~\ref{torus-fig}, the coordinate $\RR$ is thus radius-like, while the coordinates $(\KSu,\KSv)$ represent rotations about the $z$-axis and rotations around the tube of the torus respectively.

The relation between these toroidal coordinates and Kerr-Schild coordinates is
\begin{align}\label{19VII19.4}
  x &= \left[a + \RR \cos(\KSv)\right] \cos (\KSu)\,,
  \nonumber \\
  y &= \left[a + \RR \cos (\KSv)\right] \sin (\KSu)\,,
  \nonumber \\
  z &= \RR \sin (\KSv)
   \,.
\end{align}
In these coordinates the Euclidean line element takes the form
\begin{equation}\label{20VII19.1}
  \diff x^2 + \diff y^2 +  \diff z^2 = \diff \RR^2 + \RR^2 \diff \KSv ^2 + \left[a + \RR \cos(\KSv) \right]^2 \diff \KSu^2
  \,.
\end{equation}

As already mentioned, the function $H$  vanishes on the coordinate disc \xout{$x^2 + y^2 < a^2$ lying on the equatorial hyperplane $z=0$} \greenn{$\{\rho<a\,,\ z=0\}$}.
Inside the ring, and using the coordinates of Eq.~\eqref{19VII19.4}, we then find
\begin{equation}\label{22VII19.1}
 g_{\rm induced}  = -\diff t^2 + \diff \RR^2 + (a - \RR)^2 \diff \KSu^2
  \,,
\end{equation}
where the minus sign inside the parenthesis arises because inside the disk $\KSv  = \pi$.
One recognizes this metric as that of the Minkowski spacetime in $2+1$ dimensions, as already observed previously.
Outside the ring and on the equatorial hyperplane, on the other hand, one finds
\begin{align}\label{22VII19.1out}
 {g_{\rm induced}}   &= -\diff t^2 +\diff \RR^2 + (a + \RR)^2 \diff \KSu^2
\nonumber \\
&
 \red{\pm}\frac{2 m \left[\diff \RR
   \sqrt{\RR (2 a+\RR)}-(a+\RR) (a
   \diff \KSu -\diff t)\right]
   ^2}{(a+\RR)^2 \sqrt{\RR (2
   a+\RR)}}
  \,,
\end{align}
\red{with the sign in the last line coinciding with that of $\tilde r$.}
The Boyer-Lindquist radial coordinate $\tilde{r}^{2}$ has two solutions in toroidal coordinates
\begin{equation}\label{21VII19.12}
  \tilde{r}^2 =
  \frac{1}{2} \RR \left(\pm \sqrt{4 a^2+4 a \RR \cos (\KSv)+\RR^2}+2 a \cos (\KSv)+\RR\right)
   \,.
\end{equation}
Clearly the solution in Eq.~\eqref{21VII19.12} with the plus sign is relevant for small $\RR$, and hence everywhere by continuity. Alternatively, Eq.~\eqref{21VII19.12} can be rewritten as
\begin{align}\label{21VII19.12-new}
  \tilde{r}^2 =
  \frac{1}{2} \RR & \left(\pm  \sqrt{4 a^2\sin^2  (\KSv)+ ( 2a  \cos (\KSv)+\RR)^2 }
  \right.
  \nonumber \\
  & \left. +2 a \cos (\KSv)+\RR\right)
   \,,
\end{align}
which shows that the negative sign would lead to a negative $\tilde r^2$.

Some care must be taken, however, concerning the sign of $\tilde{r}$, since Eq.~\eqref{21VII19.12}
obviously leads to
\begin{equation}\label{21VII19.12+}
 \tilde{r} = \pm \sqrt{
  \frac{1}{2} \RR \left(\sqrt{4 a^2+4 a \RR \cos (\KSv)+\RR^2}+2 a \cos (\KSv)+\RR\right)
   }
\,,
\end{equation}
where we have adopted the plus sign of Eq.~\eqref{21VII19.12} as discussed above.

One way of rewriting Eq.~\eqref{21VII19.12+} is\footnote{We are grateful to J\'er\'emie Joudioux for a suggestion which led to this form of $\tilde{r}$.}
\begin{equation}\label{26VII19.31}
  \tilde{r} = \pm a \sqrt{\RR} \left( \sqrt{ e^{i \KSv} + \frac{\RR}{2a}}
+ \sqrt{ e^{-i \KSv} + \frac{\RR}{2a}}
 \right)
 \,,
\end{equation}
where the  square root is taken to have a cut on the negative real axis. The formula can be checked by comparing the square of Eq.~\eqref{26VII19.31} with Eq.~\eqref{21VII19.12}. Equation~\eqref{26VII19.31} makes clear the direct relation of the covering construction of the previous section with the  double-cover of $\C^*$ defined by the complex square-root function.

\subsection{The periodicity question}
\label{ss22VII19.2}

We already saw in Sec.~\ref{ss22VII19.1} that $\tilde{r}$ must change sign as one crosses the equatorial hyperplane in order for the metric to remain differentiable, so let us study this further with our new toroidal coordinates. After removing an overall factor $\sqrt{\RR}$, a Taylor expansion of Eq.~\eqref{21VII19.12+} for small $\RR$ presents no difficulties except at $\cos(\KSv) = -1$,  where one would need to expand the remaining square root near zero. But we can also write
\begin{equation}\label{25VII19.21}
  \tilde{r}  = \pm \sqrt{ \frac{2 a^2 \RR \sin ^2(\KSv  )}{ \sqrt{4 a^2+4 a \RR \cos (\KSv  )+\RR^2}-2 a \cos (\KSv  )-\RR}}
 \,,
\end{equation}
which one can prove is equal to Eq.~\eqref{21VII19.12+} by taking the ratio of both expressions and simplifying. After removing again a multiplicative factor of $\sqrt{\RR}$, Eq.~\eqref{25VII19.21} can be Taylor-expanded in $\RR$ near $\cos(\KSv)=-1$ in a straightforward manner.

 It turns out that  the most  useful form of $\tilde{r}$ is obtained by writing $\sqrt{\sin^2(\KSv)} =\pm  2 \cos(\KSv/2)\sqrt{\sin^2(\KSv/2)}$ and   choosing the signs  so that Eq.~\eqref{25VII19.21} becomes
\begin{align}\label{25VII19.22}
  \tilde{r}  &=   \sqrt{  2 a\RR }\cos(\KSv/2)
  \nonumber \\
  &\times \underbrace{
 \sqrt{ \frac{4a  \sin^2  (\KSv/2 )}{ \sqrt{(2a \cos (\KSv  )+ \RR)^2+4 a^2\sin^2 (\KSv)}-(2 a \cos (\KSv  )+\RR)}}
}_{=:\Psi}
 \,.
\end{align}
The point here is that the function $\Psi$ is
\begin{enumerate}
 \item manifestly $2\pi$-periodic in $\KSv$;
 \item a smooth function of all its variables near $\cos(\KSv)=-1$ for $\RR \ge  0$ (obviously so for small $\RR$), with $\Psi|_{\RR=0}=1$;
 \item and smooth in all its variables away from $\cos(\KSv)=-1$ for
$0\le \RR < 2a $.
\end{enumerate}
The third property is not obvious by simply staring at Eq.~\eqref{25VII19.22}, but it does become so after dividing Eq.~\eqref{21VII19.12+} by $\sqrt{\RR} \cos(\KSv/2)$, noting that  Eq.~\eqref{21VII19.12+} divided by $\sqrt{\RR}$ is manifestly smooth in all its variables away from $\cos(\KSv)=-1$ for $\RR\ge 0$, and that $\cos(\KSv/2)$ is smooth and has no zeros away from  $\cos(\KSv)=-1$. In fact, $\Psi$ has extrema at $\KSv =2n \pi$ where it equals $\sqrt{2a+\RR}/2>0$ for $\RR>a$, and $\KSv =(2n +1)\pi$ where it equals $a/{\sqrt{2a-\RR}} >0$ for $0<\RR<2a$. Physically, of course, we want to restrict the range of $ \RR < a$, as otherwise the coordinates in Fig.~\ref{torus-fig} do not make sense.

The mechanism underlying point 3  above can be illustrated by the following alternative analysis of Eq.~\eqref{25VII19.21}. Let us write
\begin{equation}
\label{18IX19.1}
\Psi^{2}  = \frac{\sin^{2}(\omega )}{\frac{w}{2} \left(1 + \sqrt{1 + h \cos^{2}(\omega ) }\right) - \cos^{2}(\omega )}\,,
\end{equation}
where
\begin{align}
\omega := \KSv/2\,,
\quad
w := 1 - \frac{\RR}{2 a}\,,
\quad
h := \frac{2 \RR}{a \left(1 - \frac{\RR}{2 a}\right)^{2}}\,.
\end{align}
When $\RR \to 0$, we have $h \to 0$ and $w \to 1$, so  the denominator in Eq.~(\ref{18IX19.1}) goes to $1 - \cos^{2}(\omega ) = \sin^{2}(\omega)$, which is compensated nicely by the numerator.

For further reference, we note that the equalities
$$
 \Psi(\RR, \KSv +2\pi)= \Psi(\RR, \KSv) = \Psi(\RR, -\KSv)\,,
$$
imply  that
\begin{equation}\label{21IX19.1}
  \tilde{r}(\RR, \KSv +2\pi) = -\tilde{r}(\RR, \KSv)
\,,
 \quad
  \tilde{r}(\RR,-\KSv) = \tilde{r}(\RR, \KSv)
\,.
\end{equation}

Let us then summarize the above results. When viewing $\tilde{r} $ as an independent variable, in  Kerr-Schild coordinates the metric functions are rational functions of $(x,y,z,\tilde{r})$,  and smooth in their arguments away from $\{\tilde{r}=0 \}$. Moreover, the function $\tilde{r}^2$ is a smooth function of $(x,y,z)$ away from $\{\RR=0\}$, and the function $\tilde{r}$ factors out as $\sqrt{2a \RR}\cos(\KSv/2)\Psi$, where $\Psi$ is a smooth function of $(x,y,z)$ away from $\{\RR=0\}$
for, say, $\rho<a$, and Eq.~\eqref{21IX19.1} holds.

 After writing the metric function $H$ as
\begin{equation}\label{2X19.1}
H =  \tilde{r} \times \frac{2 m \tilde{r}^2}{\tilde{r}^4 + a^2 z^2}
\,,
\end{equation}
we conclude that, in Kerr-Schild coordinates, all metric functions can either be written in the form $\sqrt{\RR} \cos (\KSv/2) $ times  a function  which is manifestly smooth in  $(x,y,z)$ away from $\{\RR=0\}$, or are directly smooth in  $(x,y,z)$ away from $\{\RR=0\}$.

\red{As such, in the original definition \eqref{H-eq}, the metric function $H$ was defined only up to a sign, depending upon whether the negative of positive sign of $\tilde r$ has been chosen. From now on  we view the function $H$ as a well-defined function on the double covering space, given by \eqref{2X19.1}, with $\tilde r$ defined there by \eqref{25VII19.22}.}

In view of Eq.~\eqref{21IX19.1} we have
\begin{equation}\label{21IX19.2}
  H (\RR, \KSv +2\pi)= -H(\RR, \KSv)
\,,
 \quad
  H (\RR,-\KSv )= H (\RR, \KSv)
\,.
\end{equation}

A geometric understanding of the global structure of the spacetime is obtained by passing to the covering space mentioned earlier, but the behavior of the metric \emph{near the ring} $\{\RR=0\}$ becomes transparent after declaring that
\begin{equation}
\label{21IX19.4}
\nonumber
 \mbox{{$\KSv$ is a coordinate which is $4\pi$, rather than $2\pi$, periodic.}}
\end{equation}
The careful reader at this point may ask: ``wait, what?'' Before, we defined the coordinate $\KSv$ as a periodic quantity that measured the interior angle of the torus in Fig.~\ref{torus-fig}, and now we are saying that this angle is $4 \pi$-periodic instead of $2 \pi$-periodic? To elucidate this seeming contradiction, consider the following thought experiment. Imagine an observer in {a}
 spacecraft, at a fixed angle $\KSu$ and fixed distance $\RR$, going around the torus on a trajectory that starts at $\KSv  = 0$ at some time $t_{0}$ (event A), reaches $\KSv  = 2 \pi$ at some time $t_{1}$ (event B), and continues on to end at $\KSv  = 4 \pi$ at some time $t_{2}$ (event C). When the observer reaches event B, she is not back where she started, i.e.~the spatial part of event $B$ is not the same as the spatial part of event A. This is because as we go around the torus once, the Boyer-Lindquist coordinate $\tilde{r}$ flips sign once. To get back to where she started, the observer needs to go around a full $4 \pi$ in the $\KSv$ coordinate, forcing $\tilde{r}$ to flip sign twice.

Mathematically, the source of this $4 \pi$-periodicity can already be seen in Eq.~\eqref{25VII19.3}, back when we were working in Kerr-Schild coordinates. As we saw in that section, the function $H$ actually vanishes on the disk $\rho < a$, which implies that the induced metric is flat there, and thus, the observer can easily traverse through this region as she goes from event A to event B and then event C. However, as the observer goes through the disk (from $z>0$ to $z<0$), the $\tilde{r}$ coordinate flips sign due to the absolute value operator in Eq.~\eqref{25VII19.3}. Since the metric function $H$ (and thus the line element) depends on (an odd power of) $\tilde{r}$, the only way to extend smoothly (i.e.~not just continuously but also differentially) through $z =0$ is to flip the sign of $\tilde{r}$ once every time one goes through the disk. The realization that the metric is $4 \pi$-periodic in $\KSv$ guarantees that this flipping-sign condition on $\tilde{r}$ is automatically enforced.

\rout{Moreover,} \red{We further note that} the second equalities in Eqs.~\eqref{21IX19.1} {and} \eqref{21IX19.2} make it clear that the map
$  \KSv\mapsto - \KSv $ is an isometry \rout{both of the original metric, and} of the metric on the covering manifold.

While the Kerr-Schild coordinates have an obvious geometric character in the asymptotic region, where $r=\sqrt{x^2 + y^2 + z^2}\to \infty$, they are completely misleading near the ring, where the natural periodicity of $\KSv$ is $4\pi$.

\begin{figure*}
\begin{center}
\includegraphics[width=0.9\columnwidth]{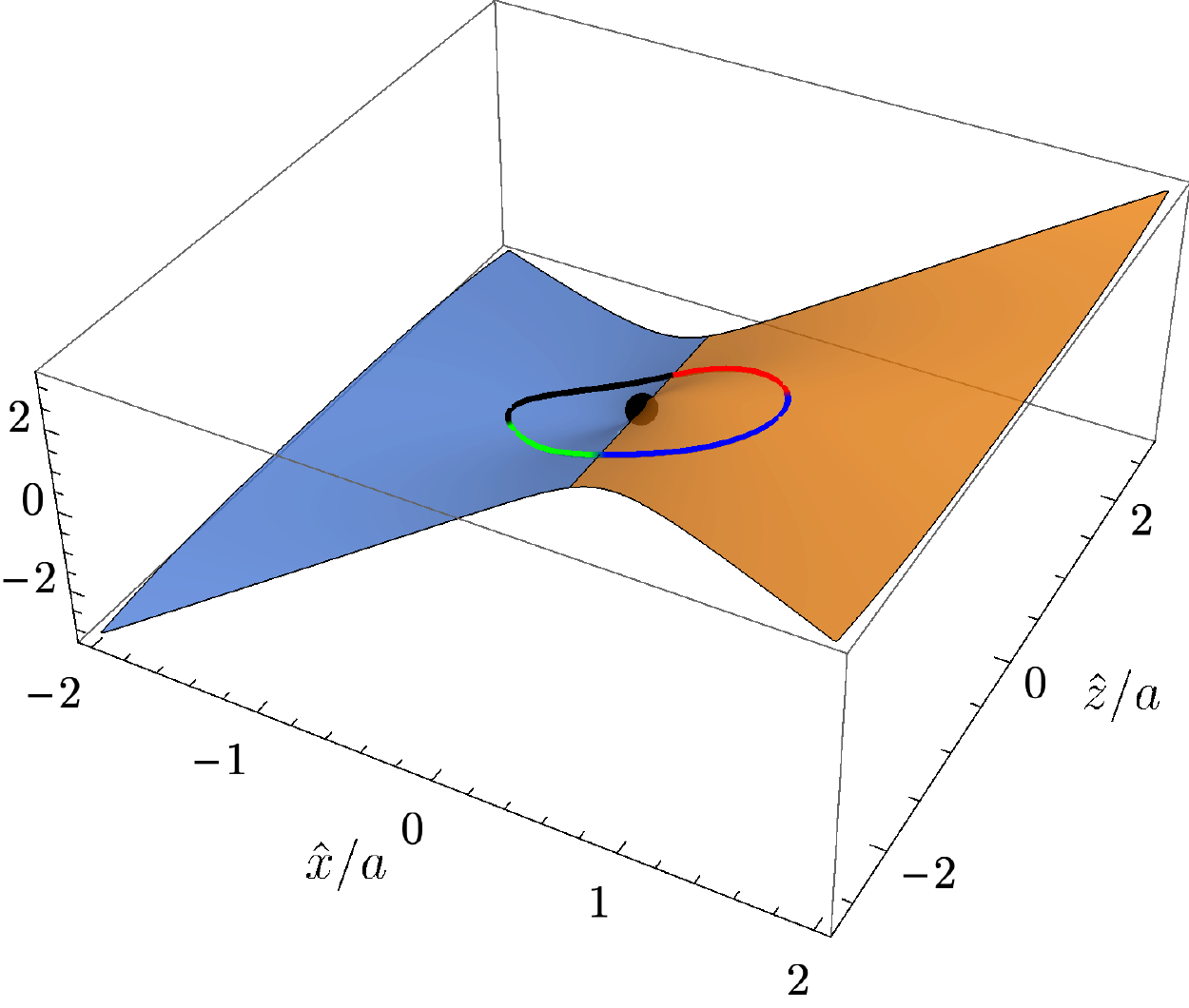}
\hspace{10ex}
\includegraphics[width=0.9\columnwidth]{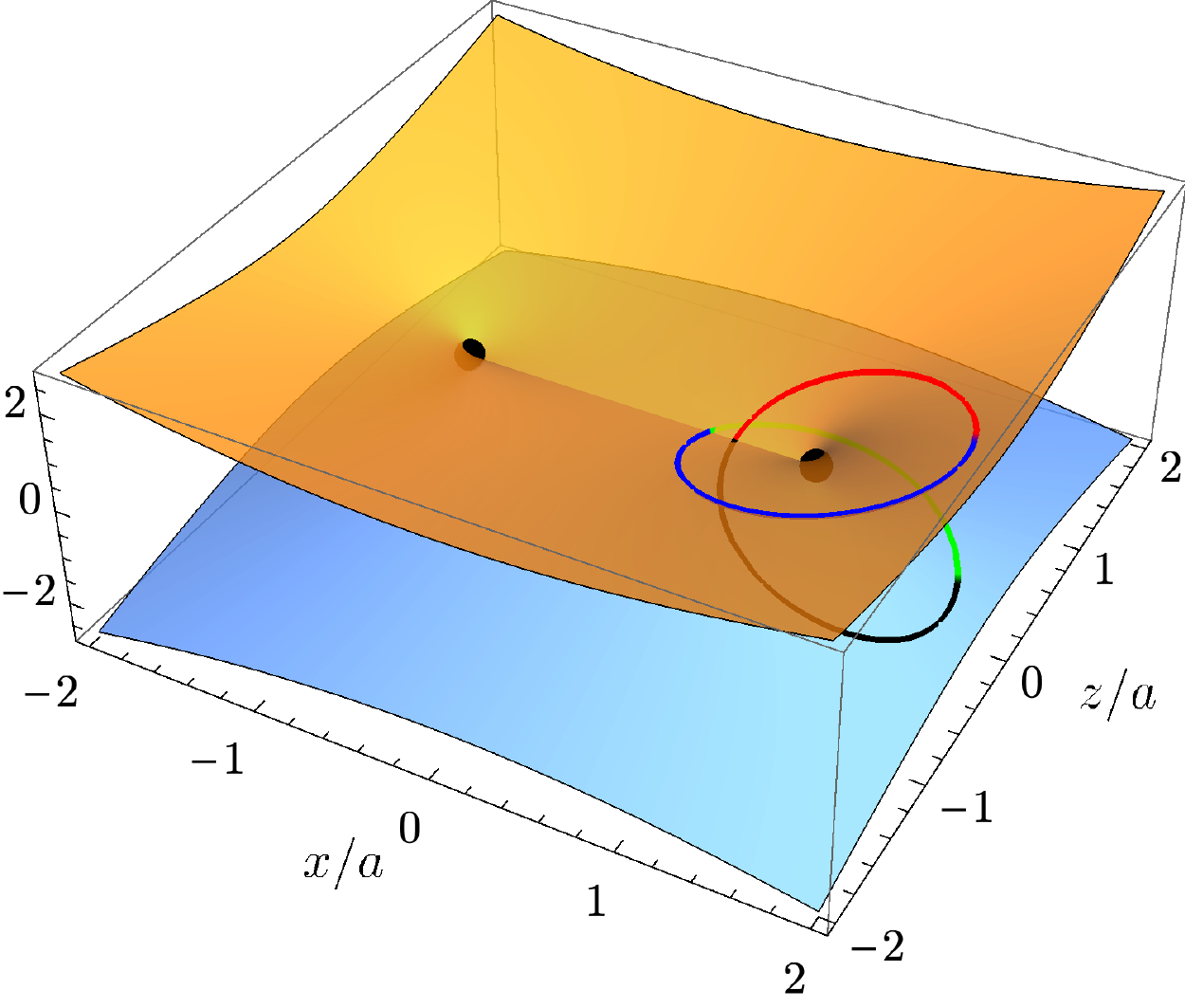}
\caption{Left: Graph of $\tilde{r}/a$ over the image of the region $\{x>0,y=0\}$ in $(\hat x,\hat z)$ coordinates. Right: Graphs of  $\pm \tilde{r}$, parametrized by $(x, z)$ coordinates, over the hyperplane $\{t=0,\, y=0\}$. The location of the ring is indicated by black dots. The red curve corresponds to $\KSv \in [0, \pi)$ ($\tilde{r}>0$), the black curve to $\KSv \in (\pi,2 \pi]$ ($\tilde{r}<0$), the green curve to $\KSv \in (2 \pi,3 \pi)$ ($\tilde{r}<0$) and the blue curve to $\KSv \in (3\pi,4 \pi]$ ($\tilde{r}>0$). Observe that the closed curve around the ring in the $(\hat x,\hat z)$ coordinates on the left panel lifts to the curve shown on the right panel. The right plot incorrectly suggests that there is only one equatorial disc: a faithful visualization is provided by the left figure, which shows  clearly  that the extension contains two copies of the equatorial disc, which correspond to the sets $\{\hat x=0,\, \hat z>0\}$ and $\{\hat x=0,\, \hat z<0\}$, as well as two copies of the exterior equatorial hyperplane, which correspond to the sets $\{\hat z=0,\, \hat x>0\}$ and $\{\hat z=0,\, \hat x<0\}$.}
\label{F25IX19.1}
\end{center}
\end{figure*}

\rout{The geometrically correct picture can be captured}
\red{Yet another way of capturing the geometrically correct picture proceeds} by introducing on the hyperplane $\{y=0\}$, near the ring, coordinates $(\hat x, \hat z)$ defined as
\begin{equation}\label{23IX19.5}
  \hat x = \RR \cos (\KSv/2)
  \,,
  \quad
  \hat{z} = \RR \sin (\KSv/2)
  \,.
\end{equation}
In these coordinates the function $\tilde{r}$ is positive on the half-space $\{\hat x>0\}$ and negative on $\{\hat x< 0\}$. The Kerr-Schild coordinates $(x,z)$ provide separate coordinate systems on the half-planes $\{t=\const,\, x>0,\, y=0\}$ for both regions $\tilde{r}>0$ and $\tilde{r}<0$. Each of them corresponds  to half of the region
\begin{eqnarray}
\nonumber
 \lefteqn{
 -   \sqrt{\frac{  a^2 + 2\hat z ^2 -   a \sqrt{a^2+8\hat z ^2 }}{2}}
}
&&
\\
 &&
< \hat x <    \sqrt{\frac{  a^2 + 2\hat z ^2 -   a \sqrt{a^2+8\hat z ^2 }}{2}}
\label{25IX19.31}
\end{eqnarray}
on the  $(\hat x, \hat z)$ plane. In Fig.~\ref{F25IX19.1} we show how a curve circling around the origin of the $(\hat x, \hat z)$ coordinates lifts to the two branches $\pm \tilde{r}$ in the $(x,z)$ coordinates.
%

\subsection{The curvature singularity}
\label{s19VII19.2}

The standard way to show that the Kerr solution is $C^2$-inextendible across the ring $\RR=0$ proceeds by inspection of the Kretschmann scalar, which in Boyer-Lindquist equals
\begin{align}
  K &:= R_{\alpha\beta\gamma\delta} R^{\alpha\beta\gamma\delta}
\nonumber   \\
  &= 48 m^{2} \left(\tilde{r}^{2} - a^{2} \cos^{2}(\BLtheta) \right)
  \nonumber \\
  & \times \frac{\left[\left(\tilde{r}^{2} + a^{2} \cos^{2}(\BLtheta)\right)^{2} - 16 a^{2} \tilde{r}^{2} \cos^{2}(\BLtheta) \right]}{\left(\tilde{r}^{2} + a^{2} \cos^{2}(\BLtheta)\right)^{6}}\,,
\end{align}
while in the toroidal coordinates introduced above, it equals
\begin{align}
\nonumber
  K &= \frac{48 m^2}{\RR^{3}} (2 a \cos (\KSv)+\RR)
  \nonumber \\
  &\times \frac{\left(8 a^2 \cos (2 \KSv)-4 a^2+4 a \RR \cos (\KSv)+\RR^2\right)}{\left(4 a^2+4 a \RR \cos (\KSv)+\RR^2\right)^3}
  \,.
  \label{19VII19.1}
\end{align}
The invariant $K$ clearly tends to infinity on generic curves approaching the set $\{\RR=0\}$. However, there exist curves approaching this set on which it does not. For example,
on any curve lying on the hypersurface $2 a \cos (\KSv)+\RR =0$ we have $K \equiv 0$. Clearly then, $K$ by itself does not establish the singular character of the set $\{\RR=0\}$.

The  usual resolution of this  is by showing  that the only causal geodesics which accumulate at the singular ring lie
  entirely in the equatorial hyperplane $ {\cos}(\KSv)=0$
  (cf.~\cite[Proposition~4.15.9]{BONeillK}), which is a rather heavy exercise. Here we note that this issue can be settled in a much simpler way by using the invariant
\begin{align}
\nonumber
  P &:=\epsilon_{\alpha\beta\mu\nu}R^{\mu\nu}{}_{\gamma\delta} R^{\alpha\beta\gamma\delta}\,,
  \\
\nonumber
   &=  \!- 192 a m^2 \tilde{r} \cos (\BLtheta)\frac{
   \left(\!\tilde{r}^{2} \! - \! 3 a^{2} \cos^{2}(\BLtheta)\! \right) \!
   \left(\!3 \tilde{r}^{2} \!-\! a^{2} \cos^{2}(\BLtheta) \!\right)
   }{\left(\!\tilde{r}^2+a^2 \cos^{2}(\BLtheta)\!\right)^6}\,,
  \\
  &= \!- \frac{192 a m^{2}}{\RR^{3}} \frac{\left(4 a^2 \!+ \!3 \RR^2 \!+ \!12 a \RR \cos{(\KSv)} \!+\! 8 a^2 \!\cos(2\KSv)\right)}{\left(4 a^2 + 4 a \RR \cos{(\KSv)} + \RR^2 \right)^3} \sin(\KSv) \,,
\label{19VII19.2}
\end{align}
where the second line is in Boyer-Lindquist coordinates and the third line in toroidal coordinates.

A relatively simple calculation  then reveals that
\begin{align}\label{19VII19.3}
 K^2 + P^2 &=
  \frac{2304 m^4}{\RR^6 \left(4 a^2+4 a \RR \cos (\KSv)+\RR^2\right)^6}
  \times
  \nonumber \\
  &
  \left[\left( 8 a^3 \cos (3 \KSv) +6 a \RR (2 a \cos (2 \KSv)+\RR \cos(\KSv))+\RR^3\right)^2
  \right.
  \nonumber \\
  &\left.
  +16 \left(4 a^3 \sin (3 \KSv)+3 a \RR \sin (\KSv) (4 a \cos
   (\KSv)+\RR)\right)^2\right]
  \,.
\end{align}
An asymptotic expansion about $\RR = 0$ yields
\begin{align}
K^{2} + P^{2} &=
 \frac{18 m^4}{a^6 \RR^6} \left[5 - 3 \cos{(6 \KSv)}\right]
\nonumber \\
&+ \frac{54 m^4}{a^7 \RR^5} \left[-5 \cos{(\KSv)} + 3 \cos{(7 \KSv)}\right] + \bigO{\RR^{-4}}
\end{align}
Clearly, the first term diverges when $\RR \to 0$ unless $\cos{(6 \KSv)} = 5/3$, which is not in the reals.
This establishes explicitly the uniform divergence of the curvature as the singular set is approached.

We note here in passing that the above analysis does not necessarily prove that the
spacetime is geodesically incomplete, as we have not yet shown that geodesics can reach the curvature
singularity in finite affine time, a point that will be revisited in Sec.~\ref{s1X19}.

\section{The Kerr metric near the singular ring}
\label{sec:near-the-ring}

We would now like to \red{further} understand the geometry of the Kerr metric near the singular ring.
For this we need to study the small-$\RR$ behavior of the metric. The first step in this direction is provided by an analysis of the small-$\RR$ behavior of $\tilde{r}$ and $H$.
We follow this with a calculation of the behavior of the Killing vectors near the ring, as well as of the quotient space metric, and an analysis of some properties of two different natural families of hypersurfaces.

\subsection{Small-$\RR$ expansions of $\tilde{r}$ and $H$}

We wish to  investigate several of the quantities introduced in previous sections, but in a small $\RR$ expansion. Before proceeding, however, let us first discuss which form of the function $\tilde{r}$ to adopt. The prescription of the previous section, which explained how to extend $\tilde{r}$ to the covering manifold, leads to an analytic metric. Hence, whenever an expression of the form $\sqrt{(1+\cos(\KSv))/2}$ occurs in explicit calculations (which it often does), it should be replaced by $\cos(\KSv/2)$, because this is correct for $\cos(\KSv/2)>0$, and is therefore the correct formula for all $\KSv$ by analyticity of the metric. The alternative systematic replacement by $-\cos(\KSv/2)$ produces an isometric spacetime, with the isometry provided by a shift of $\KSv$   by $2\pi$.

With this in mind, one can readily check that
\begin{equation}\label{27VII19.1}
  \tilde{r} = \sqrt{2a \RR}\cos(\KSv/2) + \bigO{\RR^{3/2}}
 \,.
\end{equation}
Furthermore, the function $H$ appearing in Eq.~\eqref{21VI119.14} can be written in the following form, which becomes manifestly smooth near $\KSv =\pi$ for small $\RR $ after multiplying by $\sqrt{\RR}$, \red{keeping in mind the already-established smoothness of the function $\Psi$ defined in \eqref{25VII19.22}}:
\begin{align}\label{3VIII19.11}
  H &= \frac{  \cos (\KSv/2) }{\sqrt{\RR  }}
 \frac{4 a m   }{\sqrt{ 4 a^2+4 a \RR \cos (\KSv)+\RR^2 }}
 \nonumber \\
 &\times \sqrt{\frac{2  \sin^2 (\KSv/2)}{\sqrt{4 a^2+4 a \RR \cos (\KSv)+\RR^2}-2 a
   \cos (\KSv)-\RR}}
\,.
\end{align}
Taylor expanding for small $\RR$, one obtains
\begin{align}\label{27VII19.2}
  H  &=  m  \sqrt{\frac 2 { a\RR}} \cos(\KSv/2) \left(1+ \bigO{\RR} \right)
  \nonumber \\
  &= \frac{m (\cos(\KSv) + 1)}{\tilde{r}} \left(1 + \bigO{\RR}\right)
 \,.
\end{align}
The leading-order behavior of each component of the one-form $\KStheta$ is
\begin{equation}\label{27VII19.2-b}
 \KStheta \approx
  (-\diff t+a \diff \KSu )-\diff \RR  \sqrt{\frac{2\RR}{a}}\cos(\KSv/2)
 + \sqrt{\frac{2 \RR^{3}}{a}} \diff \KSv   \sin(\KSv/2)
 \,.
\end{equation}
Expansions to one order higher of $\KStheta$, as well as of the remaining metric functions, can be found in Appendix~\ref{A19IX19.1}.

\subsection{Killing vectors near the ring}
\label{s6IX19.1}

The Kerr metric possesses  two Killing vector fields. In Boyer-Lindquist coordinates, the Killing vector $\partial_\BLphi$ is defined uniquely by the requirement that its orbits are closed, with period $ 2 \pi$, while the Killing vector $\partial_\tildet $ is defined uniquely by the requirement that it generates time translations in the asymptotically flat region. Hence, the scalar products $g(\partial_\tildet, \partial_\tildet)$, $g(\partial_\BLphi,\partial_\BLphi)$ and $g(\partial_\tildet,\partial_\BLphi)$ are geometrically-defined scalar functions on the Kerr spacetime. It is therefore of interest to inquire how these functions behave near the ring singularity. The reader will note that we have
\begin{align}
\partial_{\tildet} &= \partial_{t}\,,
\quad
\partial_{\BLphi} = \partial_{\KSu}\,.
\end{align}

Let us begin by considering the causal character of the orbits of the isometry group of the metric,
generated by time translations and by rotations.
If the two-dimensional determinant of the metric induced on the orbits of the group
vanishes, a horizon must be present, while if the metric is Lorentzian,
one can always find a Killing vector that is timelike, so the metric is stationary in this region.
Alternatively, if the induced metric were Riemannian, then there would be no timelike Killing
vectors there, and the spacetime would be dynamical in this region.

In Boyer-Lindquist coordinates, the determinant of the two-dimensional matrix of scalar products of the Killing vectors equals
$$
  g_{\tildet \tildet } g_{\BLphi \BLphi} - g_{\tildet \BLphi}^2 = - \DeltaBL \sin^2(\BLtheta) = -(\tilde{r}^2 -2m \tilde{r}+a^2)\sin^2(\BLtheta)
\,.
$$
Since $\tilde{r}$ tends to zero and $\sin^2 (\BLtheta)$ tends to one as the singular ring is approached, we have
\begin{equation}\label{4VIII19.90}
 \lim_{\tilde{r} \to 0, \, \sin^2(\BLtheta) \to 1} \left(g_{\tildet \tildet } g_{\BLphi \BLphi} - g_{\tildet \BLphi}^2
 \right)= -a^2
 \,,
\end{equation}
and thus the orbits of the group are timelike near the ring.
This can be confirmed by a direct calculation in our toroidal coordinates:
\begin{align}
\nonumber
   g_{tt} g_{\KSu\KSu}-g_{t\KSu}^2  &= (-1+H)(\eta_{\KSu\KSu} + (\KStheta_{\KSu})^2 H) -  (\KStheta_{\KSu})^2 H^2
\nonumber \\
 &= -\rho^2 +H (\rho^2 -  (\KStheta_{\KSu})^2)
\nonumber
\\
\label{18X19.01}
 & \approx   -a^2 + 2m\sqrt{2 a \RR} \cos(\KSv/2) - 2 \RR a \cos(\KSv)\,,
\end{align}
where the last line has remainders of $\bigO{\RR^{3/2}}$.
This result implies then that \emph{in the region near the singular ring there exist timelike Killing vectors that render the spacetime stationary in this region.}

We continue with an analysis of the periodic Killing vector field $\partial_{\KSu}$, which changes causal character, for small $\RR$, depending upon $\KSv$. To see this, we note that
\begin{equation}\label{3VIII19.41}
  \KStheta_{\KSu} = \frac{-\RR \sqrt{4 a^2+4 a \RR \cos (\KSv)+\RR^2}+2 a^2+2 a \RR \cos (\KSv)+\RR^2}{2 a}
\,,
\end{equation}
which can be used to calculate
\begin{equation}\label{3VIII19.42}
  g_{\KSu\KSu} = \eta_{\KSu\KSu} + (\KStheta_{\KSu})^2 H = (a+\RR \cos (\KSv))^2+  (\KStheta_{\KSu})^2 H
 \,.
\end{equation}
Since $H$ tends to minus infinity as $\RR$ tends to zero when $\cos(\KSv/2)<0$, $g_{\KSu\KSu}$ will indeed be negative for such $\KSv$'s and sufficiently small $\RR$.
This is made clear by the expansion
\begin{equation}\label{28VII19.2}
 g_{\KSu\KSu}
= \frac{a m \sqrt{2 a} \cos(\KSv/2)}{\sqrt{\RR}}+a^2+\bigO{\sqrt{\RR}}
 \,,
\end{equation}
which also shows that, for small $\RR$,  the hypersurface $\{g_{\KSu\KSu}=0\}$, on which the Killing vector $\partial_{\KSu}$ becomes null, asymptotes to the equatorial disc $D$:
\begin{equation}\label{28VII19.2-u}
 \cos    (\KSv/2) =
- \frac{1}{m}\sqrt{\frac{a\RR}{2}} + \bigO{\RR}
 \,.
\end{equation}

\begin{figure}[htb]
\begin{center}
\includegraphics[width=0.69\columnwidth,clip=true]{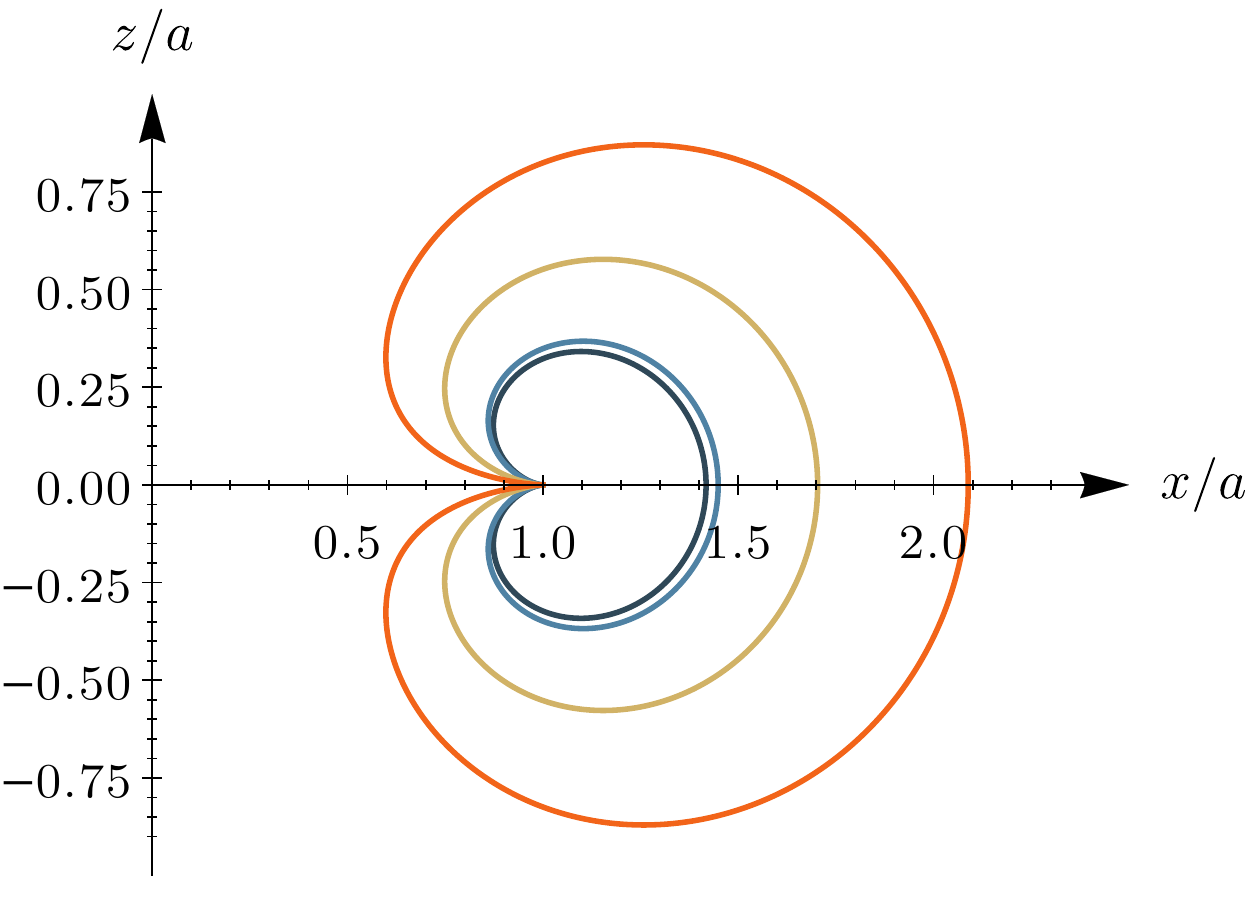}
\includegraphics[width=0.69\columnwidth,clip=true]{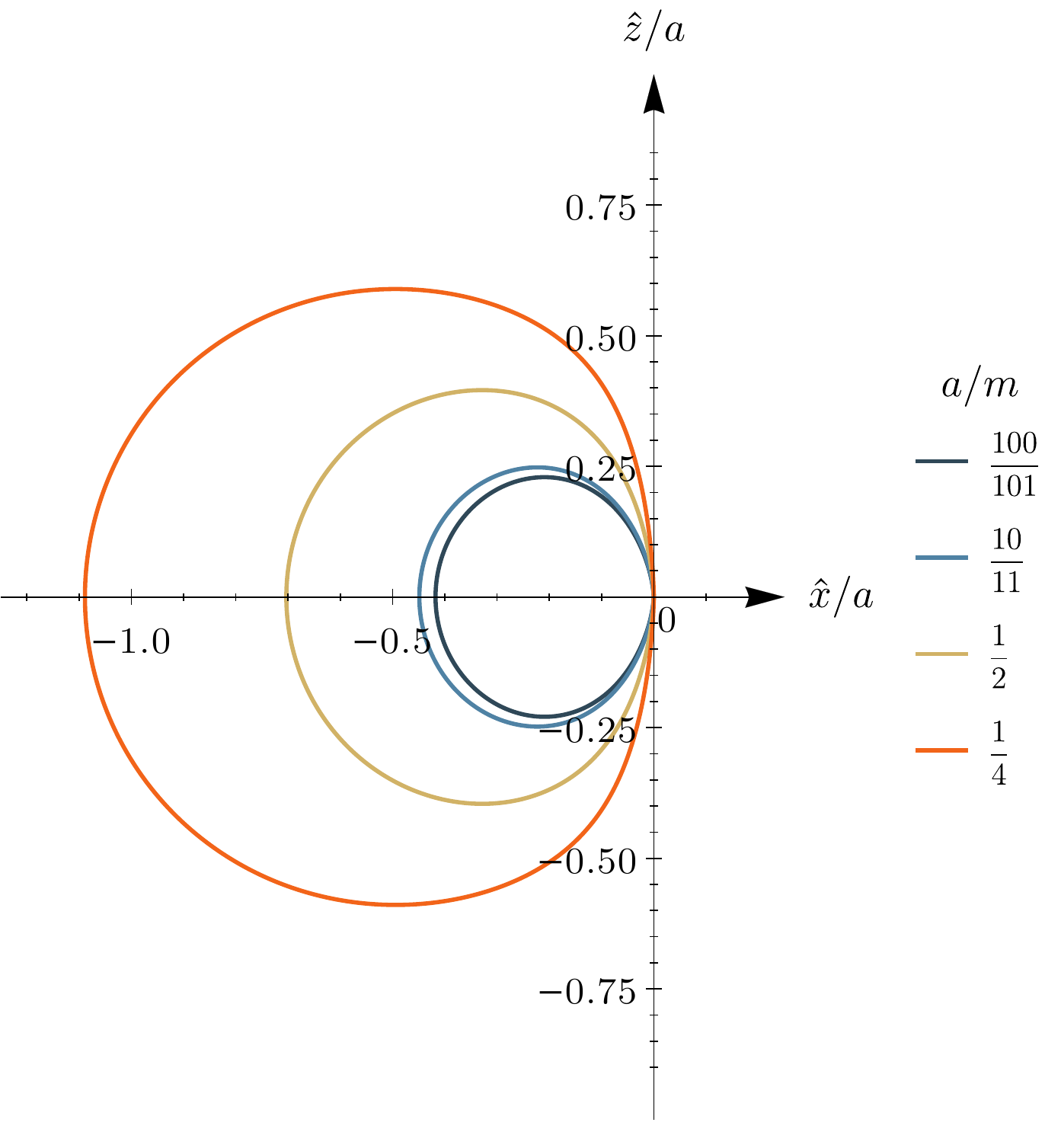}
\includegraphics[width=0.69\columnwidth,clip=true]{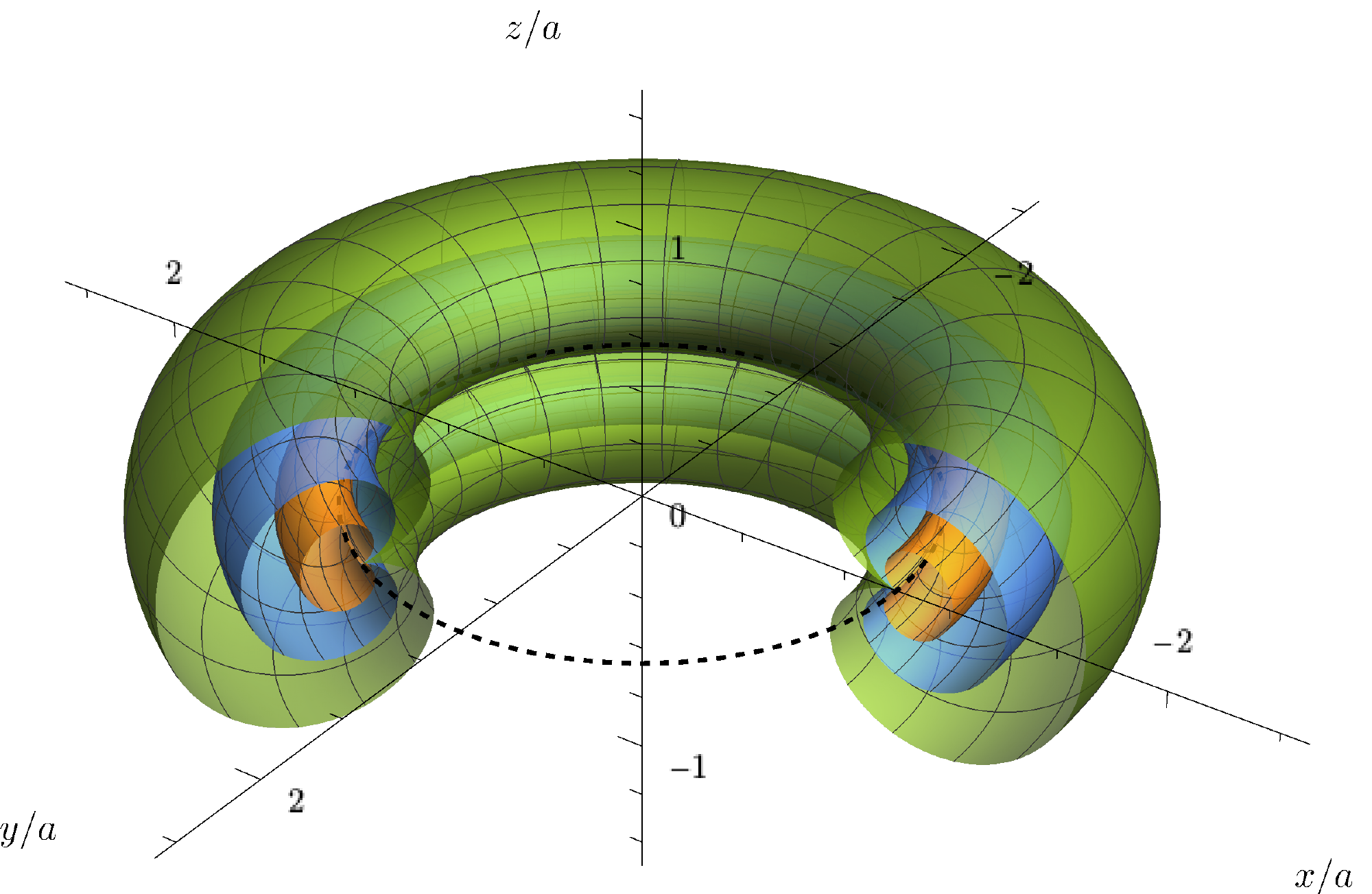}
\caption{Intersection of the boundary of the causality-violating region (i.e.~the region with closed timelike\red{-or-lightlike} curves) with the half-plane $\{ x>0,\, t=0,\, y=0\}$
in Kerr-Schild coordinates (top) and the coordinates \eqref{23IX19.5} (middle), and the three-dimensional boundary close to the singular ring in Kerr-Schild coordinates (bottom),
all in the negative-$\tilde{r}$ region for different values of $a/m$ ($a/m=1/2$ corresponds to green, $1$ to blue, and $2$ to orange/yellow).
The Killing vector $\partial_{\KSu}$ is timelike in the region bounded by the curve and null on the boundary. The ring is located at $(1,0)$ on the top panel, at the origin in the middle panel and on the dotted line in the bottom panel. Observe that the cusp in the top panel disappears in the middle panel.
}
\label{F30VIII19.1}
\end{center}
\end{figure}

The  connected component of the region $g_{\KSu\KSu}<0$ adjacent to the ring, as well as the boundary of the region where $g_{\KSu\KSu}=0$ are shown in Fig.~\ref{F30VIII19.1} for some values of $a/m$, using both the Kerr-Schild coordinates and the coordinates of \eqref{23IX19.5}. From the above analysis, we conclude that \emph{any neighborhood of the ring contains causality-violating, closed timelike curves}.

Na\"ive physical thinking might suggest that, \red{even for macroscopic objects}, an observer should be a Planck {length} away from the ring in order to find closed timelike curves. But for astrophysically relevant black holes this is not the case. Indeed, once the observer is at a distance
\begin{align}
\RR & \approx 2 m \left(\frac{m}{a}\right) \cos^{2}(\KSv/2)\,,
\nonumber \\
& \approx 15 \times 10^{6} \; {\rm{km}} \left(\frac{m}{M_{{\rm SagA}*}}\right) \left(\frac{0.8}{a/m}\right) \cos^{2}(\KSv/2) \,,
\end{align}
where $M_{{\rm SagA}*}$ is the mass of Sag A*, then closed timelike curves are possible, and clearly this distance is macroscopic.
Note also that taking the $a \to 0$ limit of the above expression is not allowed since it derives from a $\RR \ll a$ expansion.
 The quantity $\RR$ is obviously coordinate dependent, but as we will show later in Sec.~\ref{subsec:RRlevels}, 
 this
coordinate does represent the physical distance from the ring singularity for \red{a natural class of} observers close to the ring.

Next, we  continue with an analysis of the Killing vector field
$\partial_{t}$. We have
\begin{equation}\label{28VII19.3-}
 g_{t t}
 = \eta_{t t} + H  = -1 + H
 \,,
\end{equation}
and hence the set where $\partial_{t}$ becomes null satisfies
\begin{equation}\label{28VII19.3-2}
 \{g_{t t}=0\}
 = \{ H  = 1 \}
 \,.
\end{equation}
The set where $\partial_{t}$ becomes spacelike is known as the \emph{ergoregion}, a region inside which timelike  observers cannot help but co-rotate with the singularity. Recall that the part of the ergoregion which lies outside of the event horizon plays a key role in the Penrose process. We conclude from this analysis that \emph{the ergoregion of the Kerr metric has at least two components, one outside the event horizon, and a second one near the singular ring.}

\begin{figure}
\begin{center}
\includegraphics[height=0.25 \textwidth]{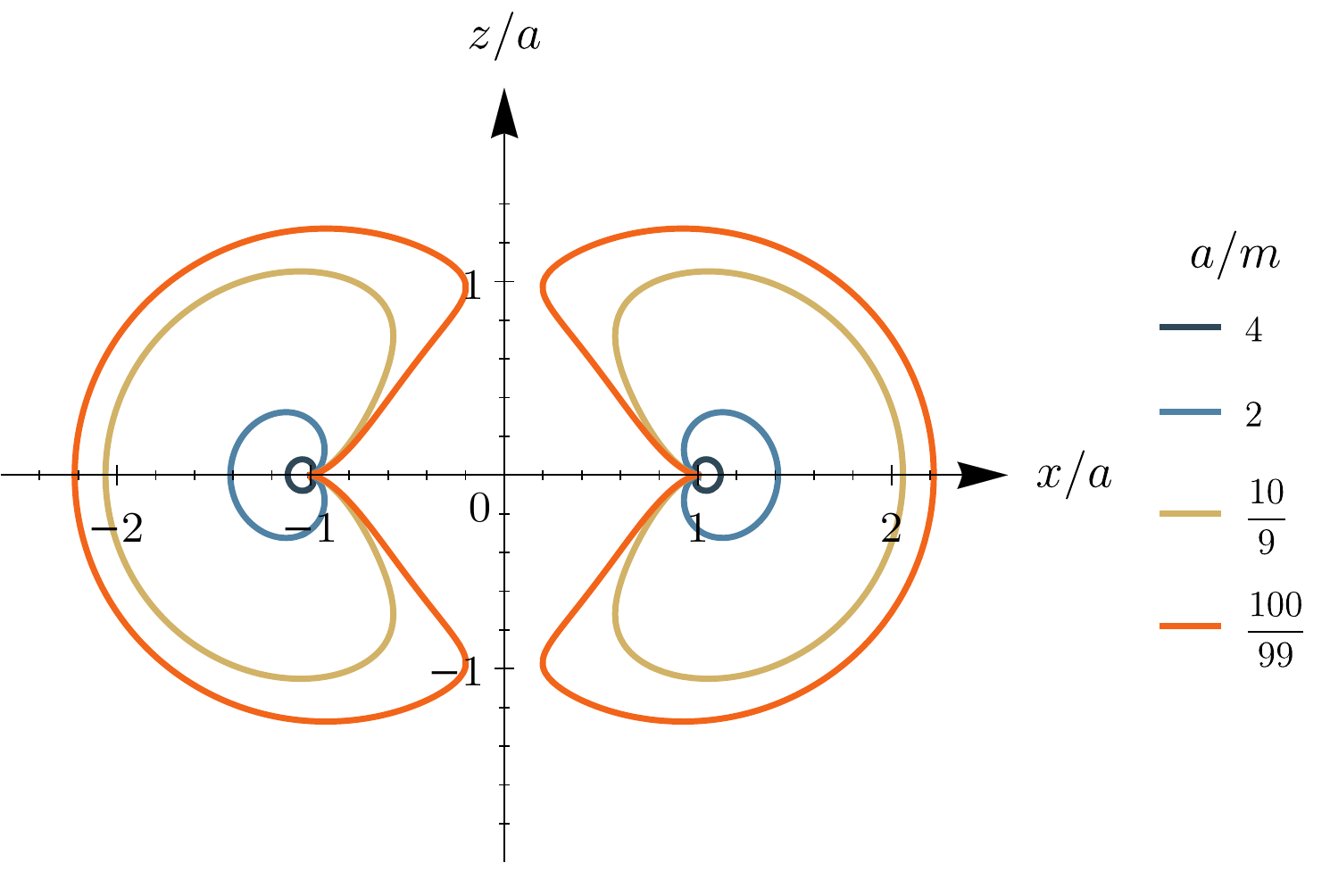}
\hspace{10ex}
\includegraphics[height=0.25 \textwidth]{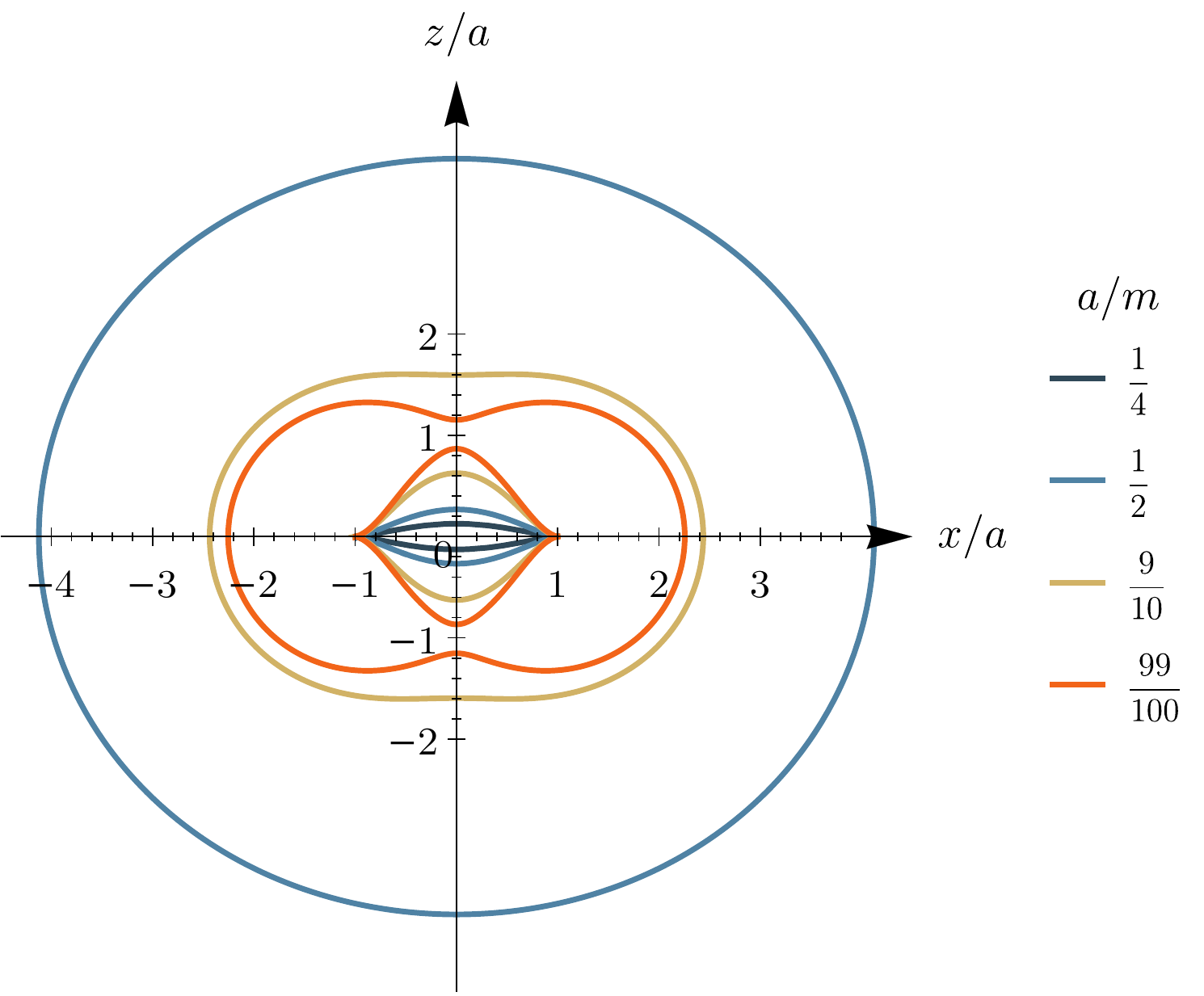}%
\caption{The intersection of the boundary of the ergoregion (i.e.~the region where the Killing vector $\partial_{\tildet}$ changes causal type) with the hyperplane $y=0$,  in Kerr-Schild coordinates in the positive-$\tilde{r}$ region, for $a/m\in\{4,\,2,\,10/9,\,100/99\}$ (top) and $a/m\in\{1/4,\, 1/2,\, 9/10,\, 99/100\}$ (bottom).  The Killing vector $\partial_\tildet $ is spacelike in the region bounded by the curves.  We see that, near the ring, the boundary of the region where $\partial_\tildet$ is spacelike is connected for $a \geq m$ (\emph{not embedded} when $a=m$), while it has two components for $a<m$.
}
\label{F20IX19.1}
\end{center}
\end{figure}

What is the geometry of this second ergoregion? Recall that $H$ is negative in the negative-$\tilde{r}$ region, and therefore $\partial_t$ is timelike throughout that region. But $\partial_{t}$
 changes causal type in the positive-$\tilde{r}$ region for sufficiently small $\RR$:
\begin{equation}\label{28VII19.3}
 g_{tt} =
-1 + H
 =
-1 + m  \sqrt{ \frac{2}{a \RR}} \cos(\KSv/2) + \bigO{\sqrt{\RR}}
 \,,
\end{equation}
with a {coordinate} cusp of the hypersurface $\{g_{tt}=0\}$ at the equatorial hyperplane on the exterior of the disc
\begin{equation}\label{28VII19.2-t}
 \cos    (\KSv/2) =
 \frac 1 m \sqrt{\frac{a \RR}{2 }} + \bigO{\RR}
 \,,
\end{equation}
as shown in Fig.~\ref{F20IX19.1}.  We note that  $\partial_t$ is timelike on the disc itself.
The figure suggests strongly that \emph{the component of the ergoregion adjacent to the ring has the topology of a solid torus for $a<m$, and of a
 hollow marble
when $a>m$}, but we have not attempted  a formal proof of this.

\bigskip

\subsection{Flowing along isometries near the ring}

\rout{The isometry structure of the spacetime was already discussed in the Introduction.}
Recall that $G$ is the connected component of the identity of the isometry
group of the Kerr metric, \red{as discussed in the Introduction}. Given $G$, a useful two-dimensional geometry is that
which is defined by the tensor field $\chi$ induced by the spacetime metric $g$ on the orbits of $G$.

\begin{figure}[htb]
\begin{center}
\includegraphics[width=0.7\columnwidth]{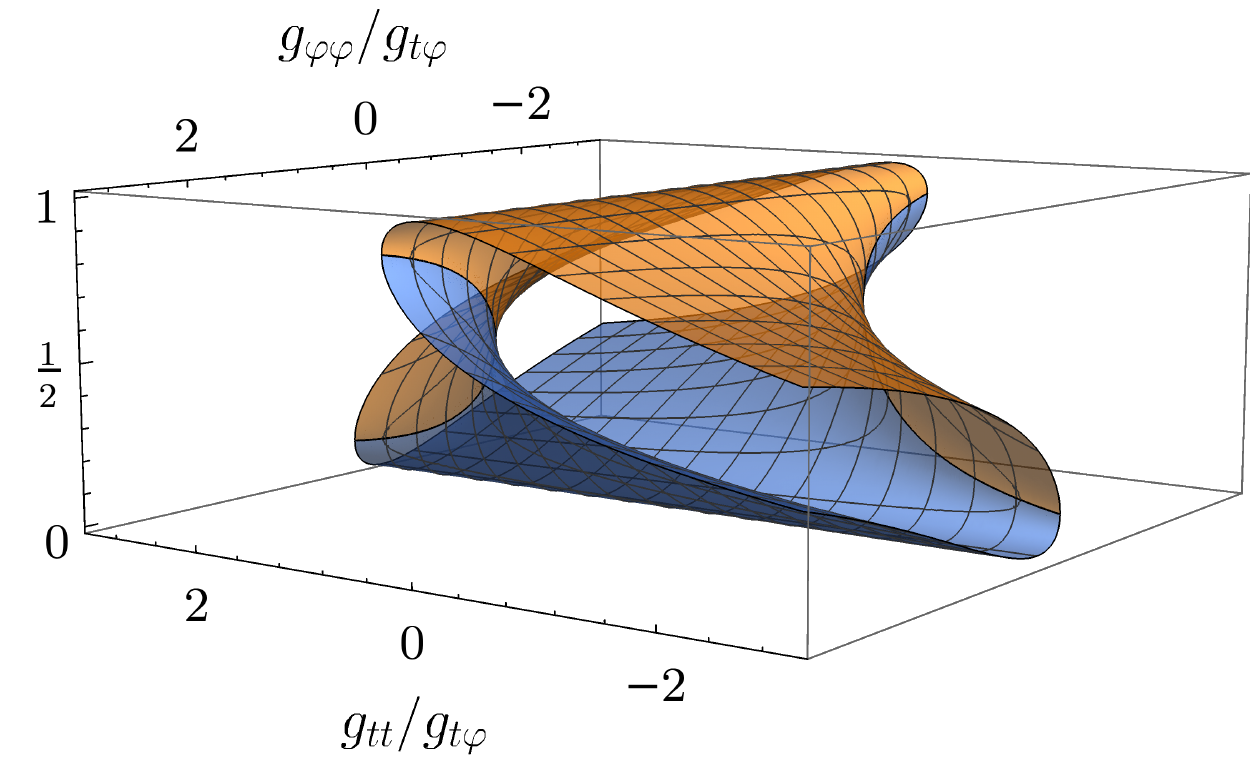}
\includegraphics[width=0.7\columnwidth]{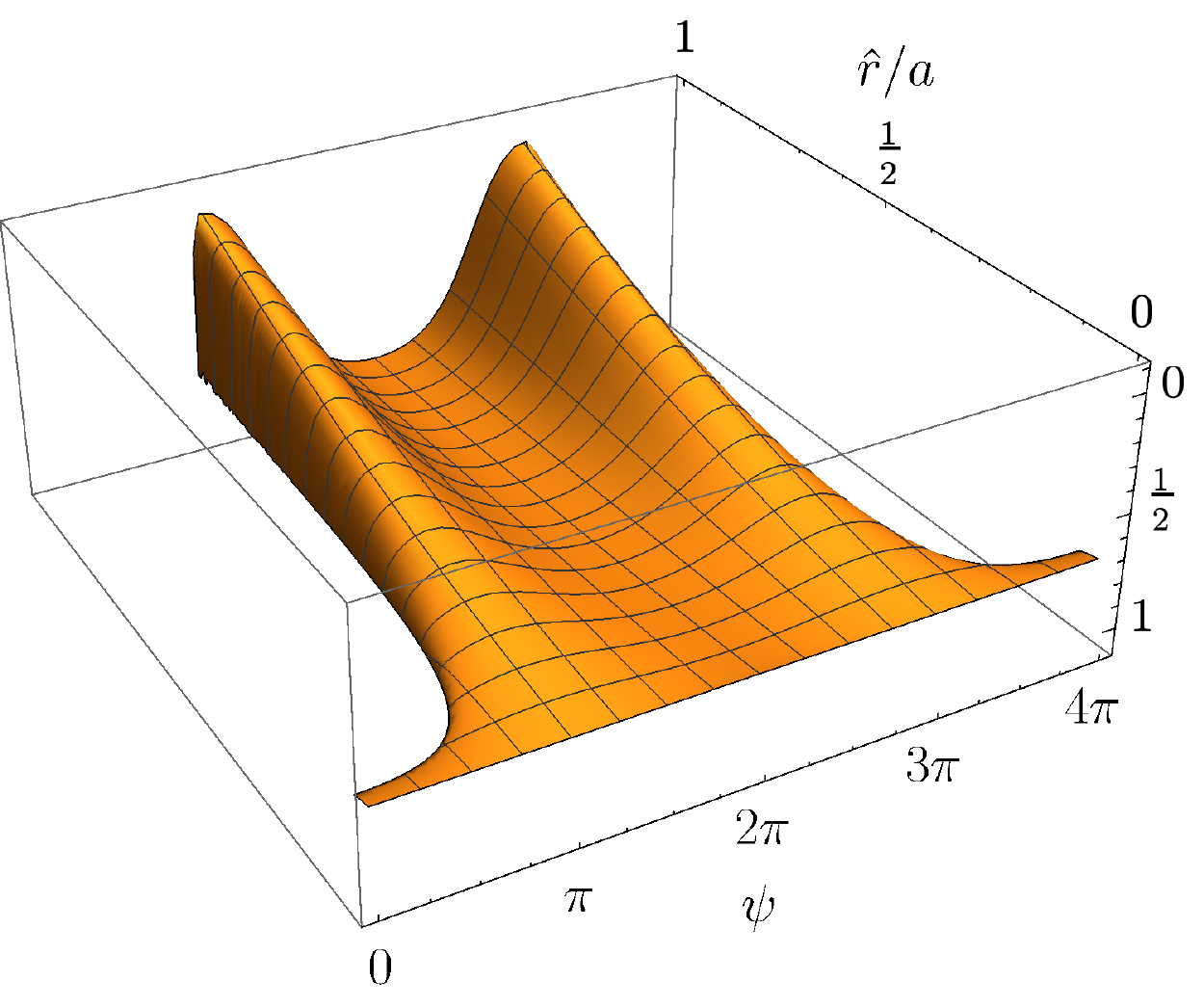}
\includegraphics[width=0.7\columnwidth]{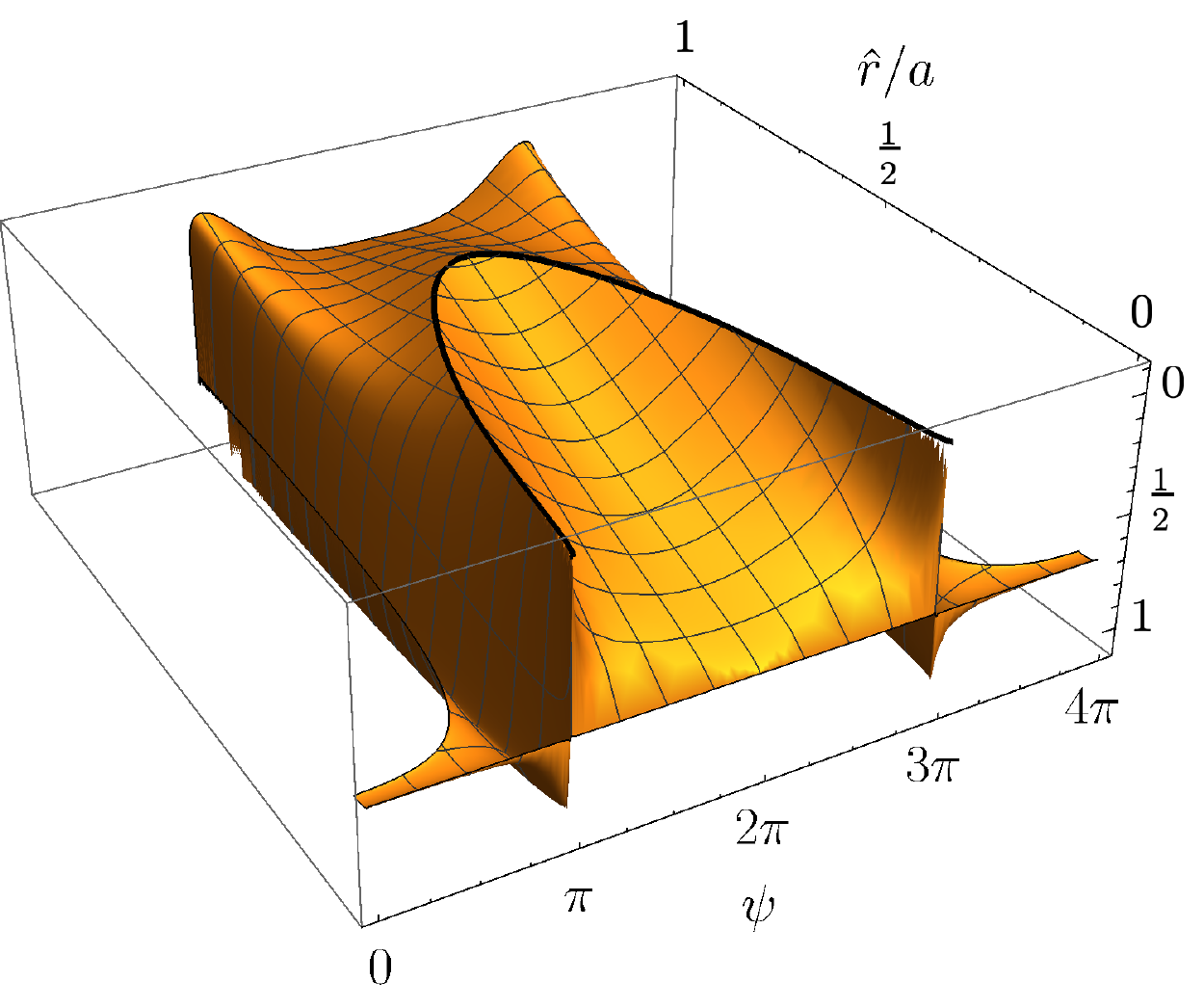}
\caption{
The functions $\cos^2(\alpha_+)\ge \cos^2(\alpha_-)$ as functions of $g_{tt}/g_{t\KSu}$ and $g_{\KSu\KSu}/g_{t\KSu}$ plotted over the region $g_{tt} g_{\KSu\KSu}-g_{t\KSu}^2<0$ and $g_{t\KSu}\ne 0$ (top) as well as the functions $\cos (\alpha_+)$ (middle) and $ \cos(\alpha_-)$ (bottom),  with the ordinate axis increasing downwards, as functions of $\RR/a$ and $\KSv$ with $a/m = 1/2$. The set $\cos(\alpha_-)=0$ coincides with the hypersurface $\{g_{\KSu\KSu}=0\}$, on which the rotational Killing vector $\partial_\BLphi=\partial_{\KSu}$ becomes null. The metric on the space of the orbits becomes spacelike beyond the curves at which the plots disappear.
}
\label{F23IX19.1}
\end{center}
\end{figure}

\rout{Given the above considerations, let us}
\red{We} begin by studying the flow along the isometry group of the Kerr metric near the singular ring.
Let us then define the vector field in toroidal coordinates
\begin{equation}\label{14VIII19.31}
 X = \cos(\alpha) \partial_t + \sin (\alpha) \partial_{\KSu}
\,,
\end{equation}
with $\alpha\in [0,\pi/2]$, and find the angles at which this vector field becomes null, namely
\begin{eqnarray}
\nonumber
 0 &= & g(X,X)= g_{tt} \cos^2(\alpha)
 + 2 g_{t\KSu} \cos(\alpha)\sin(\alpha)
\\
 &&   + g_{\KSu\KSu} \sin^2(\alpha)
 \,.
\label{14VIII19.31b}
\end{eqnarray}
 This equation is equivalent to
$$
 2 g_{t\KSu} \cos(\alpha)\sqrt{1-\cos^2(\alpha)} = g_{\KSu\KSu}( \cos^2(\alpha) -1) -   g_{tt} \cos^2(\alpha)
 \,,
$$
which can be solved algebraically for $\cos^2(\alpha)$ after squaring both sides. Keeping in mind that $\alpha\in [0,\pi/2]$, the solution is
\begin{align}
\cos{(\alpha_\pm)}
 &=\sqrt{ \frac{\left(g_{t\KSu}\pm\sqrt{g_{t\KSu}^2-g_{tt} g_{\KSu\KSu}}\right)^2+g_{\KSu\KSu}^2}{(g_{tt}- g_{\KSu\KSu})^2 +4 g_{t\KSu}^{2} } }\,.
 \label{23IX19.35}
\end{align}
The right-hand side is real, and we show in Appendix~\ref{A23IX19.1} that it is smaller than or  equal to one when the matrix of scalar products of the Killing vectors is Lorentzian  (as is the case in the region
$\{\tilde{r}^2 +a^2 - 2 m \tilde{r} >0 \}$);
see also Fig.~\ref{F23IX19.1}.

The small $\RR$ expansion of the above distinct branches of solutions leads to
\allowdisplaybreaks[4]
\begin{align}
\cos{(\alpha_-)}
 & =
  \left\{
                     \begin{array}{ll}
 \displaystyle
\frac{a}{\sqrt{1 + a^{2}}}  + \frac{\sqrt{2 \RR} \; a^{3/2}}{(1 + a^{2})^{3/2} m \cos(\KSv/2)} + \bigO{\RR} \,, &
\\
\\
\hbox{for $\cos(\KSv/2) \ne 0$\,;} &
\\
\\
\\
             \displaystyle
       =  \frac{a}{\sqrt{1 + a^{2}}} - \frac{\RR}{(1 + a^{2})^{3/2}} + \bigO{\RR^{2}},
  &
  \\
  \\
  \hbox{for $\cos(\KSv/2) = 0$\,,} &
                     \end{array}
                   \right.
 \label{19IX19.19}
\end{align}
and
\begin{align}
\cos{(\alpha_+)}
 & =
  \frac{a}{\sqrt{1 + a^{2}}}  +  \frac{\RR \cos{(\KSv)}}{(1 + a^{2})^{3/2}} + \bigO{\RR^{3/2}}
\,.
\end{align}
The vector $X$ with an angle between $\alpha_{-}$ and $\alpha_{+}$ is spacelike.
Keeping the leading order term only, the null version of the linear combination of Killing vectors $X$
is
\begin{equation}\label{14VIII19.31-new}
 X =  \frac{1}{\sqrt{1 + a^{2}}}  \left(a\,\partial_t + \partial_{\KSu} \right)
\end{equation}
as the ring is approached, independently of the direction of approach.
This implies that the spacetime metric cannot be extended through the ring as a $C^2$-Lorentzian metric.
Indeed, arguing by contradiction, let us  suppose that such an extension exists. It then follows from the equations
\begin{equation}\label{16XI19.1}
  \nabla_\alpha\nabla_\beta X_\gamma = R^\sigma {}_{\alpha\beta{\gamma}}X_\sigma
  \,,
\end{equation}
which are satisfied by every Killing vector field $X$, that the Killing vector fields extend differentiably to the ring.
The metric on the orbits therefore extends continuously to a Lorentzian two-dimensional metric by Eq.~\eqref{4VIII19.90},
which guarantees the existence of two distinct null directions at the ring spanned by the Killing vectors. But this  contradicts Eq.~\eqref{14VIII19.31-new}.

We note by passing that this gives the simplest proof of a $C^2$ singularity of the Kerr metric at the ring, independently of the direction of approach, without having to calculate any curvature invariants. A similar argument, based on the overdetermined system of equations satisfied by conformal Killing vector fields, shows that there exists no conformal extension of the metric at the ring of $C^3$ differentiability class.

\subsection{Two-dimensional geometry near the ring}
\label{ss6IX19.4}

\newcommand{\mychi}{{\chi}}

\rout{As we discussed before, the metric $g$ has two Killing vectors, which for simplicity we here label
$X_1:=\partial_t$ and $X_2:=\partial_{\KSu}$.
 Given this} A natural two-dimensional geometry associated with the four-dimensional metric is the \emph{quotient-space metric $q$}, familiar from the Kaluza-Klein reduction of metrics with symmetries~\cite{Coquereaux:1988ne,Overduin:1998pn}. The physical importance of the quotient-space metric is that, loosely speaking, it is the part of the metric orthogonal to Killing vectors, and thus, it is the quantity that could show interesting curvature behavior.

The metric $q$ is  defined as follows.  \red{Let us label the Killing vectors of $g$ as
$X_1:=\partial_t$ and $X_2:=\partial_{\KSu}$, and } for $a,b,\in \{1,2\}$, let us define
\begin{align}
  \left(\mychi_{ab}\right)
 &:=
\left( g(X_a, X_b)
 \right)
 \equiv \left(
     \begin{array}{cc}
       g_{tt} & g_{t\KSu} \\
       g_{\KSu t} & g_{\KSu\KSu} \\
     \end{array}
   \right)
   \nonumber \\
 &= \left(
     \begin{array}{cc}
       \eta_{tt} + H & - H \KStheta_{\KSu} \\
       - H \KStheta_{\KSu} & \eta_{\KSu\KSu} + H (\KStheta_{\KSu})^2 \\
     \end{array}
   \right)
   \nonumber \\
 &= \left(
     \begin{array}{cc}
       -1 + H &  -\frac{a H \rho^2}{\tilde{r}^2 + a^2}  \\
       -\frac{a H \rho^2}{\tilde{r}^2 + a^2}  & \rho^2  + \frac{a^2 H \rho^4}{(\tilde{r}^2 + a^2)^2} \\
     \end{array}
   \right)\label{20VIII19.1}
\end{align}
to be the two-by-two matrix of scalar products of the Killing vectors, and denote by $\mychi^{ab}$ the matrix inverse\footnote{Note that this requires $\mychi_{ab}$ to be non-degenerate, so that the two-covariant tensor field $q$ provides an invariantly defined geometric field associated with $g$ only away from the Killing horizons, where the metric function $
\DeltaBL$ vanishes; this is not an issue near the singular ring, cf. Eq.~\eqref{18X19.01}.} to $\mychi_{ab}$.  Next, let $ x^A  $ be any coordinates on the space of orbits of the isometry group $\R\times U(1)$ near the ring (with the $\R$ factor corresponding to $t$-translations and the $U(1)$ factor to rotations),
 e.g.~in Boyer-Lindquist coordinates $(x^A) = (\tilde{r}, \BLtheta)$. With this at hand, the quotient-space metric $q$ is defined via
\begin{equation}\label{20VIII19.2}
  q_{AB} = g_{AB} - \mychi^{ab} (X_a)_A (X_b)_B
\end{equation}
where choosing e.g.,\ $(x^A)=(\RR, \KSv )$ we have
\begin{equation}\label{5V20}
 \left( (X_a)_A \right) :=
 \left( (X_a)^\mu g_{A\mu}
 \right) =
\left(
  \begin{array}{cc}
    g_{t \RR }  &g_{t \KSv  } \\
   g_{\KSu  \RR } & g_{\KSu \KSv  } \\
  \end{array}
\right)
 \,.
\end{equation}
Note that in Boyer-Lindquist coordinates $(x^A) = (\tilde{r}, {\BLtheta})$, one has directly that $q_{AB}=g_{AB}$ because the second term in Eq.~\eqref{20VIII19.2} vanishes identically;
this is a consequence of the Kerr metric in Boyer-Lindquist coordinates being block diagonal (a fact that is not true in for example Kerr-Schild coordinates).

\begin{figure}[thb]
\begin{center}
\includegraphics[width=0.9 \columnwidth,clip=true]{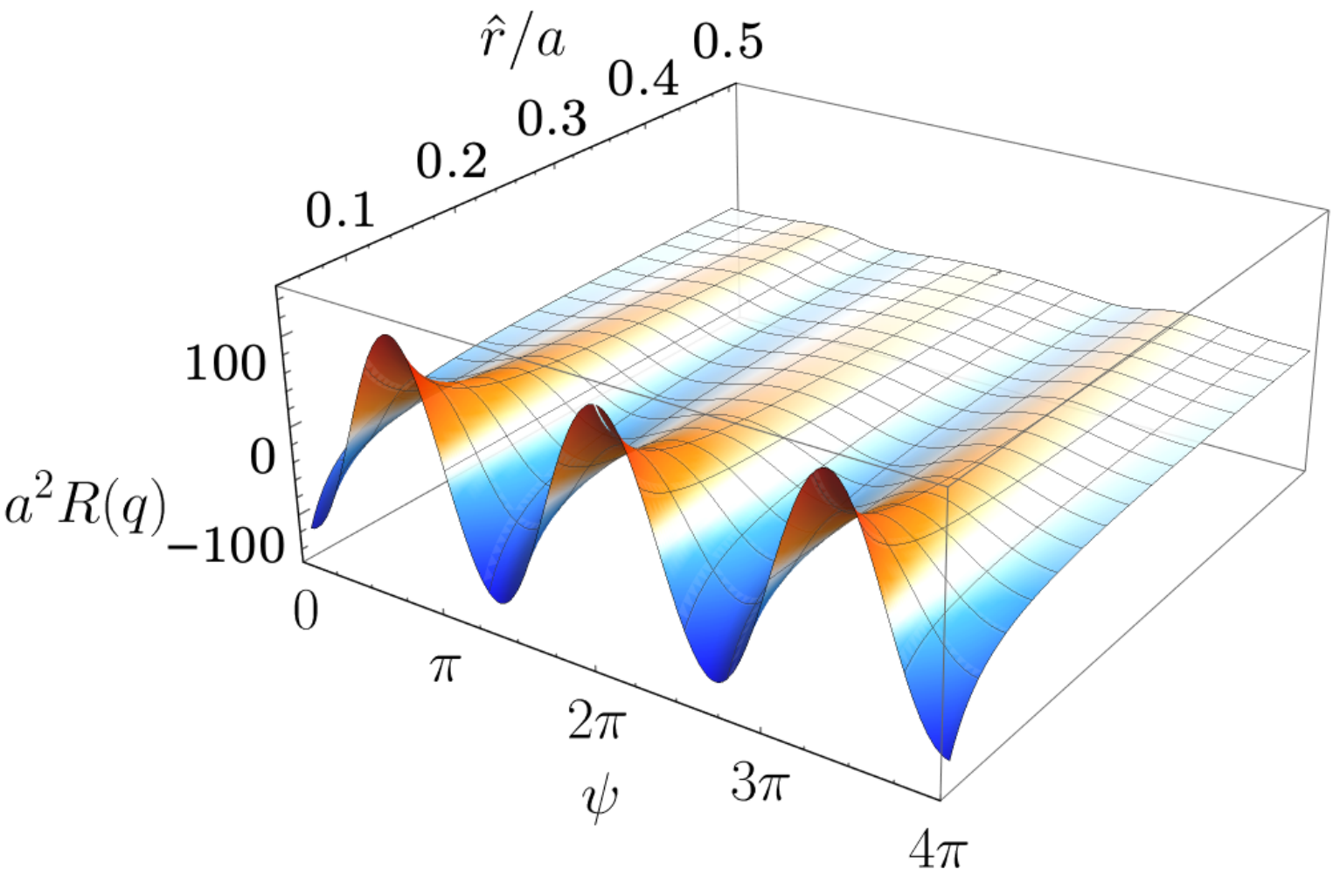}
\caption{Representative graph of the Ricci scalar  $a^2R(q)$
as a function of $\RR/a\,\in[1/20,\,1/2]$ and $\KSv\in[0,\,4\pi]$; here  $a/m=1/2$.}
\label{F14VIII19.2}
\end{center}
\end{figure}

All the curvature information about $q$ is contained in its Ricci scalar, which we will denote by $R(q)$. It turns out simplest to calculate $R(q)$ in Boyer-Lindquist coordinates   and transform the result to toroidal coordinates. Doing so, one finds:
\begin{widetext}
\begin{align}
\label{21IX19.5}
	R(q) &= -\frac{2m\tilde{r}\left(\tilde{r}^{2}-3 a^{2}\cos^2(\BLtheta)\right)}{(\tilde{r}^{2}+a^{2}\cos^{2}(\BLtheta))^{3}}
	= -m \RR^{-3/2}\frac{e^{-3/2i\KSv}\left(2a+e^{i\KSv}\RR\right)^{3/2} +
    e^{3/2i\KSv}\left(2a+e^{-i\KSv}\RR\right)^{3/2}}
  {\left(4 a^{2}+\RR^{2}+4 a \RR\cos(\KSv)\right)^{3/2}}
  \\
	&{=-\frac{m}{(a\RR)^{3/2}}\sqrt{\frac{2}{\pi}}\sum_{k\geq 0}\frac{\Gamma(k+3/2)}{\Gamma(k+1)}
	\left(-\frac{\RR}{2a}\right)^{k}\cos\left((k+3/2)\KSv \right)}
\\
\label{21IX19.5-b}
 &= -m\frac{\cos(3\KSv/2)}{\sqrt{2}a^{3/2}\RR^{3/2}} +
	\bigO{\RR^{-1/2}}
	\,.
\end{align}
\end{widetext}
To derive these expressions, some care has to be taken concerning the choice of the root-branches in Eq.~\eqref{21IX19.5}.
A representative plot of $R(q)$ can be found in Fig.~\ref{F14VIII19.2}.
Notice that $R(q)$ is invariant under reflection of $\KSv$ around $2\pi$; this follows of course from the fact that the maps $\KSv \mapsto -\KSv $ and $\KSv \mapsto \KSv+ 4\pi$ are isometries of $g$, and hence of $q$. We conclude then that \emph{although the full Ricci scalar $R(g)$ obviously vanishes everywhere, the scalar curvature $R(q)$ diverges as one approaches the ring.}

For small $\RR$, the expansion of $q$ reads
\allowdisplaybreaks[4]
\begin{align}
q & =  \left(
 1 + \frac{2\sqrt{2}m\sqrt{\RR}\cos^{3}\left(\KSv/2\right)}{a^{3/2}} + \bigO{\RR^{}}
 \right)\diff{\RR^2}
 \nonumber \\
 &+2\left(-\frac{4\RR^{3/2}m\sin(\KSv/2)\cos^{2}(\KSv/2)}{\sqrt{2}a^{3/2}} + \bigO{\RR^{2}}
\right)\diff{\RR}\,\diff{\KSv}
 \nonumber
\\
 &
 +
 \left(
 \RR^2 + \frac{\sqrt 2 m \RR^{5/2}\sin(\KSv)\sin(\KSv/2)}{a^{3/2}} + \bigO{\RR^{3}}
 \right) \diff{\KSv^2}
 \,.\label{7VIII19.1-a}
\end{align}
Since $\KSv$ ranges over $4\pi$, we see that $q$ describes a metric on a cone with opening angle $4\pi$. There does not seem to be a consensus in the literature whether, and in which sense, one can assign a distributional curvature, or else, to the tip of this cone~\cite{VickersStrings,GerochTraschen,IsraelKerr3}.

\subsection{Hypersurfaces}

\subsubsection{Level sets of $t$}

The level sets of $t$ have a geometric character since they have vanishing mean extrinsic curvature. They therefore form maximal hypersurfaces (hypersurfaces that locally maximize the induced Riemannian volume) in the region where they are \rout{timelike} \red{spacelike}, and minimal hypersurfaces (hypersurfaces that locally minimize the induced Lorentzian volume) in the regions where they are \rout{spacelike} \red{timelike}. The  causal character of these level sets is determined by the sign of $g(\nabla t, \nabla t)$. For small $\RR$,  this is revealed by a Taylor expansion about $\RR = 0$:
\begin{align}\label{27VII19.21}
  g(\nabla t, \nabla t) \equiv g^{tt} &= -1 - H
  \nonumber \\
  &= -1 - m\sqrt{\frac{2}{a}}\cos\left(\KSv/2\right)\RR^{-1/2} + \bigO{\RR^{1/2}}
\,,
\end{align}
Recall that $g_{tt}=-1+H$, and that $H(\RR, \KSv +2\pi)=-H(\RR, \KSv )$. It follows that the hypersurface on which $\nabla t$ becomes null is obtained by shifting by $2\pi$ in $\KSv$ the hypersurface on which $\partial_t$ becomes null. Hence the level sets of $t$ are spacelike throughout the positive-$\tilde{r}$ region, but change type in the negative-$\tilde{r}$ region.
The boundary of the region where $\nabla t$ is timelike can therefore be read off from Fig.~\ref{F20IX19.1}, where now the curves show the set where $\nabla t$ becomes null in the negative-$\tilde{r}$ region.
Overall then, we find that \emph{the character of these hypersurfaces is not unique near the ring}.

\subsubsection{Level sets of $\RR$}
\label{subsec:RRlevels}

Equation~\eqref{4VIII19.90}
 implies that the level sets of $\RR$ must be timelike, since they are foliated by two-dimensional timelike submanifolds, namely the orbits of the isometry group.  This can be checked by a direct calculation,
\begin{align}
  g(\nabla \RR, \nabla \RR)
   & \equiv
     g^{\RR\RR}
  =
1-\frac{4 \sqrt{\RR}
 m \cos    ^3\left(\KSv/2\right)}{a^{3/2} \sqrt{2}} + \bigO{\RR^{3/2}}
\,,
\label{28VII19.1}
\end{align}
confirming timelikeness of the level sets of $\RR$ for sufficiently small $\RR$.

Consider the family of spacelike curves orthogonal to the level sets or $\RR$. Equation~\eqref{28VII19.1} shows that  each member of this congruence will reach the ring $\RR=0$ after a distance equal to
\begin{equation}\label{4VII19.51}
  \int_0^{\RR}
  \sqrt{  g(\nabla \RR, \nabla \RR) }\,
  \diff{\RR} =
  \RR +
  \bigO{\RR^{3/2}}
 \,.
\end{equation}
This gives a precise sense to the statement that the singular set lies a distance approximately equal to $\RR$ from the level sets of the function $\RR$.

The vector field  $\partial_{\KSv}$ is tangent to both the level sets of $t$ and the level sets of $\RR$. Because of its periodicity it is of interest to enquire about its causal character.  It turns out that the fast fall-off of $\KStheta_\KSv$ near $\RR=0$ compensates the divergence of $H$, so that the periodic vector field $\partial_{\KSv}$ is spacelike for small $\RR$. Indeed, a Taylor expansion about $\RR = 0$ returns
\begin{align}\label{3VIII19.1}
  g_{\KSv\KSv} &= \RR^2 + (\KStheta_\KSv)^2 H
  \nonumber \\
  &= \RR^2 + 2 \sqrt{2} \frac{\RR^{5/2} m}{a^{3/2}} \sin^{2}{(\KSv/2)}  \cos{(\KSv/2)} + \bigO{\RR^{3}}
 \,,
\end{align}
which is clearly positive near the ring, regardless of the direction of approach.
From this, we conclude that \emph{there are no causality violations along the integral curves of $\partial_{\KSv}$ near the ring}.

\section{Tidal forces on geodesics}
\label{s1X19}

The tidal forces along geodesics $s\mapsto \gamma^\mu(s)$ are determined by the right-hand side of the geodesic-deviation equation:
\begin{equation}\label{20VIII19.15a}  \frac{D^2 Z^\mu}{ds^2}
 = R^\mu{}_{\alpha \beta \nu} \dot \gamma^\alpha Z^\beta \dot \gamma^\nu
 \,,
\end{equation}
where $Z^{\mu}$ is the vector describing the infinitesimal separation between neighboring geodesics and $\dot \gamma$ is the tangent to the four-trajectory of the geodesic, i.e., the four-velocity of the geodesic.
The question then arises: how disruptive are these forces? Clearly, some coordinate components of the Riemann tensor tend to infinity when the singular ring is approached in any coordinate system. However, the \rout{sums} \red{contractions} in the right-hand side of Eq.~\eqref{20VIII19.15a} could conspire to produce a  finite result. Alternatively, the rate of blow-up of the right-hand side along a geodesic could be sufficiently slow so that the net effect of the resulting forces \rout{is} \red{would be} innocuous. That something like that could happen is suggested by the curvature scalars of Sec.~\ref{s19VII19.2}: an invariant blow-up rate $\RR^{-3/2}$ of the Riemann tensor, suggested by the results there, could lead to finite effects when integrated twice in $\RR$, which would then lead to finite distortions of test bodies. In this section, we will thus study the tidal forces experienced by timelike geodesics as they approach the ring singularity.

\subsection{Hitting and avoiding the ring singularity}

The geodesics in the Kerr metric are complicated enough~\cite{BONeillK} to prevent one to give a general definitive answer. We will therefore consider only a few special families of geodesics. We start by noting that
\begin{equation}\label{4VIII19.11}
  \partial_z H|_{ {\{\rho > a,\, z=0\}}} =0
 \,.
\end{equation}
This implies immediately (the well-known fact) that the exterior  component of the \emph{equatorial hyperplane},
\begin{equation}\label{1XI19.21}
 D':= \R\times
 \{(x,y,z)\in\R\,|\ x^{2}+y^{2}>a^{2},\ z=0\}
 \subset \R \times \R^3
 \,,
\end{equation}
is \emph{totally geodesic}:
 a geodesic starting  at $D'$ and initially tangent to $D'$ is entirely contained in $D'$. Equivalently, the second fundamental form of $D'$ (``extrinsic curvature tensor'') vanishes.  We will refer to $D'$ as the \emph{exterior equatorial hyperplane}. Here one needs to keep in mind that there are \emph{two} exterior equatorial hyperplanes, namely one on which $\tilde r>0$, and another one where $\tilde r<0$. As discussed by Carter in~\cite{CarterKerr},  the only causal geodesics which accumulate at the ring are those entirely contained in $D'$. (Compare~\cite[Proposition~4.5.1]{BONeillK}).

In order to obtain a measure of the ``strength'' of the singularity, we calculate the tidal forces along a class of timelike geodesics contained in $D'$. For this we need to determine the asymptotics of $\dot \gamma$ for those exterior equatorial geodesics $\gamma$ which reach the ring.
The usual polar coordinates  associated with the Kerr-Schild coordinates, which we will denote by $(\rho, \KSu)$ with $\rho$ defined back in Eq.~\eqref{eq:rho},
appear to be most convenient for the calculations to follow:
$$
 x=\rho \cos (\KSu)
 \,,
 \quad
 y = \rho \sin(\KSu)
 \,.
$$
Letting $\epsilon$ denote the sign of $\tilde{r}$, on the exterior equatorial plane we then have
\begin{align}\label{20X19.1}
  \tilde{r} &= \epsilon \sqrt{\rho^2 - a^2}
  \,,
  \quad
   H = \frac{2 m }{\tilde{r}}= \frac{2\epsilon m }{\sqrt{\rho^2 - a^2}}
   \,,
   \nonumber \\
   \KStheta &= - \diff t + a \diff\KSu  -\frac{\tilde{r} \diff\rho}{\rho}
    \,,
\end{align}
where we recall that $\KStheta$ is the one-form associated with one of the principal null directions.

The components of the four-velocity of the geodesics we consider can most
easily be written down in terms of the constants of the motion associated with its Killing vectors.
Let us then denote by
$$
 \mcE:=- g(\partial_t,\dot \gamma)\,,
 \quad
 \mcL := g(\partial_\varphi, \dot \gamma)
 \,,
$$
the usual specific energy and the specific  ($z$-component of)  angular momentum of the orbit. In terms of these we find
\begin{eqnarray}
  \dot\KSu  &=& \frac{2 a {m} ({\mcE} \rho +\dot \rho  \tilde{r})+\mcL
   \rho  (\tilde{r}-2 {m})}{2 a^2 {m} \rho +\rho ^3
   (\tilde{r}-2 {m})}
   \,, \\
  \dot t &=& \frac{2 a {m} (a {\mcE}-\mcL)+\rho  \tilde{r}
   ({\mcE} \rho +2 {m} \dot \rho )}{2 a^2 \epsilon
   m+\rho ^2 (\tilde{r}-2 {m})}
   \,.
\end{eqnarray}
The above equations depend on $\dot{\rho}$ because we are here using Kerr-Schild coordinates,
instead of Boyer-Lindquist coordinates in which the geodesic equations decouple.
The equation $g(\dot \gamma, \dot \gamma) = -1$ can then also
be used to find
\begin{align}
  \dot \rho ^2  & =
  \frac{2   {m}}{\tilde{r}\rho ^2 } \left( \mcL-a {\mcE}\right)^2
   - \frac{1}{\rho ^2 } \left(a^2 + \mcL^2- \left(a{\mcE}\right)^2\right)
    \nonumber \\
    &+ \frac{1}{\tilde{r}\rho ^2 } \left(  \rho ^2-a^2\right)\left(2{ m} + \tilde{r}( \mcE^2-1)\right)
    \,.
     \label{24X19.11}
\end{align}

With this in mind, we then make the following claims:
\begin{enumerate}
  \item no geodesic with either  $\epsilon=-1$ or $\mcL=a{\mcE}$
   reaches the ring, and
   \item all geodesics with $\mcL\ne a{\mcE}$ and $\epsilon=1$ that start close enough to the ring at $s=0$ and with $\dot \rho(0) <0$ will reach the ring in finite proper time.
\end{enumerate}
Indeed, suppose first that $\mcL= a{\mcE}$, in which case Eq.~\eqref{24X19.11} gives (recall that $\rho^2-a^2 $ behaves as $\RR$ for small $\RR$, and  $\tilde r $ behaves as $\epsilon \sqrt{\RR}$ there)
\begin{eqnarray}
  \dot \rho ^2  & = & -1 + O(\sqrt{\RR})
    \,.
     \label{24X19.13}
\end{eqnarray}
Clearly, no such geodesics can reach the ring.

Next, if $\mcL\ne a{\mcE}$, then for small $\RR$ we obtain from Eq.~\eqref{24X19.11}
\begin{eqnarray}
   \dot \rho ^2  & = &\frac{2   {m}\left( \mcL-a {\mcE}\right)^2  +O(\sqrt{\RR}) }{\tilde{r}\rho ^2 }
    \,.
     \label{24X19.12}
\end{eqnarray}
This is not possible if $\tilde{r}$ is negative, or equivalently if $\epsilon=-1$; thus, again, no such geodesics will reach the ring.

Another way of obtaining the same conclusion proceeds by rewriting Eq.~\eqref{24X19.11} as
\begin{equation}\label{21XI19.101}
 \frac 12  \dot \rho ^2 + V = \frac 12 (\mcE^2-1)
 \,,
\end{equation}
where
\begin{align}
 V(\rho) & = \frac{1}{2 \rho^2}
 \left\{ \mcL^2 - 2m \epsilon \left[\frac{\left(\mcL-a\mcE\right)^2}{\sqrt{\rho^2-a^2}} + \sqrt{\rho^2-a^2}
 \right]\right\}
    \,.
     \label{21XI19.102}
\end{align}
This reduces the problem to the study of the motion $s\mapsto \rho(s)$ of a particle of unit mass  moving in the potential $V $ with  {total} energy $\frac 12 (\mcE^2-1)$.  Inspection of this potential reveals that when $\epsilon=-1$ and $\mcL\ne a\mcE$, the potential is positive and it acts as a repelling barrier that forces turning points before geodesics reach the ring. When $\epsilon=+1$ and $\mcL = a \, \mcE$, the potential allows (unstable) circular orbits, and it asymptotes to $\mcE^{2}/2$ as $\rho \to a$, which is clearly larger than $(\mcE^2-1)/2$. This then means that no \red{such}
geodesic can ever reach $\rho = a$, and instead there is a turning point in the motion.

But is there a physical reason \rout{for} why these geodesics cannot reach the ring?
The negative $\tilde{r}$ region can be mapped to the positive $\tilde{r}$ region if one also takes $m \to -m$. From this point of view, a ``negative mass'' acts as a repeller that prevents geodesics from reaching the ring. The fact that no geodesic can reach the ring when  $\mcL=a{\mcE}$ regardless of the sign of $\tilde{r}$ is a bit more curious. First, note that the ratio of the specific energy to the specific orbital angular momentum is the same as the ratio of the orbital energy to the orbital angular momentum, because $E = \mu {\cal{E}}$ and $L = \mu {\cal{L}}$, where $\mu$ is the mass of the small particle. Given this, one then has that $L = S (E/m)$, where we recall that $m$ is the ADM mass of the black hole, and $S = a \, m$ is  (the $z$-component of)  its ADM (spin) angular momentum. Moreover, one can also show that the angular velocity of such geodesics, defined via $\Omega := \dot{\KSu}/\dot{t}$, asymptotes to $\Omega = 1/a +\bigO{ {\RR}^{1/2}}$ in the limit of small $\RR$. This angular velocity for timelike geodesics coincides with the angular velocity of the ring, defined through the $X$ vector in Eq.~\eqref{14VIII19.31-new}. The latter, however, is null, while geodesics of massive particles are timelike. This seems to suggest then that only massless particles would be able to acquire the angular velocity of the ring as the ring is approached when $\mcL=a{\mcE}$ and $\epsilon = +1$.

On the other hand,
 suppose that $\epsilon =  1$ \red{but $\mcL\ne a{\mcE}$ }, and consider a geodesic directed, at some proper time $s_0$, toward the ring  in the region where the error term in the numerator is dominated by the constant.
Differentiating Eq.~\eqref{24X19.12} with respect to $s$, and using that
\begin{align}
\partial_{\RR} (\tilde{r} \rho^2 ) &= \partial_{\RR} \left(\sqrt{\RR (2 a+\RR )}(a+\RR )^2 \right)
\nonumber \\
&= \frac{(a+\RR ) \left(a^2+6 a \RR+3 \RR^2\right)}{\sqrt{\RR (2 a+\RR )}}
>0
\,,
\end{align}
one finds that the derivative $ \dot \rho(s)$ is a decreasing function of $s$, and hence it will remain negative, and smaller than $\dot \rho(s_0)$ for $s> s_0$; in fact, it will tend to minus infinity as $s$ increases and the ring is approached. It follows then that $\rho=a$ will be reached in finite time. The physical interpretation here is much clearer: if an object is in a generic orbit on the exterior equatorial hyperplane (on the $\tilde{r}>0$ sheet), and if it gets close enough to the ring during its trajectory, then it will be attracted by the ring and it will hit the curvature singularity in finite time.

From the above analysis, it also follows that the statement that ``the ring is repulsive no matter what'' is not correct. On the exterior equatorial plane, the ring is repulsive to geodesics  only when $\epsilon = - 1$ (or equivalently when $\tilde{r} < 0$), while the ring is attractive when $\epsilon = + 1$ (or equivalently when $\tilde{r} > 0$) and $\mcL \neq a{\mcE}$.

Incidentally, note that the Carter constant, $\mcK$, is irrelevant for the problem at hand. To see this, recall that in Boyer-Lindquist coordinates $\mcK$ is given by~\cite{CarterKerr}
 \begin{equation*}
   \mcK = p_{\tilde{\theta}}^2 + \left( a \mcE \sin(\tilde{\theta})  -\sin(\tilde{\theta})^{-1}\mcL  \right)^2 + a^2 \cos(\tilde{\theta})^2
    \,,
 \end{equation*}
 where
 \begin{equation*}
  p_{\tilde{\theta}} =  g_{ {\tilde{\theta}}\nu } \frac{d{\tilde{x}^\nu}}{ds}
 = g_{ {\tilde{\theta}}{\tilde{\theta}}} \frac{d{\tilde{\theta}}}{ds}  =g_{\mu\nu} \frac{\partial x^\mu}{\partial{\tilde{\theta}}} \dot \gamma^\nu
    \,.
\end{equation*}
For geodesics on the equatorial plane we have $p_{\tilde{\theta}} =0$, $\sin(\tilde{\theta})=1$, and so no new information about these geodesics can be obtained, since then \rout{$\mcK$} $\red{\sqrt{\mcK}}$ is just a linear combination of the energy $\mcE$ and of the angular momentum  $\mcL $. Perhaps interestingly, the Carter constant vanishes identically on the equatorial plane for those orbits which have $a \mcE = \mcL$, which are also precisely those that never hit the ring.

\subsection{Tidal forces on geodesics that hit the ring singularity}
\label{ss1XI19.1}

Now that we understand which geodesics hit the ring singularity, we
will focus on the tidal force equation in Eq.~\eqref{20VIII19.15a},
which requires us to evaluate the Riemann tensor contracted with the
four-velocity of the geodesic. This calculation is simplest in
Kerr-Schild coordinates, which means we will have to perform some coordinate
transformations.

Let us then begin by investigating the four-velocity. In toroidal
coordinates, the geodesics hitting the ring satisfy [see
Eq.~\eqref{24X19.11}]
\begin{eqnarray}
\label{27X19.2}
  \dot \rho &=& -\frac{\sqrt[4]{2} \sqrt{m (a {\mcE}-\mcL)^2}}{a^{5/4}
   \sqrt[4]{\RR}} +O\left(\RR^{1/4}\right)
   \,,
\\
\label{27X19.3}
  \dot\KSu   &=& \frac{\sqrt{2} m (a {\mcE}-\mcL)}{a^{5/2} \sqrt{\RR}}
   +O\left(\RR^{-1/4}\right)
  \,,
\\
\label{27X19.4}
  \dot t &=& \frac{\sqrt{2} m (a {\mcE}-\mcL)}{a^{3/2} \sqrt{\RR}}
  +O\left(\RR^{-1/4}\right)
  \,.
\end{eqnarray}
With this at hand, the components of the four-velocity
$\dot{\gamma}^{\mu}$ in Kerr-Schild coordinates are
\begin{align}
  \dot{\gamma}^{t} &=  \frac{\sqrt{2} m (a {\mcE}-\mcL)}{a^{3/2} \sqrt{\RR}}
  +O\left(\RR^{-1/4}\right)\,,
  \\
   \dot{\gamma}^{x} &= - \sin{\KSu} \frac{\sqrt{2} m (a {\mcE}-\mcL)}{a^{3/2} \sqrt{\RR}}
   +O\left(\RR^{-1/4}\right)
   \,,
     \\
   \dot{\gamma}^{y} &=  \cos{\KSu} \frac{\sqrt{2} m (a {\mcE}-\mcL)}{a^{3/2} \sqrt{\RR}}
   +O\left(\RR^{-1/4}\right)
   \,,
     \\
   \dot{\gamma}^{z} &= 0\,,
\end{align}
where $\RR$ is to be understood as a function of Kerr-Schild
$(x,y,z)$.

Let us now carry out the necessary contractions to compute the
right-hand side of Eq.~\eqref{20VIII19.15a}.
Using the formulae in Appendix~\ref{sA1XI19.1}, one finds
\begin{widetext}
\begin{equation}
\label{27X19-1}
	{R^{\mu}{}_{\alpha\nu \beta}{\dot \gamma}^{\alpha} \dot \gamma^{\beta}} =
	\frac{3 m^2 (\mcL-a \mcE)^2}{4 a^3 \RR^3}
\left(
\begin{array}{cccc}
 1 & \sin (\varphi ) & -\cos (\varphi ) & 0 \\
 -\sin (\varphi ) & -\sin ^2(\varphi ) & \sin (\varphi ) \cos (\varphi ) & 0 \\
 \cos (\varphi ) & \sin (\varphi ) \cos (\varphi ) & -\cos ^2(\varphi ) & 0 \\
 0 & 0 & 0 & 0 \\
\end{array}
\right) + \bigO{\RR^{-11/4}}
 \,.
\end{equation}
\end{widetext}
The $\mu=\nu=z$ component of the tidal force sits in the error terms above:
\begin{equation}
\label{27X19-2}
	R^{\mu}{}_{ \alpha z\beta}\dot \gamma^{\alpha}\dot \gamma^{\beta} = \left(0,0,0,\frac{3 (\mcL - a\mcE)^2 m}{4 \sqrt{2} a^{5/2} \RR^{5/2}} + \bigO{\RR^{-3/2}}\right)
 \,.
\end{equation}

In order to determine the effect of the tidal force we proceed as follows. First, using $\RR=\rho-a$  and keeping only the dominant terms, we find from Eq.~\eqref{27X19.2} that
\begin{equation}\label{27X19.1}
  \frac{d\RR}{ds} \approx - c_1 \RR^{-1/4}
   \,, \ \mbox{ where } \  c_1:=
   \frac{\sqrt[4]{2} \sqrt{m (a {\mcE}-\mcL)^2}}{a^{5/4}}
    \,.
\end{equation}
It is convenient to change the proper time $s$ to its negative, and to shift,  so that the geodesics emanate from the singularity at $s=0$. Then, for $s>0$,
\begin{equation}\label{27X19.5}
  \RR \approx \left(\frac {5 c_1} 4 s \right)^{4/5}
   \,.
\end{equation}

Next, we shall show that the equation for the $z$-component of $Z^{\mu}$ decouples.
The left-hand side of Eq.~\eqref{20VIII19.15a} depends on covariant derivatives, but these can be replaced
with simple partial derivatives because
\begin{equation}
\label{27X19-4}
	\Gamma^{\mu}{}_{ z\beta}  = 0
 \ \mbox{ for } \ \beta \ne 3
 \,.
\end{equation}
Moreover, the right-hand side of the $z$-component of Eq.~\eqref{20VIII19.15a} does depend on the $x$- and $y$-components of $Z^{\mu}$ because
on the exterior equatorial hyperplane we have
\begin{equation}\label{1XI19.1}
  R^\alpha{}_{\beta\gamma z} \equiv 0
  \ \mbox{ unless  }\ \alpha = 3
   \,.
\end{equation}
From the arguments above, it follows that there exists a consistent solution of the geodesic deviation equation of the form
$Z^\mu \partial_\mu = f(s) \partial_z$, with
\begin{equation}\label{27X19.6}
  \frac{DZ^ \mu}{ds} =
   \left(0,0,0,\frac{df}{ds}
   \right)
   \,,
   \quad
  \frac{D^2Z^ \mu}{ds^2} =
   \left(0,0,0,\frac{d^2f}{ds^2}
   \right)
   \,.
\end{equation}

Our calculations so far show that $f$ satisfies
\begin{align}\label{27X19.7}
  \frac{d^2 f}{ds^2} &\approx c_2 s^{-2} f
\end{align}
with
\begin{align}
   c_2 &:= \frac{3 (\mcL - a\mcE)^2 m}{4 \sqrt{2} a^{5/2} }\left(\frac{4}{5c_1}\right)^2
    = \frac{6}{25}    \,.
\end{align}
This leads to
\begin{align}\label{27X19.8}
  f(s) &=c_+ s^{\alpha_+}
  + c_- s^{\alpha_-}
  \,,
  \nonumber \\
  & \mbox{ with } \
  \alpha_{\pm} = \frac{1 \pm \sqrt{1+4c_2}}{2}=\left\{
                                                 \begin{array}{ll}
                                                   \frac{6}{5} \\
                                                   -\frac{1}{5}
                                                 \end{array}
                                               \right.
    \,.
\end{align}
Since the metric length of $Z^\mu$ equals $|f|$, it follows that the
tidal forces will tear physical objects apart as $s \to 0$, or
equivalently as $\RR \to 0$ by Eq.~\eqref{27X19.5}. The physical
interpretation of this result is that \emph{as massive bodies approach
  the ring on the exterior equatorial plane, they will experience a
  force in the direction perpendicular to the plane that would push
  adjacent particles away from each other by an infinite amount if the
  object could reach the ring}.  This tidal force will break the
object apart presumably well before the ring is reached, depending on
its extent and other physical properties.

\subsection{Tidal forces on geodesics that approach but never hit the ring singularity}
\label{sec:ss1XI19.1}

We have seen in the previous section that no timelike geodesic  in the region $H<0$  ($\Leftrightarrow \epsilon=-1$), can hit the ring. But one can instead consider what happens on sequences of geodesics which come arbitrarily close to the ring.

This situation is made simpler by the fact that when $H<0$ the vector field $\partial_t$ is timelike everywhere, since $g(\partial_t, \partial_t) \equiv g_{tt} = -1 +H$. To get a glimpse into the tidal forces one can  {therefore} consider those  timelike geodesics, parameterised by proper time, which at $s=0$ meet the plane $\{y=0\}$ with tangent
\begin{equation}\label{20X19.11}
{\dot \gamma}^{\mu}(0)=\left(\dot t(0),0,0, 0\right)
\end{equation}
and with $ x(0)>a$. We emphasise that this form of $\dot \gamma$ is \emph{not} valid for all $s$, but only at $s=0$. We normalise $s$ so that $g(\dot \gamma, \dot \gamma) = -1$, leading to
\begin{align}\label{7X19.1}
	\dot{t}^{2} &= \frac{1}{1-H} = \frac{1}{1-2m/\tilde{r}} = \frac{1}{1-\frac{2m\epsilon}{\sqrt{\RR(2 a+\RR)}}}
	\nonumber \\
	&= -\epsilon\frac{\sqrt{a}\sqrt{\RR}}{\sqrt{2}m} + \bigO{\RR}
\end{align}
for small $\RR$. (It clearly follows that timelike geodesics as in Eq.~\eqref{20VIII19.11} cannot exist close to the ring if $\epsilon =1$.)

Let us now consider the tidal forces experienced by a sequence of such geodesics that approach (but never hit) the ring.
Using the formulae in Appendix~\ref{sA1XI19.1}, one finds that the leading-order expansion in $\RR$ of the right-hand side of Eq.~(\ref{20VIII19.15a}) on a geodesic with tangent given by  Eq.~\eqref{20X19.11} at $s=0$ takes the form
\begin{equation}
	\label{eq:402}
R^{\mu}{}_{\alpha\beta\nu}{\dot \gamma}^{\alpha} \dot \gamma^{\nu} \approx
-\frac{3 }{4 \RR^{2}}
\left(
\begin{array}{cccc}
 0 & -\frac{m}{a} & 0 & 0 \\
 0 & -\frac{1}{2} & 0 & 0 \\
 0 & -\frac{m}{a} & 0 & 0 \\
 0 & 0 & 0 & \frac{1}{2} \\
\end{array}
\right)
\,,
\end{equation}
keeping in mind that $\epsilon=-1$.
Taking $Z^\mu(0)$ to be the unit vector in the direction $\partial_z$, which is orthogonal to the equatorial hyperplane, we find that the resulting tidal acceleration is also orthogonal to the equatorial hyperplane and equals
\begin{equation}\label{7X19.1b}
  \left| \frac{d^2Z}{ds^2}(0)\right|
   \approx \frac{3}{8 \RR^{2}}
   \,,
\end{equation}
where $|\cdot |$ denotes the  length of a vector with respect to the metric $g$.

Consider, then, a sequence of geodesics $\gamma_{(i)}$ as above passing through points $(0,a+\RR_{(i)},0,0)$, with $\RR_{(i)}\to 0$.
(Here we put the sequence index $_{(i)}$ in brackets, to avoid confusion with vector components.)
If we denote by $|\ddot{Z}^\perp_{(i)}|$ the length of the part of the tidal acceleration vector, which is orthogonal to the equatorial hyperplane, we will have
\begin{equation}\label{2XI19.2}
 \lim_{\RR_{ {(i)}}\to0}|\ddot{Z}^\perp_{(i)}|\ge \lim_{\RR_{ {(i)}}\to0} \frac{ {3}}{ {8 \RR^{2}_{(i)}} } =  \infty
\,.
\end{equation}
This provides another sense in which  the tidal forces are unbounded near the ring.

We note that the energy $\mcE$ of the geodesics satisfying Eq.~\eqref{20X19.11} equals
\begin{equation}\label{2XI19.1}
 \mcE = (1-H) \dot t(0)
 \,.
\end{equation}
Together with Eq.~\eqref{7X19.1}, this implies that
$$
 \mcE_{(i)}\to_{ {(i)}\to\infty}\infty
$$
for any sequence of geodesics for which $\RR_{(i)}\to 0$. The physical interpretation here is that \emph{such geodesics coming in from infinity, if any exist, need a tremendous amount of initial energy to come close to the ring, and when they do, they experience a very large acceleration tangential to the exterior equatorial hyperplane}. We expect that something similar occurs for  geodesics more general than the ones considered above.

\section{Tidal forces for accelerated observers}
\label{s1XI19.11}

One would also like to quantify the forces felt by objects hitting the singular ring from directions other than from the exterior equatorial hyperplane. The usual way of doing this is to calculate the tidal forces along geodesics. But, as already pointed out, only very special geodesics accumulate at the ring. The question then arises as to how to describe forces on \emph{non-geodesic} trajectories approaching the ring.

For this we, start by revisiting the usual geodesic-deviation calculation for a family of non-geodesic curves. Let
$\lambda\mapsto x^\mu(s,\lambda)$ be a one parameter family of timelike curves, each of them parameterised by proper time $s$.  Set
\begin{equation}\label{20VIII19.11}
 \dot \gamma^\mu := \frac{dx^\mu}{ds}
 \,, \;\;
 a^\mu := \frac{D \dot \gamma^\mu}{ds} \equiv
 \frac{d \dot \gamma^\mu}{ds } + \Gamma^\mu _{\alpha\beta} \dot \gamma^{\alpha} \dot \gamma^\beta
 \,, \;\;
 Z^\mu := \frac{dx^\mu}{d\lambda}
 \,.
\end{equation}
The calculation leading to the geodesic deviation equation leads instead to the following equation:
\begin{equation}\label{20VIII19.15}  \frac{D^2 Z^\mu}{ds^2}
 = R^\mu{}_{\alpha \beta \gamma} \dot \gamma^\alpha Z^\beta \dot \gamma^\gamma
 +  \frac{D a^\mu}{d\lambda}
 \,.
\end{equation}
The first term is the usual tidal force due to the gravitational field. The second term has the interpretation of the supplementary force that needs to be applied to a nearby object so that it follows the  trajectory \red{$x^\mu(s,\lambda)+ Z^\mu (s,\lambda) \, d\lambda$.
}

As in the previous section, we will only analyze what happens on a specific family of curves, namely radial outward-directed curves contained inside the equatorial disc. One could think of this disk as an \emph{inner equatorial hyperplane}, the complement of the exterior equatorial hyperplane discussed in the previous section. Since the metric induced there is the Minkowski metric, these are geodesics of three-dimensional Minkowski space-time. However, these are not geodesics of the Kerr spacetime.

In order to check this, let $v\in (0,1)$ and  consider the radial curves
\begin{eqnarray}
&&[0,a)\ni   s\mapsto
\nonumber
\\
&&\left(
 t=\frac s{\sqrt{1-v^2}},\, x= \frac {vs}{\sqrt{1-v^2}}  ,\, y= 0,\, z=0
  \right)
   \,,
    \phantom{xxx}
    \label{28IX19.7}
\end{eqnarray}
which are proper-time-parameterised timelike geodesics of the  metric induced on $D$.
Letting the Latin indices $a$, $b$, $c$, and $d$ range over $\{0,1,2\}$, the  
acceleration four-vector equals
\begin{align}
  a^\mu \partial_\mu
   & :=
   \left(
     \frac{d^2x^\mu}{ds^2}
   + \Gamma^\mu _{\alpha\beta} \frac{dx^\alpha}{ds } \frac{dx^\beta}{ds }
    \right) \partial_\mu
   =
   \Gamma^\mu _{ab} \frac{dx^a}{ds } \frac{dx^b}{ds } \partial_\mu
   \,.
   \label{28IX19.8-}
\end{align}
To continue, we need to calculate the relevant Christoffel symbols. For this we start by noting that
the extrinsic curvature tensor of the disc, say $k_{ab}$, as defined with respect to the field of unit normals  $N=\partial_z$,  reads
\begin{equation}\label{27IX19.1}
  k_{ab} = \frac 12 \partial_z(H \KStheta_a \KStheta_b)|_{z=0}
  \,.
\end{equation}
Next we use that
%
\begin{align}
 &
 \displaystyle
  \tilde{r} |_D = 0\,, \quad
  H|_D = 0\,, \quad
 \partial_z \tilde{r} |_D = \frac{ a }{\sqrt{a^2-\rho^2}}\,,
 \nonumber \\
 &
  \partial_z H|_D = \frac{2 a m}{(a^2-\rho^2)^{3/2}}\,,
  \nonumber \\
 &
 \displaystyle
  \KStheta_\mu |_D\diff x^\mu  =
-
\diff{t} +  a^{-1} (x\diff{y}-y\diff{x})
-\frac{ \sqrt{a^2-\rho^2}}{a} \diff z
 \,,
\nonumber \\
&  \partial_z\KStheta_a|_D \diff x^a =
-  \frac{x\diff{x}+y\diff{y} }{a\sqrt{a^2-\rho^2}}
 \,,
 \label{28IX19.1}
\end{align}
to find
\begin{align}\label{28IX19.1b}
  k_{ab} \diff x^a \diff x^b
  &= \frac 12 \partial_z H|_{z=0}  (\KStheta_a  \diff x^a)^2
   \nonumber \\
   &=
 \displaystyle
   \frac{  m}{a^2(1-\frac{\rho^2}{a^2})^{3/2}}
   \left(
   -
\diff{t} +  a^{-1} (x\diff{y}-y\diff{x})
 \right)^2
 \,.
\end{align}
The non-vanishing Christoffel symbols of $g$ on $D$ then read
%
\begin{align}\label{28IX19.4}
  \Gamma^z_{ab}|_D& = - k_{ab}
  \,,
  \quad
  \Gamma^a_{zb}|_D =   k^a{}_{b}
  \,,
  \nonumber \\
  \Gamma^a_{zz}|_D &=
  - \partial_z H \eta^{ab}\KStheta_b \KStheta_z|_D = {-}\frac{2m}{a^2 -\rho^2}\eta^{ab}\KStheta_b
  \,,
  \nonumber \\
  \Gamma^z_{zz}|_D &=
  \frac 12 \partial_z H     \KStheta_z^2|_D = \frac{{m}}{a^2\sqrt{1 -\frac{\rho^2}{a^2}}}
  \,.
\end{align}
Returning to Eq.~\eqref{28IX19.8-}, we finally obtain
\begin{align}
  a^\mu \partial_\mu
   & =
   \Gamma^\mu _{ab} \frac{dx^a}{ds } \frac{dx^b}{ds } \partial_\mu = -k_{ab}\frac{dx^a}{ds } \frac{dx^b}{ds } \partial_z
   \nonumber \\
   &=
   -
 \displaystyle
   \frac{ a m}{(a^2-\rho^2)^{3/2}}\left(\frac{dt}{ds }\right)^2 \partial_z
   \nonumber
\\
   &  =
   -
 \displaystyle
   \frac{ a m }{(a^2-\rho^2)^{3/2}(1-v^2)}  \partial_z
   \,.
   \label{28IX19.8}
\end{align}

The above results show  that the acceleration needed to remain on the trajectory of Eq.~\eqref{28IX19.7} grows without bound when the ring is approached. Moreover, since $(a^2-\rho^2)^{-3/2}$ is not integrable in $\rho$ near $a$, an infinite amount of energy would be needed
to remain on this trajectory. The physical interpretation here is that at some stage the observer would need to switch off his or her engine and continue in free fall. The geodesic deviation equation will describe the forces experienced from that time on.

 The question then arises, what forces will be acted upon the observer when he or she runs out of fuel.
The evaluation of the geodesic deviation equation requires the Riemann tensor on the disc. Here the Gauss-Codazzi-Mainardi embedding equations are especially convenient, since the metric induced on the disc is flat.
Using these equations we find
\begin{align}
 \label{8XI19.1}
  R^a{}_{bcd} &= k^a{}_d k_{bc} - k^a{}_{c} k_{bd}\,,
\\
  R_{zabc} &= R^{z}{}_{abc}=\partial_c k_{ba} - \partial_b k_{ca}
  \,,
\\
  R_{zazb} &=R^{z}{}_{azb}= \underbrace{R^{\mu}{}_{a\mu b}}_0 -R^{c}{}_{ac b}
 = k^c{}_{c} k_{ab} -k^c{}_a k_{cb}
  \,.
   \label{8III19.2}
\end{align}
Using the product structure  of $k$, namely $k_{ab}\sim \KStheta_a \KStheta_b$,  Eq.~\eqref{8XI19.1} shows immediately that $ R^a{}_{bcd} \equiv 0 $, and then the second equality in Eq.~\eqref{8III19.2}  gives $
  R_{zazb}\equiv 0$.
With some work one further finds
\begin{align}
	\label{eq:405}
	R_{ztti} &= \frac{3am}{\left(a^{2}-\rho^{2}\right)^{5/2}}\,x^{i}
	\,,
	\nonumber \\
	R_{ztij} &= \frac{m(2a^{2}+\rho^{2})}{\left(a^{2}-\rho^{2}\right)^{5/2}}\,\epsilon_{ij}
	\,,
	\nonumber \\
	R_{zijl} &= \frac{3am}{\left(a^{2}-\rho^{2}\right)^{5/2}}\,\epsilon_{in}x^{n}\epsilon_{jl}
	\,,
\end{align}
where $i,j,l,n\in\{1,2\}$ and $\epsilon_{ij}$ is antisymmetric with $\epsilon_{12}=1$.
The non-vanishing components of the gravitational tidal force in Eq.~(\ref{20VIII19.15}) along the curves of Eq.~(\ref{28IX19.7}) therefore read
\begin{align}
	\label{eq:500}
	R^z{}_{\mu t \nu} \dot \gamma^{\mu} \dot \gamma^{\nu}
 &= \frac{3  a m  \rho}{(a^{2}-\rho^{2})^{5/2}} \times \frac{v}{1-v^2}
	\,,
	\\
	\label{eq:501}
	R^z{}_{\mu x \nu} \dot \gamma^{\mu} \dot \gamma^{\nu} &= -\frac{3  am\rho}{(a^{2}-\rho^{2})^{5/2}}\times \frac{1}{1-v^2}
	\,,
	\\
	\label{eq:502}
	R^z{}_{\mu y \nu} \dot \gamma^{\mu} \dot \gamma^{\nu} &= -\frac{3a^2 m }{(a^{2}-\rho^{2})^{5/2}} \times \frac{v}{1-v^2}
	\,.
\end{align}
As before, we conclude that tidal forces grow without bound for \emph{generic} $Z^{\mu}$ as the ring singularity is approached.



\section{Adding an electric charge and a cosmological constant}
\label{sec:electric-CC}

Most considerations so far
generalise to charged metrics with a cosmological constant, as discovered independently by Carter~\cite{CarterSeparable} and by Demia\'nski~\cite{Demianski}.
The metrics can be written in a form~\cite{CarterSeparable,Exactsolutions2} formally identical to their vacuum asymptotically flat counterparts,

\begin{align}
  \label{14VIII19.11}
    g &= \rhoSquareBL \left(\frac{1}{\Deltar } \diff \tilde{r}^2+\frac{1}{\Deltatheta} \diff \BLtheta^2\right)
      \nonumber \\
      &+\frac{\sin^2(\BLtheta)}{\rhoSquareBL \Xi^{2}}\Deltatheta \left(a\, \diff \tildet -(\tilde{r}^2+a^2)\, \diff \BLphi\right)^2
\nonumber \\
&-  \frac{\Deltar }{\rhoSquareBL \Xi^{2}}\left(\diff \tildet -a \sin^2(\BLtheta) \, \diff \BLphi\right)^2
    \,,
\end{align}
where only the functions $\Deltar$, $\Deltatheta$ and $\Xi$ differ from the corresponding functions in vacuum:
\begin{eqnarray}
  \label{14VIII19.12}
    \rhoSquareBL  &=& \tilde{r}^2+a^2\cos^2(\BLtheta) \vphantom{\frac11}
     \,,
\\
  \label{14VIII19.13}
    \Deltar  &=& (\tilde{r}^2+a^2)\left(1-\frac{\Lambda}3 \tilde{r}^2\right)-2 m \tilde{r} + \charge ^2
     \,,
\\
  \label{14VIII19.14}
    \Deltatheta &=& 1+\frac{\Lambda}3 a^2 \cos^2(\BLtheta)
     \,,
\\
  \label{14VIII19.15}
    \Xi &=& 1+\frac{\Lambda}3 a^2
    \,.
\end{eqnarray}
The quantity $\Lambda$ here is the cosmological constant, while the quantity $\charge$ is the charge in suitable units.
Whatever the values of the constants $a$, $m$, $\Lambda$ and $\charge $, the singularity is located at $\tilde{r}=0=\cos(\BLtheta)$.

In the case $\Lambda=0$ the procedure is obvious, as the Kerr-Newman metric can also be written in the Kerr-Schild form, where the singular ring has the intuitive interpretation of a coordinate ring in the underlying Minkowski spacetime.
Nevertheless, when $\Lambda \ne 0$, one can still use the coordinate transformations of Eq.~\eqref{14VIII19.1} followed by Eq.~\eqref{19VII19.4} and inquire about the small-$\RR$ asymptotics of the metric.

\subsection{Charged metric without a cosmological constant}

Consider the Kerr-Newman metric with $a,m>0$, with charge parameter $\charge \ne 0$, and with $\Lambda = 0$.
The explicit formulae for the curvature invariants are the following
\begin{widetext}
\begin{align}
K &= \frac{48}{\left(\tilde{r}^{2} + a^{2} \cos^{2}(\BLtheta)\right)^{6}}
\left\{
m^{2} \left(\tilde{r}^{2} - a^{2} \cos^{2}(\BLtheta) \right) \left[\left(\tilde{r}^{2} + a^{2} \cos^{2}(\BLtheta)\right)^{2} - 16 a^{2} \tilde{r}^{2} \cos^{2}(\BLtheta) \right]
\right.
\nonumber \\
&\left. - 2 m \tilde{r} \charge^2 \left[\tilde{r}^4 - 10 a^2 \tilde{r}^2 \cos^2(\BLtheta) + 5 a^4 \cos^4(\BLtheta)\right]
+ \frac{\charge^4}{6} \left[7 \tilde{r}^4 - 34 a^2 \tilde{r}^2 \cos^2(\BLtheta) + 7 a^4 \cos^4(\BLtheta)\right]
\right\}
\,,
\\
P &=  - 192 a m^2 \tilde{r} \cos (\BLtheta)\frac{
   \left(\tilde{r}^{2}  -  3 a^{2} \left(1- \frac{\charge^2}{3 m \tilde{r}} \right)\cos^{2}(\BLtheta) - \frac{\charge^{2} \tilde{r}}{m} \right)
   \left(3 \tilde{r}^{2} - a^{2} \cos^{2}(\BLtheta) - 2 \frac{\charge^{2} \tilde{r}}{m} \right)
   }{\left(\!\tilde{r}^2+a^2 \cos^{2}(\BLtheta)\!\right)^6}\,.
\end{align}
\end{widetext}
Transforming from Boyer-Lindquist to Kerr-Schild coordinates, and then from these to
toroidal coordinates as we did in Sec.~\ref{s19VII19.2}, we find that close to the ring
\begin{align}
K &= \frac{\charge^{4}}{2 a^{4} \RR^{4}} \left(1 + 6 \cos(\KSv) \right)+ \bigO{\RR^{-3}}\,,
\\
P &= - \frac{6\charge^{4}}{a^{4} \RR^{4}} \sin(2\KSv) + \bigO{\RR^{-3}}\,,
\end{align}
where we recall that $\RR$ is the torus coordinate, which is related to the Kerr-Schild coordinates as in the $\charge=0$ case.
As before, there are directions in $\KSv$ along which either $K$ or $P$ vanish as one approaches the ring,
but upon computing the sum of the squares of these invariants, we find
\begin{align}
 K^2 + P^2
 &  =
  \frac{\charge^8 (12 \cos (2 \KSv)-54 \cos (4 \KSv)+91)}{4 a^8 \RR^8}
  + \bigO{\RR^{-7}}
  \,,
  \label{19VII19.3a}
\end{align}
Here there are no $\KSv$ directions along which the sum of these curvature invariants vanishes as one approaches
the ring because $91> 54 + 12$. Therefore, the above expression again establishes the singular character of the ring independently of the direction of approach.
Note moreover that the the divergent character of the curvature invariants is enhanced in the charged case, since when $e = 0$, the sum of the squares of these
invariants diverged as $\RR^{-6}$.

In transforming from Boyer-Lindquist coordinates to toroidal coordinates, we first had to go through Kerr-Schild coordinates.
The transformation between Kerr-Schild and Boyer-Lindquist coordinates is still given by Eqs.~\eqref{14VIII19.1} and \eqref{14VIII19.1+},
but now the function $\DeltaBL$ is the one appropriate for the charged case. With this transformation, the metric continues to take
the Kerr-Schild form of Eq.~\eqref{21VI119.14}, but with the metric function $H$ now
\begin{equation}
	\label{2X19.0}
	H = \frac{2m\tilde{r}^{3}-\charge^{2}\tilde{r}^{2}}{\tilde{r}^4+a^2 z^2}
	\,.
\end{equation}
This implies that one of the symmetries of $H$ is lost and only the second identity of Eq.~(\ref{21IX19.2}) holds, as now we have
\begin{equation}
	\label{2X19.1b}
	H(\RR, \KSv +2\pi) \neq -H(\RR, \KSv ),\quad H(\RR,-\KSv )=H(\RR, \KSv )
	\,.
\end{equation}
Indeed, after rewriting  $H$ in terms of toroidal coordinates one finds
\begin{equation}
	\label{2X19.2}
	H = H|_{\charge=0}
    - \frac{\charge^{2}}{\RR\sqrt{4a^{2}+\RR^{2}+4a\RR\cos{\KSv}}}
   \,,
\end{equation}
with the second term in Eq.~(\ref{2X19.2}) symmetric with respect to $\KSv \rightarrow -\KSv $, and  periodic with period $2\pi$.

The small-$\RR$ behavior of $H$ is dramatically different now,
\begin{equation}
	\label{2X19.3}
	H = -\frac{\charge^2}{2 a \RR}+\frac{\sqrt{2} m \cos \left(\KSv/2\right)}{\sqrt{a}
   \sqrt{\RR}}+
   \bigO 1
   \,,
\end{equation}
which one should compare with Eq.~(\ref{27VII19.2}). Thus, $H$ tends to minus infinity near the ring, independently of the direction of approach, and this at a rate twice as fast as before. This increased rate of divergence near the ring is responsible for the increased rate of divergence of the quadratic curvature invariants calculated above.

As already pointed out, the character of the metric induced on the orbits of the symmetry group can be understood from the determinant of this metric.
In Boyer-Lindquist coordinates, we find
\begin{equation}
	\label{2X19.4}
	g_{\tildet\tildet}g_{\BLphi\BLphi} - g_{\tildet\BLphi}^2
	= - \greenn{\DeltaBL} \sin^{2}(\BLtheta)
	\,,
\end{equation}
with a negative limit as the singular ring is approached, similar to the uncharged case. It follows that the orbit space metric remains Lorentzian near the ring, with
\begin{equation}
	\label{2X19.5}
 \lim_{\RR \to 0, \, \sin^2(\BLtheta) \to 1} \left(g_{\tildet \tildet } g_{\BLphi \BLphi} - g_{\tildet \BLphi}^2
 \right)= -a^2 -\charge^{2}
 \,,
\end{equation}
In toroidal coordinates and near the ring, we have the asymptotic expansion (cf.~Eq.~(\ref{18X19.01}))
\begin{align}
	\label{2X19.6}
	g_{tt}g_{\KSu\KSu} - \left(g_{t\KSu}\right)^{2} &=
	- a^{2} - \charge^{2} + 2\sqrt{2}m\sqrt{a\RR}\cos\left(\KSv/2\right)
	\nonumber \\
	&+ \frac{\left(\charge^2-\left(2a^2+\charge^2\right) \cos(\KSv)\right)\RR}{a}
	+ \bigO{\RR^{3/2}}
	\,.
\end{align}

\begin{figure*}
\begin{center}
\includegraphics[width=0.9\textwidth]{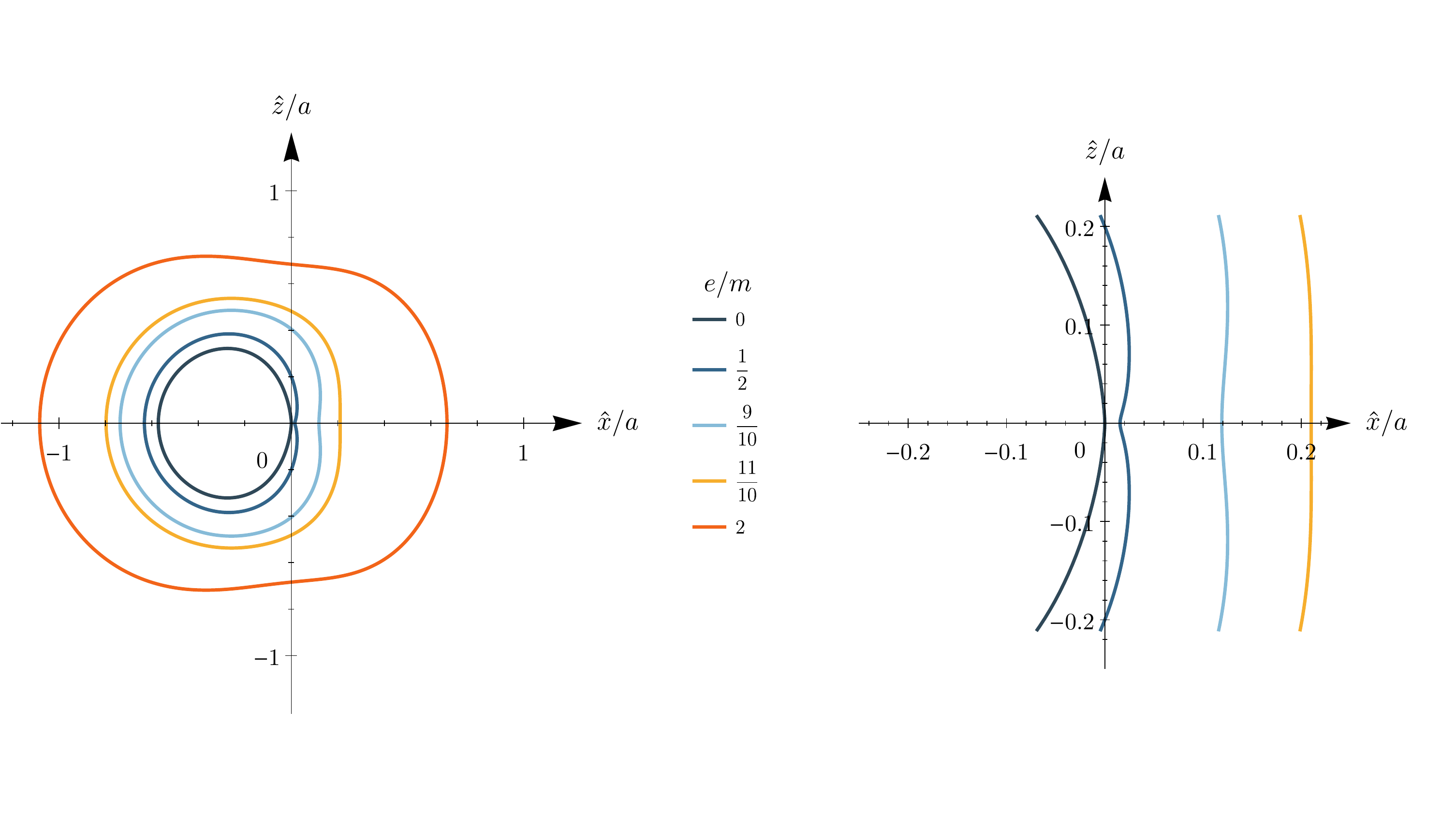}
\caption{In the hatted coordinates of Eq.~\eqref{23IX19.5}, the Killing vector $\partial_\varphi$ is timelike in the region bounded by the curves of the left panel, with a zoom to the location of the ring (which sits at the origin of the coordinates) in the right panel. Here $a=2m /3$ and $e/m\in\{0,1/2,9/10,11/10,2\}$.
 }
\label{F1XI19.1}
\end{center}
\end{figure*}

Let us now consider the causal character of the Killing vectors.
The causal character of $\partial_{\KSu }$ is determined by
\begin{equation}
	\label{2X19.7}
	g_{\KSu\KSu} = -\frac{a \charge^2}{2 \RR}+\frac{\sqrt{2}
   a^{3/2} m \cos \left(\KSv/2\right)}{\sqrt{\RR}}
   + \bigO{1}
   \,,
\end{equation}
which note recovers Eq.~(\ref{28VII19.2}) when $\charge=0$.
The right-hand side of the above equation is now negative for all $\RR$ small enough,
regardless of the direction of approach. \emph{This implies that the regions of spacetime with causality-violating,
closed time-like curves near the ring singularity is now larger than in the uncharged case.} 
This is illustrated  in Figure~\ref{F1XI19.1}.
The causal character of $\partial_{t}$ is determined by
\begin{equation}
	\label{2X19.8}
	g_{tt} = -\frac{e^2}{2 a \RR}+\frac{\sqrt{2} m \cos \left(\KSv/2\right)}{\sqrt{a}
   \sqrt{\RR}}
   + \bigO{1}
\,,
\end{equation}
which recovers Eq.~(\ref{28VII19.3}) when $\charge=0$.
The above equation shows that the $\partial_{t}$ Killing vector is
also timelike near the ring in all angular directions, as opposed to the vacuum case in Eq.~(\ref{28VII19.3}).
\emph{This implies that there are no ergoregions near the ring singularity in the charged case.}

We pass now to the question of how the distribution of null Killing vectors is affected by the charge and the cosmological constant near the ring.
The Killing vectors of the form
\begin{equation}\label{14VIII19.31+}
 X = \cos(\alpha_\pm) \partial_t + \sin (\alpha_\pm) \partial_{\KSu}
\end{equation}
which are null at a  point with coordinates  $(\RR, \KSv )$ have now the expansion
\begin{align}
	\label{3X19.0+}
	\cos(\alpha_{\pm}) &= \frac{a}{\sqrt{1+a^2}}
	\nonumber \\
	&+\frac{\left(2 a \left(-a \pm \sqrt{a^2+e^2}\right)+e^2
   (-1+\cos (\KSv))\right)}{\left(1+a^2\right)^{3/2} e^2}\RR
   \nonumber \\
   &+ \bigO{\RR^{3/2}}\,
\end{align}
with a seemingly problematic limit when $\charge\to 0$ for $\alpha_-$, which is due to the non-uniformity in $\charge$ of the error terms. At fixed $\charge$ we obtain a $\KSv$-independent limit as $\RR$ goes to zero, identical to the uncharged case, for  both  angles.

 We address now the question of  the behavior of the Ricci scalar of the quotient space metric $q$ defined by Eq.~\eqref{20VIII19.1} and \eqref{20VIII19.2}.
The quotient space-metric with $\charge\ne 0$ in  coordinates
\begin{equation}\label{4X19.1b}
 (x,z)=(\RR \cos(\KSv),\RR \sin(\KSv))
  \end{equation}
centered at the ring is
\begin{align}
	\label{4X19.0}
	q_{xx} &= \!\frac{a^2\!+\!e^2 \sin ^2\left(\KSv/2\right)}{a^2+e^2}\!
	\nonumber \\
	&+\!\frac{2^{3/2}
   a^{5/2} m \cos ^3\left(\KSv/2\right) \sqrt{\RR}}{\left(a^2+e^2\right)^2}
   \!+\! \bigO{\RR}
\\
	\label{4X19.1}
	q_{xz} &= -\frac{e^2 \sin (\KSv)}{2 \left(a^2+e^2\right)}
	\nonumber \\
	&+\frac{2\sqrt{2} a^{5/2} m \sin\left(\KSv/2\right) \cos^2\left(\KSv/2\right) \sqrt{\RR}}{\left(a^2+e^2\right)^2}
   + \bigO{\RR}
   \,,
   \\
   q_{zz} &= \frac{2 a^2+e^2+e^2 \cos (\KSv)}{2 \left(a^2+e^2\right)}
 \nonumber \\
 &+\frac{\sqrt{2} a^{5/2} m
   \sin \left(\KSv/2\right) \sin (\KSv)
   \sqrt{\RR}}{\left(a^2+e^2\right)^2}
   + \bigO{\RR}
   \,,
\end{align}
From this, we can \rout{easily} compute the determinant to be
\begin{align}
	\label{4X19.3}
	\textrm{det}(q) &= \frac{a^2}{a^2+e^2}
	+ \frac{2 a^{5/2} m \left(\cos(\KSv) + 1\right)^{1/2} \RR^{1/2}}{(a^{2}+e^{2})^{2}}
	+ \bigO{\RR}\,.
\end{align}

In calculating the above, or the Ricci scalar $R(q)$, either one can
work directly in Boyer-Lindquist coordinates in which $q_{AB}=g_{AB}$,
or one can work in Kerr-Schild coordinates (where
$q_{AB} \neq g_{AB}$) and one must compute the full expression in
Eq.~\eqref{20VIII19.1}. Doing the latter to leading order in an
$\RR \ll 1$ expansion, we find
\begin{align}
\mychi^{ab} (X_a)_{\RR} (X_b)_{\RR} &= - \frac{\charge^{4} \left(\cos(\KSv) + 1\right)}{2 a^{2} \left(a^{2} + \charge^{2}\right)} + \bigO{\RR^{1/2}}\,,
\\
\mychi^{ab} (X_a)_{\RR} (X_b)_{\KSv} &=
 \frac{\charge^{4} \sin(\KSv)   \RR} {2 a^{2} \left(a^{2} + \charge^{2}\right)} + \bigO{\RR^{{3/2}  }}\,,
\\
\mychi^{ab} (X_a)_{\KSv} (X_b)_{\KSv} &= \frac{\charge^{4} \left(\cos(\KSv) - 1\right) {\RR^{2}}}{2 a^{2} \left(a^{2} + \charge^{2}\right)} + \bigO{\RR^{5/2}}\,,
\end{align}
while in the $\charge = 0$ case the above expressions are all $\bigO \RR$ higher.

\begin{figure}[thb]
\begin{center}
\includegraphics[width=0.9\columnwidth]{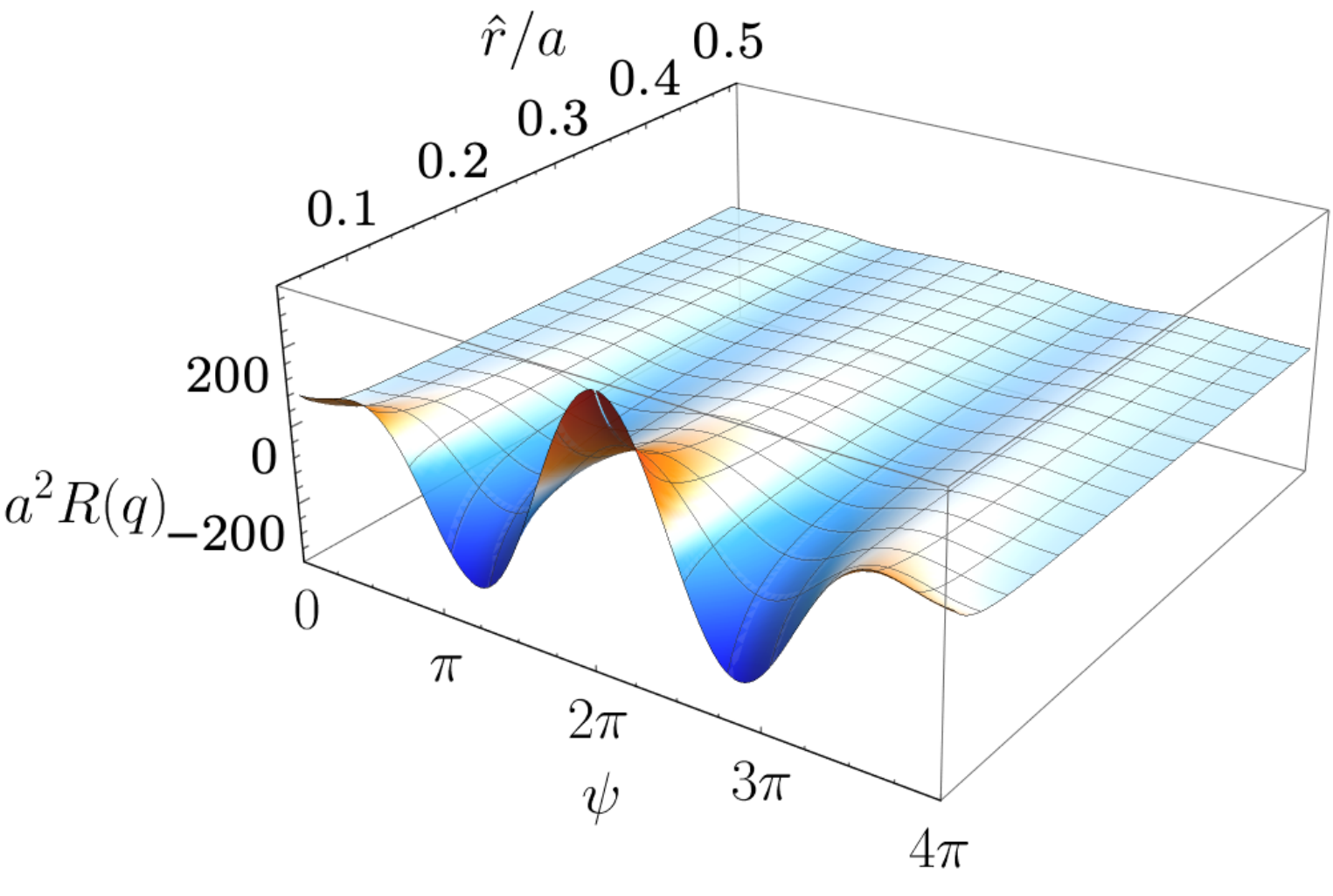}
\caption{The asymptotic form of the Ricci scalar of the quotient space metric $q$ close to the singular ring changes for $e\neq 0$, see Eq.~(\ref{2X19.10}). Here we plot a representative graph of the Ricci scalar  $a^2R(q)$
as a function of $\RR/a\,\in[1/20,\,1/2]$ and $\KSv\in[0,\,4\pi]$, with  $a/m=1/2$ and $e/m=1$. Compare Fig.~\ref{F14VIII19.2}.}
\label{F7XII19.1}
\end{center}
\end{figure}

The asymptotic expansion of the Ricci scalar about the ring singularity is more singular when compared to the vacuum case in Eq.~(\ref{21IX19.5-b}),
since now we find
\begin{align}
	R(q)
	\label{2X19.10}
	&=
   \frac{e^2\cos(\KSv)}{2a^2\RR^2}
 - \frac{m\cos\left(3\KSv/2\right)}{\sqrt{2}a^{3/2}\RR^{3/2}}
	 + \bigO{1}
   \,.
\end{align}
See Figure~\ref{F7XII19.1}. 
We can conformally transform the metric $q$ to get rid of the most-singular part of $R(q)$.
Letting $\tilde{q} = e^{2 \sigma} q$, with
\begin{equation}
 e^{2 \sigma} = 	2a^{2}+e^{2}\left(1-\cos(\KSv)\right)\,,
\end{equation}
the (2-dimensional) Ricci scalar transforms as
\begin{equation}
\tilde{R}(q) = e^{-2\sigma} \left[R(q) - 2 \nabla^2 \sigma\right]\,.
\end{equation}
The expansion of the conformally transformed Ricci scalar is then
\begin{align}
	R(\tilde{q}) &= - \sqrt{2} \sqrt{a} \; m \cos \left(\KSv/2\right)
	\nonumber \\
	&\times  \frac{-2 a^2-3 e^2+\left(4
   a^2+3 e^2\right) \cos(\KSv)}{\left(2 a^2+e^2-e^2 \cos(\KSv)\right)^3 \RR^{3/2}}
+ \bigO{\RR^{-1 }}
\,.
\end{align}
This singularity is mild-enough to obtain existence of coordinates in which $\tilde q$ asymptotes, as $\RR$ tends to zero, to a flat cone-metric of opening angle $4\pi$, which can be justified using, e.g.,\ Eqs.~(3.12) and (3.21) of \cite{MoncriefNI}.

Let us now consider the geometry of the level sets of $\RR$, which requires us to return to the full four-dimensional metric.
This geometry becomes more interesting than that following from Eq.~(\ref{28VII19.1}),
\begin{align}
	\label{2X19.14}
	g(\nabla \RR,\nabla\RR) &\equiv g^{\RR\RR} =
	\frac{2 a^2+e^2+e^2 \cos (\KSv)}{2 a^2}
	\nonumber \\
&	-\frac{2  \sqrt{2} m \cos
   ^3\left(\KSv/2\right) \sqrt{\RR}}{a^{3/2}}
   + \bigO{\RR}
   \,.
\end{align}
 It follows that $\RR$ does not tend to the distance from the ring to the level sets of $\RR$ as the ring is approached, though it remains equivalent to it up to a $\KSv$-dependent factor, with Eq.~(\ref{4VII19.51}) taking now the form
\begin{equation}
  \label{2XI19.15}
  \int_0^{\RR}
  \sqrt{  g(\nabla \RR, \nabla \RR) }\,
  \diff{\RR} =
  \frac{\sqrt{a^2+e^2 \cos ^2\left(\KSv/2\right)}}{a} \RR +
  \bigO{\RR^{3/2}}
 \,.
\end{equation}
The causal character of $\partial_{\RR}$ is determined by
\begin{equation}
	\label{2X19.16}
	g_{\RR\RR} = \frac{a^2-e^2 \cos ^2\left(\KSv/2\right)}{a^2}+\frac{2 \sqrt{2} m \cos
   ^3\left(\KSv/2\right) \sqrt{\RR}}{a^{3/2}}
   + \bigO{\RR}
   \,,
\end{equation}
and so $\partial_{\RR}$ can change type near the ring for solutions with $e^2 > a^2$. Note that such solutions might be nakedly singular, but do not have to since the condition for the existence of naked singularity in the Kerr-Newman spacetime is $a^{2} + e^{2}  > m^2$.

Neither  $\partial_{\RR}$ nor $\partial_{\KSv}$  have a clear geometric character because they are tied to the Boyer-Lindquist coordinates $\BLphi$ and $\tildet$ which are somewhat arbitrary. But since the orbits of $\partial_{\KSv}$ are periodic, their causal character is of interest. This is determined by the sign of
\begin{align}
	\label{2X19.17}
	g_{\KSv\KSv} &= \left(1-\frac{e^2 \sin ^2\left(\KSv/2\right)}{a^2}\right)
   \RR^2
   \nonumber \\
&   +\frac{\sqrt{2} m \sin \left(\KSv/2\right) \sin (\KSv)
   \RR^{5/2}}{a^{3/2}}
	+ \bigO{\RR^3}
	\,,
\end{align}
which shows that $\partial_{\KSv}$ is timelike in some angular sectors
near the ring when $e^2 > a^2$. This does not necessarily lead to a
new way of violating causality, since $\partial_{\KSv}$ is always
spacelike in the remaining angular sectors.  If a causal vector field
has periodic orbits, then there is an obvious causality violation, but
otherwise, as in the case here, further study is needed to determine
whether there is a violation of causality or not.

\subsection{Charged metric with a cosmological constant}

As already mentioned, our analysis extends to a non-zero cosmological constant $\Lambda$. Both in the charged and in the uncharged cases we obtain the surprising identities
\begin{align}\label{14IV19}
  R_{\alpha\beta\gamma\delta}
  R^{\alpha\beta\gamma\delta}  \Big| _{\Lambda \ne  0}
   &=
  R_{\alpha\beta\gamma\delta}
  R^{\alpha\beta\gamma\delta}
  \Big|_ {\Lambda = 0}   + \frac{8}{3}\Lambda^{2}
  \,,
   \nonumber \\
  \epsilon_{\alpha\beta\gamma\delta}
  R^{\alpha\beta}{}_{\mu\nu}
    R^{\gamma\delta \mu \nu}\Big| _{\Lambda \ne  0}
   &=
  \epsilon_{\alpha\beta\gamma\delta}
  R^{\alpha\beta}{}_{\mu\nu}
    R^{\gamma\delta \mu \nu}
  \Big| _{\Lambda = 0}
  \,.
\end{align}
The leading order asymptotics of $K^2+P^2$ remains therefore the same, and so do the conclusions concerning the singular character of the ring.
The formulae to follow look somewhat nicer if one sets
$$ \lambda := {\Lambda}/{3}
\,,
$$
and we do so henceforth.

The determinant of the metric induced on the orbits of the group is straightforward to calculate in the Boyer-Lindquist coordinates and reads
\begin{align}
	\label{3X19.8}
	g_{\tildet\tildet}g_{\BLphi\BLphi} - g_{\tildet\BLphi}^2
	&= -\frac{\DeltaBL  \Delta _{\BLtheta} \sin ^2(\BLtheta)}{\Xi ^4}\,,
	\\
\lim_{\RR \to 0, \, \sin^2(\BLtheta) \to 1} g_{\tildet\tildet}g_{\BLphi\BLphi} - g_{\tildet\BLphi}^2 &= -\frac{a^2+e^2}{\left(1+a^2 \lambda \right)^4}
 \,.
\end{align}

The causal character of $\partial_{\KSu }$ is determined by
\begin{align}
	\label{3X19.3}
	g_{\KSu\KSu}
 &=
 -\frac{a \charge ^2}{2 \RR  \left(a^2 \lambda +1\right)^2}+\frac{\sqrt{2} a^{3/2} m \cos
   \left(\KSv/2\right)}{\sqrt{\RR } \left(a^2 \lambda +1\right)^2}
   \nonumber \\
&   +\frac{4 \left(a^4 \lambda
   +a^2+\charge ^2\right)-3 \charge ^2 \cos (\KSv)}{4 \left(a^2 \lambda +1\right)^2}+O\left(\sqrt{\RR }\right)
   \nonumber
\\
 &\ezero
 \frac{\sqrt{2} a^{3/2} m \cos \left(\KSv/2\right)}{\sqrt{\RR } \left(a^2 \lambda
   +1\right)^2}+\frac{a^2}{a^2 \lambda +1}+O\left(\sqrt{\RR }\right)
   \,.
\end{align}
The causal character of $\partial_{t}$ is determined by
\begin{align}
	\label{3X19.4}
	g_{tt}
& =
-\frac{\charge ^2}{2 \RR  a \left(a^2 \lambda +1\right)^2 }+\frac{\sqrt{2} m \cos
   \left(\KSv/2\right)}{\sqrt{a} \sqrt{\RR } \left(a^2 \lambda +1\right)^2}
   \nonumber \\
&   +\frac{4 a^4 \lambda -4
   a^2+\charge ^2 \cos (\KSv)}{4 a^{2} \left(a^2 \lambda +1 \right)^2}+O\left(\sqrt{\RR }\right)
   \nonumber
\\
 & =|_{\charge=0} \
 \frac{\sqrt{2} m \cos \left(\KSv/2\right)}{\sqrt{a} \sqrt{\RR } \left(a^2 \lambda
   +1\right)^2}+\frac{a^2 \lambda -1}{\left(a^2 \lambda +1\right)^2}+O\left(\sqrt{\RR }\right)
   \,.
\end{align}
The null Killing vectors of the form
\begin{equation}
\label{3X19.5}
	X = \cos(\alpha_\pm) \partial_t + \sin (\alpha_\pm) \partial_{\KSu}
\end{equation}
have the same leading order expansion as in the $\Lambda=0$ case, namely Eq.~\eqref{3X19.0+}.

The quotient-space metric is unaffected by $\Lambda$ in the first two leading terms. In the same coordinates
as those of Eq.~\eqref{4X19.1b} centered at the ring, $q_{xx}$ and $q_{yy}$ are given by Eqs.~\eqref{4X19.0} and
\eqref{4X19.1}.
The shortest expression where a contribution of $\Lambda$ matters in the expansions reads
\begin{align}
	q_{zz} &= \frac{2 a^2+e^2+e^2 \cos (\KSv)}{2 \left(a^2+e^2\right)}
 +\frac{\sqrt{2} a^{5/2} m
   \sin \left(\KSv/2\right) \sin (\KSv)
   \sqrt{\RR}}{\left(a^2+e^2\right)^2}
   \nonumber
\\
&
- \frac{e^6-a^4 \left(e^2+8 m^2\right)+2 a^2 e^2 \left(a^2+e^2\right) \left(2
   a^2+e^2\right) \lambda}{4 a \left(a^2+e^2\right)^3}
   \nonumber \\
 &\times  \sin^2(\KSv)\RR
   + \bigO{\RR^{2}}
	\label{4X19.2}
   \\
  &=\hspace{-.06cm} |_{e=0}\hspace{.2cm}
   1+\frac{\sqrt{2} m \sqrt{\RR} \sin \left(\KSv/2\right) \sin (\KSv)}{a^{3/2}} + \bigO{\RR^{2}}
   \,.
\end{align}
With this metric, we can now compute the Ricci scalar of the quotient space metric $q$, which reads
\begin{equation}
	\label{3X19.6}
	R(q) = R(q)\big|_{\Lambda=0} + \frac{2}{3}\Lambda
	\,.
\end{equation}
Thus, its leading order behaviour follows from the asymptotically flat case $\Lambda=0$ (\ref{21IX19.5}).

The level sets of $\RR$ remain spacelike,  affected by $\Lambda$ only in subleading terms:

\begin{align}
	\label{3X19.7}
	g^{\RR\RR}
 & =
 \frac{2 a^2+\charge ^2 \cos (\KSv)+\charge ^2}{2 a^2}-\frac{2 \sqrt{\RR } \left(\sqrt{2} m \cos
   ^3\left(\KSv/2\right)\right)}{a^{3/2}}
   \nonumber \\
   &-\frac{\RR  \left(8 \cos (\KSv) \left(2 a^4 \lambda
   +\charge ^2\right)+\charge ^2 (3 \cos (2 \KSv)+5)\right)}{8 a^3}
   \nonumber
\\
  &+O\left(\RR ^{3/2}\right) =\hspace{-.06cm} |_{e=0}\hspace{.2cm}
   1-\frac{2 \sqrt{\RR } \left(\sqrt{2} m \cos ^3\left(\KSv/2\right)\right)}{a^{3/2}}
   \nonumber \\
   &-2 \RR  (a
   \lambda  \cos (\KSv))+O\left(\RR ^{3/2}\right)
   \,.
\end{align}

\section{Conclusions}

Let us \rout{now} summarize the main results obtained in this paper regarding the properties of the curvature singularity of extensions of the Kerr family of spacetimes

The singularity can be surrounded by a family of timelike hypersurfaces $\{\mcT_{\hat R}\}$, the level sets $\{\RR={\hat R}\}$ of a natural coordinate $\RR$, with $\RR$ approaching zero as the singularity is approached. The coordinate $\RR$ can be thought of as ``the distance to the singularity'' because the distance to the singularity along the maximally extended integral curves of $\nabla \RR$ tends to $\RR$ as the singularity is approached in the Kerr and Kerr-(A)dS cases (in the charged cases, $\RR$ remains commensurate with that distance).  There exists spacetime-curvature invariants that tend to infinity as $\RR$ tends to zero, uniformly over $\mcT_{\hat R}$, although the Kretschmann scalar alone, or the Pontryagin scalar alone, do not have this property.

The timelike hypersurfaces $\mcT_{\hat R}$ have topology $\R\times \T^2$, where $\T^2=S^1\times S^1 $ is a two-dimensional torus.  One of the $S^1$ factors of the two-torus $\T^2$ is spacelike, with a period that is twice that resulting from the Minkowskian picture.
Approximately half of the other $S^{1}$ factors form closed causal curves, almost all of them timelike, close to (but, \red{for macroscopic objects}, still a macroscopic distance away from) the ring when the charge is zero, but regardless of the value of the cosmological constant; when the charge is not zero, the causality-violating region forms a complete neighborhood of the ring.

There exists an ergoregion close to the ring (i.e.~a region inside which observers must rotate with the singularity), with the topology of a solid torus for subcritical spins and that of a hollowed   marble  for supercritical spins, at least for the range of values of $a/m$ explored here.

The Killing vectors of the spacetime allow us to define two-dimensional geometries that provide further insight about the nature of the curvature singularity discussed above. The distribution of the metrics $\chi$ induced by the spacetime on the orbits of the connected component of the identity of the isometry group becomes singular as $\RR$ tends to zero, with the same type of singularity independent of the direction of approach. The two-dimensional geometry $q$, induced by the spacetime away from the set where the isometry orbits are null, \rout{however,} is \red{likewise} $C^{2}$-singular. In fact, when the charge is zero, the metric $q$ approaches a conical metric with negative deficit angle $-2\pi$ (in other words, a total angle of $ 4\pi$), as $\RR$ tends to zero, regardless of the value of the cosmological constant; in the charged case, the metric $q$ asymptotes to a conical metric as before up to an angle-dependent conformal factor.

We have also analyzed the tidal forces experienced by nearby geodesics as they approach the curvature singularity. We have shown that these forces lead to infinite displacements, and thus, infinite stresses as $\RR$ tends to zero, destroying any observer that is unfortunate enough to fall toward the singularity. We have shown that this occurs in a finite proper time. Not all geodesics on the equatorial plane, however, will reach the singularity, and we found that those that do not but still approach it require a tremendous amount of initial energy to get close to the singularity. When they do, these geodesics experience a very large transversal acceleration, \red{which would  kick a physical object on such a trajectory out} of the equatorial plane.

While finding these results, we also re-establish other known properties of the Kerr-like family of metrics. Of these, perhaps the most noteworthy is the development of an intuitive understanding for the need to introduce a double-covering space for the geometry near the singularity. As already noted by Kerr and by Newman~\cite{Newman:1965tw},
certain components of the Kerr metric in Kerr-Schild coordinates become double-valued as one considers  curves that \rout{pierce} \red{loop around}  the ring singularity. In our work, we introduce toroidal coordinates adapted to the topology of the $\cal T_{\hat R}$ hypersurfaces, which show that the toroidal angle (associated with loops that pierce the ring singularity) is indeed $4 \pi$ periodic. This provides a simple explanation for the need to introduce the double-covering space for the geometry near the singular set.

The purpose of the study presented here was to elucidate the nature of the singular behavior of Kerr-like metrics near the spacetime regime where there are curvature singularities, but in doing so, we have also provided a proposal for how to approach these issues in generic spacetimes with two Killing vectors. This is important because in this new era of gravitational wave observations and of black-hole shadow observations, many proposals have been put forward to constrain the possibility of the existence of non-Kerr compact objects, such as Manko-Novikov spacetimes~\cite{Manko_1992} or bumpy black-hole metrics~\cite{Collins:2004ex,Glampedakis:2005cf,Vigeland:2009pr,Vigeland:2011ji}.  Some of these compact objects are ``known'' to have naked curvature singularities, but the studies that establish their nature could benefit by following some of the steps laid out in this paper.

Imagine the following example. Let us say that one were able to show that nearby geodesics in these non-Kerr spacetimes experience infinite stresses as they hit a naked singularity, well outside the event horizons. It then follows that charged particles (say in an accretion disk) that follow such geodesics will emit an infinite amount of electromagnetic radiation when they hit the naked singularity. Since such an infinite amount of light is not observed in Nature, one would then have strong and rigorous arguments to state that such non-Kerr spacetimes are unphysical and not worth studying further with gravitational waves or black hole shadow observations. The analysis laid out in this paper could help in such a calculation.

\acknowledgements
We would like to thank J\'er\'emie Joudioux for comments and suggestions, and   Lars Andersson and Herbert Balasin for bibliographical advice.
PTC was supported in part by the Polish National Center of Science (NCN) under the grant 2016/21/B/ST1/00940  and  by the Austrian Science Foundation under the FWF project P 29517-N27.  NY was supported in part by NSF grant PHY-1759615 and NASA grant 80NSSC18K1352. All three authors are grateful to the ICTP in Trieste for hospitality and   financial support during the initial phase of this collaboration.  This work was also supported in parts by the Swedish Research Council under grant no.\ 2016-06596 while PTC and MM were in residence at Institut Mittag-Leffler in Djursholm, Sweden in the fall of 2019.

\appendix

\section{Expansions near the ring}
\label{A19IX19.1}

For reference, we give the asymptotic expansions of quantities of
interest at the ring in the vacuum case with vanishing cosmological
constant, keeping the first two non-trivial corrections to the
Minkowski metric.

\begin{widetext}
The components of the covector field $\KStheta$ \greenn{defined in Eq.~\eqref{8V20.01}} read
\begin{align}
	\KStheta_{t} &= -1\,,
	\\
	\KStheta_{\KSu } & = a + (\cos(\KSv)-1)\RR + \bigO{\RR^{2}}\,,
	\\
	\KStheta_{\RR} & = - \sqrt{\frac{2}{a}}\cos(\KSv/2)\RR^{1/2}
				     + \frac{1}{2\sqrt{2}a^{3/2}}\left(2\cos(\KSv/2)+\cos(3\KSv/2)\right)\RR^{3/2}
				     + \bigO{\RR^{5/2}}\,,
	\\
	\KStheta_{\KSv} & = \sqrt{\frac{2}{a}}\sin(\KSv/2)R^{3/2}
				   - \frac{1}{2\sqrt{2}a^{3/2}}\left(2\sin(\KSv/2)+\sin(3\KSv/2)\right)\RR^{5/2}
				   + \bigO{\RR^{7/2}}\,.
\end{align}
We have an exact formula for the determinant of the \greenn{Kerr} metric \greenn{\eqref{8X19.0}},
\begin{equation}
	\mathrm{det}\,g = -\RR^{2}(a+ \RR \cos(\KSv  ))^{2}
	\,.
\end{equation}
\greenn{Using \eqref{25VII19.22} and \eqref{H-eq} we find the following expansions for the functions $\tilde{r}$ and $H$:} \xout{The functions $\tilde{r}$ and $H$ have the following expansions:}
\begin{eqnarray}
	\tilde{r}
  & = &
    \sqrt{2 a}\cos(\KSv/2)\sqrt{\RR} + \frac{\cos{\KSv/2}}{2\sqrt{2a}}\RR^{3/2} +
	\bigO{\RR^{5/2}}
	\,,
\\
	H
 & = &
 m\left(\frac{\sqrt{2}\cos(\KSv/2)}{\sqrt{a\RR}} - \frac{\cos(3\KSv/2)}{2\sqrt{2}a^{3/2}}\RR^{1/2}
	+ \bigO{\RR^{3/2}}\right)
	\,.
\end{eqnarray}
The components of the metric read
\begin{align}
	g_{tt} &= -1 + m\left(\sqrt{\frac{2}{a}}\cos(\KSv/2)\RR^{-1/2}
			  - \frac{1}{2\sqrt{2}a^{3/2}}\cos(3\KSv/2)\RR^{1/2} + \bigO{\RR^{3/2}}\right)\,,
	\\
	g_{t\RR} &= m\left(\frac{1}{a}\left(1+\cos(\KSv)\right) - \frac{2}{a^{2}}\cos^{2}(\KSv/2)\cos(\KSv)\RR
			    + \bigO{\RR^{2}}\right)\,,
	\\
	g_{t\KSv} &= -m\left(\frac{1}{a}\sin(\KSv)\RR
			  + \frac{1}{a^{2}}\cos(\KSv/2)\sin(\KSv/2)\left(1+2\cos(\KSv)\right)\RR^{2}
			  + \bigO{\RR^{3}}\right)\,,
	\\
	g_{t\KSu} &= m\left(-\sqrt{2a}\cos(\KSv/2)\RR^{-1/2}
			  - \frac{1}{2\sqrt{2a}}\left(-2\cos(\KSv/2)+\cos(3\KSv/2)\right)\RR^{1/2}
			  + \bigO{\RR^{3/2}}\right)
	\\
	g_{\RR\RR} &= 1 + m\left(\frac{2\sqrt{2}}{a^{3/2}}\cos^{3}(\KSv/2)\RR^{1/2}
				  - \frac{1}{\sqrt{2}a^{5/2}}\cos^{3}(\KSv/2)\left(1+6\cos(\KSv)\right)\RR^{3/2}
				  + \bigO{\RR^{5/2}}\right)\,,
	\\
	g_{\RR \KSv } &= -\left(\frac{2\sqrt{2}}{a^{3/2}}\sin(\KSv/2)\cos^{2}(\KSv/2)\RR^{3/2}
				 + \frac{3}{\sqrt{2}a^{5/2}}\cos^{2}(\KSv/2)\sin(\KSv/2)\left(1+2\cos(\KSv)\right)\RR^{5/2}
				 + \bigO{\RR^{5/2}}\right)\,,
	\\
	g_{\RR \KSu } &= -m\left(\left(1+\cos{\KSv}\right) + \frac{1}{a}\left(1+\cos(\KSv)\right)\RR + \bigO{\RR^{2}}\right)\,,
	\\
	g_{\KSv\KSv} &= \RR^{2} + m\left(\frac{2\sqrt{2}}{a^{3/2}}\sin^{2}(\KSv/2)\cos(\KSv/2)\RR^{5/2}
			  - \frac{1}{\sqrt{2}a^{5/2}}\cos(\KSv/2)\sin^{2}(\KSv/2)\left(5+6\cos(\KSv)\right)\RR^{7/2}
			  + \bigO{\RR^{9/2}}\right)\,,
	\\
	g_{\KSv\KSu} & = m\left(\sin(\KSv)\RR - \frac{3}{2a}\sin(\KSv)\RR^{2} + \bigO{\RR^{3}}\right)\,,
	\\
	g_{\KSu\KSu} & = (a+\cos(\KSv)\RR)^{2} + m\left(\sqrt{2}a^{3/2}\cos(\KSv/2)\RR^{-1/2}
			   + \frac{\sqrt{a}}{2\sqrt{2}}\left(-4\cos(\KSv/2) + 3\cos(3\KSv/2)\right)\RR^{1/2}
			   + \bigO{\RR^{3/2}}\right)\,.
\end{align}
The inverse metric tensor has the following expansions:
\begin{align}
	g^{tt} &= -1 + m\left(-\sqrt{\frac{2}{a}}\cos(\KSv/2)\RR^{-1/2} + \frac{1}{2\sqrt{2}a^{3/2}}\cos(3\KSv/2)\RR^{1/2} + \bigO{\RR^{3/2}}\right)\,,
	\\
	g^{t\RR} &= m\left( \frac{1+\cos(\KSv)}{a} - \frac{2}{a^{2}}\cos^{2}(\KSv/2)\cos(\KSv)\RR + \bigO{\RR^{2}}\right)\,,
	\\
	g^{t\KSv} &= m\left(-\frac{1}{a}\sin(\KSv)\RR^{-1} + \frac{1}{2a^{2}}\left(\sin(\KSv)+\sin(2v)\right)
			  + \bigO{\RR}\right)\,,
	\\
	g^{t\KSu} &= m\left(-\frac{\sqrt{2}}{a^{3/2}}\cos(\KSv/2)\RR^{-1/2}
			  + \frac{3}{2\sqrt{2}a^{5/2}}\left(2\cos(\KSv/2)+\cos(3\KSv/2)\right)\RR^{1/2}
			  + \bigO{\RR^{3/2}}\right)\,,
	\\
	g^{\RR\RR} &= 1 + m\left(-\frac{2\sqrt{2}}{a^{3/2}}\cos^{3}(\KSv/2)\RR^{1/2}
				  + \frac{1}{\sqrt{2}a^{5/2}}\cos^{3}\left(1+5\cos(\KSv)\right)\RR^{3/2}
				  + \bigO{\RR^{5/2}} \right)\,,
	\\
	g^{\RR \KSv } &= m\left(\frac{2\sqrt{2}}{a^{3/2}}\cos^{2}(\KSv/2)\sin(\KSv)\RR^{-1/2}
				 - \frac{3}{\sqrt{2}a^{5/2}}\cos^{2}(\KSv/2)\sin(3\KSv/2)\RR^{1/2}
				 + \bigO{\RR^{3/2}}\right)\,,
	\\
	g^{\RR \KSu } &= m\left(\frac{1}{a^{2}}\left(1+\cos(\KSv)\right)
				 - \frac{2}{a^{3}}\cos^{2}(\KSv/2)\left(1+2\cos(\KSv)\right)\RR + \bigO{\RR^{2}}\right)\,,
	\\
	g^{\KSv\KSv} &= \RR^{-2} + m\left(-\frac{2\sqrt{2}}{a^{3/2}}\sin^{2}(\KSv/2)\cos(\KSv/2)\RR^{-3/2}
			  + \frac{1}{\sqrt{2}a^{5/2}}\sin^{2}(\KSv/2)\cos(\KSv/2)\left(5+6\cos(\KSv)\right)
			  + \bigO{\RR^{1/2}}\right)\,,
	\\
	g^{\KSv\KSu} &= m\left(-\frac{1}{a^{2}}\sin(\KSv)\RR^{-1}
			  + \frac{1}{a^{3}}\sin(\KSv/2)\cos(\KSv/2)\left(3+4\cos(\KSv)\right) + \bigO{\RR}\right)\,,
	\\
	g^{\KSv\KSv} &= \left(a+\cos(\KSv)\RR\right)^{-2} + m\left(-\frac{\sqrt{2}}{a^{5/2}}\cos(\KSv/2)\RR^{-1/2}
			  + \frac{1}{2\sqrt{2}a^{7/2}}\cos(\KSv/2)\left(7+10\cos(\KSv)\right)\RR^{1/2}
			  + \bigO{\RR^{3/2}}\right)\,.
\end{align}
\end{widetext}

\section{Slopes of null vectors of two-dimensional Lorentzian matrices}
\label{A23IX19.1}

Consider Eq.~\eqref{23IX19.35}, which we repeat here for the convenience of the reader:
\begin{equation}
\cos{(\alpha_\pm)}
 =\sqrt{ \frac{\left(g_{t\KSu}\pm\sqrt{g_{t\KSu}^2-g_{tt} g_{\KSu\KSu}}\right)^2+g_{\KSu\KSu}^2}{(g_{tt}- g_{\KSu\KSu})^2 +4 g_{t\KSu}^{2} } }\,.
 \label{23IX19.34a}
\end{equation}
We want to show that the right-hand side is smaller than or equal to
one, and therefore the angles $\alpha_\pm$ are well defined, assuming
that \emph{the matrix $\left(g_{ab}\right)$, $a,b\in \{t,u\}$ has
  Lorentzian signature} (equivalently,
$g_{tt} g_{\KSu\KSu} -g_{t\KSu}^2<0$).

We note that, under the current signature assumption,
\begin{enumerate}
\item the right-hand sides, both with the plus and the minus sign, of
  Eq.~\eqref{23IX19.34a} are real and non-negative;
\item the denominator under the radical never vanishes;
\item the product $\cos(\alpha_+) \cos (\alpha_-)$ vanishes only if
  $g_{\KSu\KSu}=0$; and
\item when $g_{t\KSu}=0$ we have
$$
  \cos(\alpha_+)= \cos(\alpha_-) = \sqrt{-\frac{g_{\KSu\KSu}}{g_{tt}-g_{\KSu\KSu}}}< 1
\,.
$$
\end{enumerate}
In particular the angles are well defined when
$g_{t\KSu}=0$. Otherwise, by homogeneity, it remains to consider the
case $g_{t\KSu}= 1$, which we assume from now on. This gives
\begin{equation}
\cos{(\alpha_\pm)}
 =\sqrt{ \frac{\left(1\pm\sqrt{1-g_{tt} g_{\KSu\KSu}}\right)^2+g_{\KSu\KSu}^2}{(g_{tt}- g_{\KSu\KSu})^2 +4  } }
 \,.
 \label{23IX19.34b}
\end{equation}
If $g_{tt}=0$, the desired inequalities for the cosine function are
clearly satisfied. It remains to consider the case $g_{tt}\ne 0$. As
such, we have
\begin{equation}\label{23IX19.32}
  \cos^{2}(\alpha_+) -\cos^{2}(\alpha_-) =
  \frac{4 \sqrt{1-g_{tt} g_{\KSu\KSu}}}{(g_{tt}-g_{\KSu\KSu})^2+4} >0
\end{equation}
so that $\cos(\alpha_+)>\cos(\alpha_-)$. Moreover it holds that
\begin{equation}\label{23IX19.33}
   1-\cos^{2}(\alpha_+) =
   \frac{g_{tt}^2}{2 + g_{tt}^2 - g_{tt} g_{\KSu\KSu} - 2 \sqrt{1 - g_{tt} g_{\KSu\KSu}}}
   \,.
\end{equation}
We wish to show that the denominator of the last expression is
positive. For this, we note the identity
\begin{align}\label{23IX19.34asdf}
g_{tt}^2 \left((g_{tt}-g_{\KSu\KSu})^2+4\right) &=
  \left(2 + g_{tt}^2 - g_{tt} g_{\KSu\KSu} - 2 \sqrt{1 - g_{tt} g_{\KSu\KSu}} \right)
  \nonumber \\
  &\times
  \left(2 + g_{tt}^2 - g_{tt} g_{\KSu\KSu} + 2 \sqrt{1 - g_{tt} g_{\KSu\KSu}}\right)
\end{align}
where the first factor is the denominator of the right-hand side of
Eq.~\eqref{23IX19.33}.  Since we have assumed that $g_{tt}\ne 0$, when
$g_{\KSu\KSu}=0$ both factors on the left-hand side of
Eq.~\eqref{23IX19.34asdf} are positive. Their product never vanishes
regardless of the value of $g_{\KSu\KSu}$. Continuity implies that
each factor is positive on the domain of interest. Hence, the
right-hand side of Eq.~\eqref{23IX19.33} is non-negative, vanishing
only if $g_{tt}=0$. We conclude that the expressions defining
$\cos(\alpha_\pm)$ are smaller than or equal to one, as
desired.

\section{The curvature tensor on the exterior equatorial hyperplanes}
\label{sA1XI19.1}

Letting $\epsilon$ denote the sign of $\tilde{r}$ in the Kerr-Schild
coordinates $(t,x,y,z)$, the leading terms of the $\RR$-expansion of
$R^{\mu}{}_{ a\nu b}$ ($a,b\in\{0,1,2\}$) read
\allowdisplaybreaks[4]
\begin{widetext}
\begin{align}
	\label{eq:400-1}
R^{\mu}{}_{ t\nu y} &=
\frac{3 m^2 \epsilon }{2 \sqrt{2} a^{3/2} \RR^{5/2}}
\left(
\begin{array}{cccc}
 \sin (\varphi ) & 1 & 0 & 0 \\
 \frac{\sin (\varphi ) (a \cos (\varphi )-2 m \sin (\varphi ))}{2 m} & \frac{a \cos
   (\varphi )-2 m \sin (\varphi )}{2 m} & 0 & 0 \\
 \frac{\sin (\varphi ) (a \sin (\varphi )+2 m \cos (\varphi ))}{2 m} & \frac{a \sin
   (\varphi )+2 m \cos (\varphi )}{2 m} & 0 & 0 \\
 0 & 0 & 0 & -\frac{a \cos (\varphi )}{2 m} \\
\end{array}
\right)
+\bigO{\RR^{-2}}
\,,
\\
R^{\mu}{}_{ x\nu y} &=
\frac{3 m^2 \epsilon }{2 \sqrt{2} a^{3/2} \RR^{5/2}}
\left(
\begin{array}{cccc}
 \frac{\sin (\varphi ) (a \cos (\varphi )+2 m \sin (\varphi ))}{2 m} & \frac{a \cos
   (\varphi )+2 m \sin (\varphi )}{2 m} & 0 & 0 \\
 -\sin ^3(\varphi ) & -\sin ^2(\varphi ) & 0 & 0 \\
 \frac{\sin (\varphi ) (a+m \sin (2 \varphi ))}{2 m} & \frac{a+m \sin (2 \varphi )}{2
   m} & 0 & 0 \\
 0 & 0 & 0 & -\frac{a \sin (2 \varphi )}{4 m} \\
\end{array}
\right)
+\bigO{\RR^{-2}}
\,,
\\
R^{\mu}{}_{ y\nu y} &=
\frac{3 m^2 \epsilon }{2 \sqrt{2} a^{3/2} \RR^{5/2}}
\left(
\begin{array}{cccc}
 \frac{\sin (\varphi ) (a \sin (\varphi )-2 m \cos (\varphi ))}{2 m} & \frac{a \sin
   (\varphi )-2 m \cos (\varphi )}{2 m} & 0 & 0 \\
 -\frac{\sin (\varphi ) (a-m \sin (2 \varphi ))}{2 m} & -\frac{a-m \sin (2 \varphi )}{2
   m} & 0 & 0 \\
 \sin (\varphi ) \left(-\cos ^2(\varphi )\right) & -\cos ^2(\varphi ) & 0 & 0 \\
 0 & 0 & 0 & \frac{a \cos ^2(\varphi )}{2 m} \\
\end{array}
\right)
+\bigO{\RR^{-2}}
 \,,
\\
R^{\mu}{}_{ t\nu t} &=
\frac{3 m^2 \epsilon }{2 \sqrt{2} a^{3/2} \RR^{5/2}}
\left(
\begin{array}{cccc}
 0 & -\cos (\varphi ) & -\sin (\varphi ) & 0 \\
 0 & -\frac{\cos (\varphi ) (a \cos (\varphi )-2 m \sin (\varphi ))}{2 m} & -\frac{\sin
   (\varphi ) (a \cos (\varphi )-2 m \sin (\varphi ))}{2 m} & 0 \\
 0 & -\frac{\cos (\varphi ) (a \sin (\varphi )+2 m \cos (\varphi ))}{2 m} & -\frac{\sin
   (\varphi ) (a \sin (\varphi )+2 m \cos (\varphi ))}{2 m} & 0 \\
 0 & 0 & 0 & \frac{a}{2 m} \\
\end{array}
\right)+\bigO{\RR^{-2}}
\,,
\\
R^{\mu}{}_{ t\nu x} &=
\frac{3 m^2 \epsilon }{2 \sqrt{2} a^{3/2} \RR^{5/2}}
\left(
\begin{array}{cccc}
 \cos (\varphi ) & 0 & -1 & 0 \\
 \frac{\cos (\varphi ) (a \cos (\varphi )-2 m \sin (\varphi ))}{2 m} & 0 & -\frac{a
   \cos (\varphi )-2 m \sin (\varphi )}{2 m} & 0 \\
 \frac{\cos (\varphi ) (a \sin (\varphi )+2 m \cos (\varphi ))}{2 m} & 0 & -\frac{a
   \sin (\varphi )+2 m \cos (\varphi )}{2 m} & 0 \\
 0 & 0 & 0 & \frac{a \sin (\varphi )}{2 m} \\
\end{array}
\right)
+\bigO{\RR^{-2}}
\,,
\\
R^{\mu}{}_{ x\nu x} &=
\frac{3 m^2 \epsilon }{2 \sqrt{2} a^{3/2} \RR^{5/2}}
\left(
\begin{array}{cccc}
 \frac{\cos (\varphi ) (a \cos (\varphi )+2 m \sin (\varphi ))}{2 m} & 0 & -\frac{a
   \cos (\varphi )+2 m \sin (\varphi )}{2 m} & 0 \\
 \sin ^2(\varphi ) (-\cos (\varphi )) & 0 & \sin ^2(\varphi ) & 0 \\
 \frac{\cos (\varphi ) (a+m \sin (2 \varphi ))}{2 m} & 0 & -\frac{a+m \sin (2 \varphi
   )}{2 m} & 0 \\
 0 & 0 & 0 & \frac{a \sin ^2(\varphi )}{2 m} \\
\end{array}
\right)
+\bigO{\RR^{-2}}
 \,.
\end{align}
\end{widetext}

\bibliography{KerrRingv3-minimal}

\begin{thebibliography}{25}%
\makeatletter
\providecommand \@ifxundefined [1]{%
 \@ifx{#1\undefined}
}%
\providecommand \@ifnum [1]{%
 \ifnum #1\expandafter \@firstoftwo
 \else \expandafter \@secondoftwo
 \fi
}%
\providecommand \@ifx [1]{%
 \ifx #1\expandafter \@firstoftwo
 \else \expandafter \@secondoftwo
 \fi
}%
\providecommand \natexlab [1]{#1}%
\providecommand \enquote  [1]{``#1''}%
\providecommand \bibnamefont  [1]{#1}%
\providecommand \bibfnamefont [1]{#1}%
\providecommand \citenamefont [1]{#1}%
\providecommand \href@noop [0]{\@secondoftwo}%
\providecommand \href [0]{\begingroup \@sanitize@url \@href}%
\providecommand \@href[1]{\@@startlink{#1}\@@href}%
\providecommand \@@href[1]{\endgroup#1\@@endlink}%
\providecommand \@sanitize@url [0]{\catcode `\\12\catcode `\$12\catcode
  `\&12\catcode `\#12\catcode `\^12\catcode `\_12\catcode `\%12\relax}%
\providecommand \@@startlink[1]{}%
\providecommand \@@endlink[0]{}%
\providecommand \url  [0]{\begingroup\@sanitize@url \@url }%
\providecommand \@url [1]{\endgroup\@href {#1}{\urlprefix }}%
\providecommand \urlprefix  [0]{URL }%
\providecommand \Eprint [0]{\href }%
\providecommand \doibase [0]{http://dx.doi.org/}%
\providecommand \selectlanguage [0]{\@gobble}%
\providecommand \bibinfo  [0]{\@secondoftwo}%
\providecommand \bibfield  [0]{\@secondoftwo}%
\providecommand \translation [1]{[#1]}%
\providecommand \BibitemOpen [0]{}%
\providecommand \bibitemStop [0]{}%
\providecommand \bibitemNoStop [0]{.\EOS\space}%
\providecommand \EOS [0]{\spacefactor3000\relax}%
\providecommand \BibitemShut  [1]{\csname bibitem#1\endcsname}%
\let\auto@bib@innerbib\@empty
\bibitem [{\citenamefont {Carter}(1968{\natexlab{a}})}]{CarterKerr}%
  \BibitemOpen
  \bibfield  {author} {\bibinfo {author} {\bibfnamefont {B.}~\bibnamefont
  {Carter}},\ }\bibfield  {title} {\enquote {\bibinfo {title} {Global structure
  of the {Kerr} family of gravitational fields},}\ }\href@noop {} {\bibfield
  {journal} {\bibinfo  {journal} {Phys.\ Rev.}\ }\textbf {\bibinfo {volume}
  {174}},\ \bibinfo {pages} {1559--1571} (\bibinfo {year}
  {1968}{\natexlab{a}})}\BibitemShut {NoStop}%
\bibitem [{\citenamefont {O'Neill}(1995)}]{BONeillK}%
  \BibitemOpen
  \bibfield  {author} {\bibinfo {author} {\bibfnamefont {B.}~\bibnamefont
  {O'Neill}},\ }\href@noop {} {\emph {\bibinfo {title} {The geometry of {K}err
  black holes}}}\ (\bibinfo  {publisher} {A K Peters, Ltd., Wellesley, MA},\
  \bibinfo {year} {1995})\BibitemShut {NoStop}%
\bibitem [{\citenamefont {Boyer}\ and\ \citenamefont
  {Lindquist}(1967)}]{BoyerLindquist}%
  \BibitemOpen
  \bibfield  {author} {\bibinfo {author} {\bibfnamefont {R.H.}\ \bibnamefont
  {Boyer}}\ and\ \bibinfo {author} {\bibfnamefont {R.W.}\ \bibnamefont
  {Lindquist}},\ }\bibfield  {title} {\enquote {\bibinfo {title} {Maximal
  analytic extension of the {K}err metric},}\ }\href@noop {} {\bibfield
  {journal} {\bibinfo  {journal} {Jour.\ Math.\ Phys.}\ }\textbf {\bibinfo
  {volume} {8}},\ \bibinfo {pages} {265--281} (\bibinfo {year}
  {1967})}\BibitemShut {NoStop}%
\bibitem [{\citenamefont {Visser}(2007)}]{VisserKerr}%
  \BibitemOpen
  \bibfield  {author} {\bibinfo {author} {\bibfnamefont {M.}~\bibnamefont
  {Visser}},\ }\bibfield  {title} {\enquote {\bibinfo {title} {{The Kerr
  spacetime: A} brief introduction},}\ }in\ \href@noop {} {\emph {\bibinfo
  {booktitle} {{Kerr Fest: B}lack Holes in Astrophysics, General Relativity and
  Quantum Gravity. {Christchurch, New Zealand, August 26-28, 2004}}}},\
  \bibinfo {editor} {edited by\ \bibinfo {editor} {\bibfnamefont {David~L.}\
  \bibnamefont {Wiltshire}}, \bibinfo {editor} {\bibfnamefont {Matt}\
  \bibnamefont {Visser}}, \ and\ \bibinfo {editor} {\bibfnamefont {Susan~M.}\
  \bibnamefont {Scott}}}\ (\bibinfo  {publisher} {Cambridge University Press},\
  \bibinfo {year} {2007})\ \Eprint {http://arxiv.org/abs/0706.0622}
  {arXiv:0706.0622 [gr-qc]} \BibitemShut {NoStop}%
\bibitem [{\citenamefont {Tahvildar-Zadeh}(2015)}]{ShadiKerr}%
  \BibitemOpen
  \bibfield  {author} {\bibinfo {author} {\bibfnamefont {A.S.}\ \bibnamefont
  {Tahvildar-Zadeh}},\ }\bibfield  {title} {\enquote {\bibinfo {title} {On a
  zero-gravity limit of the {K}err-{N}ewman spacetimes and their
  electromagnetic fields},}\ }\href {\doibase 10.1063/1.4915290} {\bibfield
  {journal} {\bibinfo  {journal} {Jour.\ Math.\ Phys.}\ }\textbf {\bibinfo
  {volume} {56}},\ \bibinfo {pages} {042501, 19} (\bibinfo {year} {2015})},\
  \Eprint {http://arxiv.org/abs/1410.0416} {arXiv:1410.0416 [math-ph]}
  \BibitemShut {NoStop}%
\bibitem [{\citenamefont {Israel}(1970)}]{IsraelKerr2}%
  \BibitemOpen
  \bibfield  {author} {\bibinfo {author} {\bibfnamefont {W.}~\bibnamefont
  {Israel}},\ }\bibfield  {title} {\enquote {\bibinfo {title} {Source of the
  {Kerr} metric},}\ }\href {\doibase 10.1103/PhysRevD.2.641} {\bibfield
  {journal} {\bibinfo  {journal} {Phys.\ Rev.\ D}\ }\textbf {\bibinfo {volume}
  {2}},\ \bibinfo {pages} {641--646} (\bibinfo {year} {1970})}\BibitemShut
  {NoStop}%
\bibitem [{\citenamefont {Israel}(1977)}]{IsraelKerr3}%
  \BibitemOpen
  \bibfield  {author} {\bibinfo {author} {\bibfnamefont {W.}~\bibnamefont
  {Israel}},\ }\bibfield  {title} {\enquote {\bibinfo {title} {Line sources in
  general relativity},}\ }\href {\doibase 10.1103/PhysRevD.15.935} {\bibfield
  {journal} {\bibinfo  {journal} {Phys.\ Rev.\ D}\ }\textbf {\bibinfo {volume}
  {15}},\ \bibinfo {pages} {935--941} (\bibinfo {year} {1977})}\BibitemShut
  {NoStop}%
\bibitem [{\citenamefont {Newman}\ and\ \citenamefont
  {Janis}(1965)}]{Newman:1965tw}%
  \BibitemOpen
  \bibfield  {author} {\bibinfo {author} {\bibfnamefont {E.~T.}\ \bibnamefont
  {Newman}}\ and\ \bibinfo {author} {\bibfnamefont {A.~I.}\ \bibnamefont
  {Janis}},\ }\bibfield  {title} {\enquote {\bibinfo {title} {{Note on the Kerr
  spinning particle metric}},}\ }\href {\doibase 10.1063/1.1704350} {\bibfield
  {journal} {\bibinfo  {journal} {J. Math. Phys.}\ }\textbf {\bibinfo {volume}
  {6}},\ \bibinfo {pages} {915--917} (\bibinfo {year} {1965})}\BibitemShut
  {NoStop}%
\bibitem [{\citenamefont {Misner}\ \emph {et~al.}(1973)\citenamefont {Misner},
  \citenamefont {Thorne},\ and\ \citenamefont {Wheeler}}]{Misner:1974qy}%
  \BibitemOpen
  \bibfield  {author} {\bibinfo {author} {\bibfnamefont {Charles~W.}\
  \bibnamefont {Misner}}, \bibinfo {author} {\bibfnamefont {K.~S.}\
  \bibnamefont {Thorne}}, \ and\ \bibinfo {author} {\bibfnamefont {J.~A.}\
  \bibnamefont {Wheeler}},\ }\href@noop {} {\emph {\bibinfo {title}
  {{Gravitation}}}}\ (\bibinfo  {publisher} {W. H. Freeman},\ \bibinfo
  {address} {San Francisco},\ \bibinfo {year} {1973})\BibitemShut {NoStop}%
\bibitem [{\citenamefont {Poisson}(2009)}]{Poisson:2009pwt}%
  \BibitemOpen
  \bibfield  {author} {\bibinfo {author} {\bibfnamefont {E.}~\bibnamefont
  {Poisson}},\ }\href {\doibase 10.1017/CBO9780511606601} {\emph {\bibinfo
  {title} {{A Relativist's Toolkit: The Mathematics of Black-Hole
  Mechanics}}}}\ (\bibinfo  {publisher} {Cambridge University Press},\ \bibinfo
  {year} {2009})\BibitemShut {NoStop}%
\bibitem [{\citenamefont {Hawking}\ and\ \citenamefont {Ellis}(1973)}]{HE}%
  \BibitemOpen
  \bibfield  {author} {\bibinfo {author} {\bibfnamefont {S.W.}\ \bibnamefont
  {Hawking}}\ and\ \bibinfo {author} {\bibfnamefont {G.F.R.}\ \bibnamefont
  {Ellis}},\ }\href@noop {} {\emph {\bibinfo {title} {The large scale structure
  of spacetime}}}\ (\bibinfo  {publisher} {Cambridge University Press},\
  \bibinfo {address} {Cambridge},\ \bibinfo {year} {1973})\ pp.\ \bibinfo
  {pages} {xi+391},\ \bibinfo {note} {{Cambridge Monographs on Mathematical
  Physics, No. 1}}\BibitemShut {NoStop}%
\bibitem [{\citenamefont {Balasin}\ and\ \citenamefont
  {Nachbagauer}(1994)}]{BalasinNachbagauer}%
  \BibitemOpen
  \bibfield  {author} {\bibinfo {author} {\bibfnamefont {H.}~\bibnamefont
  {Balasin}}\ and\ \bibinfo {author} {\bibfnamefont {H.}~\bibnamefont
  {Nachbagauer}},\ }\bibfield  {title} {\enquote {\bibinfo {title}
  {Distributional energy-momentum tensor of the {K}err-{N}ewman spacetime
  family},}\ }\href {http://stacks.iop.org/0264-9381/11/1453} {\bibfield
  {journal} {\bibinfo  {journal} {Class.\ Quantum Grav.}\ }\textbf {\bibinfo
  {volume} {11}},\ \bibinfo {pages} {1453--1461} (\bibinfo {year}
  {1994})}\BibitemShut {NoStop}%
\bibitem [{\citenamefont {Clarke}\ \emph {et~al.}(1996)\citenamefont {Clarke},
  \citenamefont {Vickers},\ and\ \citenamefont {Wilson}}]{VickersStrings}%
  \BibitemOpen
  \bibfield  {author} {\bibinfo {author} {\bibfnamefont {C.J.S.}\ \bibnamefont
  {Clarke}}, \bibinfo {author} {\bibfnamefont {J.A.}\ \bibnamefont {Vickers}},
  \ and\ \bibinfo {author} {\bibfnamefont {J.P.}\ \bibnamefont {Wilson}},\
  }\bibfield  {title} {\enquote {\bibinfo {title} {{Generalized functions and
  distributional curvature of cosmic strings}},}\ }\href {\doibase
  10.1088/0264-9381/13/9/013} {\bibfield  {journal} {\bibinfo  {journal}
  {Class.\ Quantum Grav.}\ }\textbf {\bibinfo {volume} {13}},\ \bibinfo {pages}
  {2485--2498} (\bibinfo {year} {1996})},\ \Eprint
  {http://arxiv.org/abs/gr-qc/9605060} {arXiv:gr-qc/9605060 [gr-qc]}
  \BibitemShut {NoStop}%
\bibitem [{\citenamefont {Geroch}\ and\ \citenamefont
  {Traschen}(1987)}]{GerochTraschen}%
  \BibitemOpen
  \bibfield  {author} {\bibinfo {author} {\bibfnamefont {R.}~\bibnamefont
  {Geroch}}\ and\ \bibinfo {author} {\bibfnamefont {J.}~\bibnamefont
  {Traschen}},\ }\bibfield  {title} {\enquote {\bibinfo {title} {Strings and
  other distributional sources in general relativity},}\ }\href {\doibase
  10.1103/PhysRevD.36.1017} {\bibfield  {journal} {\bibinfo  {journal} {Phys.
  Rev. D (3)}\ }\textbf {\bibinfo {volume} {36}},\ \bibinfo {pages}
  {1017--1031} (\bibinfo {year} {1987})}\BibitemShut {NoStop}%
\bibitem [{\citenamefont {Coquereaux}\ and\ \citenamefont
  {Jadczyk}(1988)}]{Coquereaux:1988ne}%
  \BibitemOpen
  \bibfield  {author} {\bibinfo {author} {\bibfnamefont {R.}~\bibnamefont
  {Coquereaux}}\ and\ \bibinfo {author} {\bibfnamefont {A.}~\bibnamefont
  {Jadczyk}},\ }\href@noop {} {\emph {\bibinfo {title} {{Riemannian geometry,
  fiber bundles, Kaluza-Klein} theories and all that}}},\ \bibinfo {series}
  {World Sci.\ Lect.\ Notes Phys.}, Vol.~\bibinfo {volume} {16}\ (\bibinfo
  {publisher} {World Scientific Publishing Co.},\ \bibinfo {address}
  {Singapore},\ \bibinfo {year} {1988})\ pp.\ \bibinfo {pages}
  {xiv+345}\BibitemShut {NoStop}%
\bibitem [{\citenamefont {Overduin}\ and\ \citenamefont
  {Wesson}(1997)}]{Overduin:1998pn}%
  \BibitemOpen
  \bibfield  {author} {\bibinfo {author} {\bibfnamefont {J.M.}\ \bibnamefont
  {Overduin}}\ and\ \bibinfo {author} {\bibfnamefont {P.S.}\ \bibnamefont
  {Wesson}},\ }\bibfield  {title} {\enquote {\bibinfo {title} {{Kaluza-Klein
  gravity}},}\ }\href {\doibase 10.1016/S0370-1573(96)00046-4} {\bibfield
  {journal} {\bibinfo  {journal} {Phys. Rept.}\ }\textbf {\bibinfo {volume}
  {283}},\ \bibinfo {pages} {303--380} (\bibinfo {year} {1997})},\ \Eprint
  {http://arxiv.org/abs/gr-qc/9805018} {arXiv:gr-qc/9805018 [gr-qc]}
  \BibitemShut {NoStop}%
\bibitem [{\citenamefont {Carter}(1968{\natexlab{b}})}]{CarterSeparable}%
  \BibitemOpen
  \bibfield  {author} {\bibinfo {author} {\bibfnamefont {B.}~\bibnamefont
  {Carter}},\ }\bibfield  {title} {\enquote {\bibinfo {title}
  {Hamilton-{J}acobi and {S}chr\"odinger separable solutions of {E}instein's
  equations},}\ }\href@noop {} {\bibfield  {journal} {\bibinfo  {journal}
  {Commun.\ Math.\ Phys.}\ }\textbf {\bibinfo {volume} {10}},\ \bibinfo {pages}
  {280--310} (\bibinfo {year} {1968}{\natexlab{b}})}\BibitemShut {NoStop}%
\bibitem [{\citenamefont {Demia\'nski}(1973)}]{Demianski}%
  \BibitemOpen
  \bibfield  {author} {\bibinfo {author} {\bibfnamefont {M.}~\bibnamefont
  {Demia\'nski}},\ }\bibfield  {title} {\enquote {\bibinfo {title} {Some new
  solutions of the {Einstein} equations of astrophysical interest},}\
  }\href@noop {} {\bibfield  {journal} {\bibinfo  {journal} {Acta Astronomica}\
  }\textbf {\bibinfo {volume} {23}},\ \bibinfo {pages} {197--231} (\bibinfo
  {year} {1973})}\BibitemShut {NoStop}%
\bibitem [{\citenamefont {Stephani}\ \emph {et~al.}(2003 (2nd
  ed.))\citenamefont {Stephani}, \citenamefont {Kramer}, \citenamefont
  {MacCallum}, \citenamefont {Hoenselaers},\ and\ \citenamefont
  {Herlt}}]{Exactsolutions2}%
  \BibitemOpen
  \bibfield  {author} {\bibinfo {author} {\bibfnamefont {H.}~\bibnamefont
  {Stephani}}, \bibinfo {author} {\bibfnamefont {D.}~\bibnamefont {Kramer}},
  \bibinfo {author} {\bibfnamefont {M.}~\bibnamefont {MacCallum}}, \bibinfo
  {author} {\bibfnamefont {C.}~\bibnamefont {Hoenselaers}}, \ and\ \bibinfo
  {author} {\bibfnamefont {E.}~\bibnamefont {Herlt}},\ }\href@noop {} {\emph
  {\bibinfo {title} {Exact solutions of {E}instein's field equations}}},\
  Cambridge Monographs on Mathematical Physics\ (\bibinfo  {publisher}
  {Cambridge University Press},\ \bibinfo {address} {Cambridge},\ \bibinfo
  {year} {2003 (2nd ed.)})\ pp.\ \bibinfo {pages} {xxx+701}\BibitemShut
  {NoStop}%
\bibitem [{\citenamefont {Moncrief}(2006)}]{MoncriefNI}%
  \BibitemOpen
  \bibfield  {author} {\bibinfo {author} {\bibfnamefont {V.}~\bibnamefont
  {Moncrief}},\ }\bibfield  {title} {\enquote {\bibinfo {title} {An integral
  equation for spacetime curvature in general relativity},}\ }in\ \href@noop {}
  {\emph {\bibinfo {booktitle} {Surveys in differential geometry. {V}ol.
  {X}}}},\ \bibinfo {series} {Surv. Differ. Geom.}, Vol.~\bibinfo {volume}
  {10}\ (\bibinfo  {publisher} {Int. Press, Somerville, MA},\ \bibinfo {year}
  {2006})\ pp.\ \bibinfo {pages} {109--146}\BibitemShut {NoStop}%
\bibitem [{\citenamefont {Manko}\ and\ \citenamefont
  {Novikov}(1992)}]{Manko_1992}%
  \BibitemOpen
  \bibfield  {author} {\bibinfo {author} {\bibfnamefont {V.S.}\ \bibnamefont
  {Manko}}\ and\ \bibinfo {author} {\bibfnamefont {I.D.}\ \bibnamefont
  {Novikov}},\ }\bibfield  {title} {\enquote {\bibinfo {title} {Generalizations
  of the {K}err and {K}err-{N}ewman metrics possessing an arbitrary set of
  mass-multipole moments},}\ }\href {\doibase 10.1088/0264-9381/9/11/013}
  {\bibfield  {journal} {\bibinfo  {journal} {Class.\ Quantum Grav.}\ }\textbf
  {\bibinfo {volume} {9}},\ \bibinfo {pages} {2477--2487} (\bibinfo {year}
  {1992})}\BibitemShut {NoStop}%
\bibitem [{\citenamefont {Collins}\ and\ \citenamefont
  {Hughes}(2004)}]{Collins:2004ex}%
  \BibitemOpen
  \bibfield  {author} {\bibinfo {author} {\bibfnamefont {N.A.}\ \bibnamefont
  {Collins}}\ and\ \bibinfo {author} {\bibfnamefont {S.A.}\ \bibnamefont
  {Hughes}},\ }\bibfield  {title} {\enquote {\bibinfo {title} {{Towards a
  formalism for mapping the space-times of massive compact objects: Bumpy black
  holes and their orbits}},}\ }\href {\doibase 10.1103/PhysRevD.69.124022}
  {\bibfield  {journal} {\bibinfo  {journal} {Phys. Rev.}\ }\textbf {\bibinfo
  {volume} {D69}},\ \bibinfo {pages} {124022} (\bibinfo {year} {2004})},\
  \Eprint {http://arxiv.org/abs/gr-qc/0402063} {arXiv:gr-qc/0402063 [gr-qc]}
  \BibitemShut {NoStop}%
\bibitem [{\citenamefont {Glampedakis}\ and\ \citenamefont
  {Babak}(2006)}]{Glampedakis:2005cf}%
  \BibitemOpen
  \bibfield  {author} {\bibinfo {author} {\bibfnamefont {K.}~\bibnamefont
  {Glampedakis}}\ and\ \bibinfo {author} {\bibfnamefont {S.}~\bibnamefont
  {Babak}},\ }\bibfield  {title} {\enquote {\bibinfo {title} {{Mapping
  spacetimes with LISA: Inspiral of a test-body in a `quasi-Kerr' field}},}\
  }\href {\doibase 10.1088/0264-9381/23/12/013} {\bibfield  {journal} {\bibinfo
   {journal} {Class.\ Quantum Grav.}\ }\textbf {\bibinfo {volume} {23}},\
  \bibinfo {pages} {4167--4188} (\bibinfo {year} {2006})},\ \Eprint
  {http://arxiv.org/abs/gr-qc/0510057} {arXiv:gr-qc/0510057 [gr-qc]}
  \BibitemShut {NoStop}%
\bibitem [{\citenamefont {Vigeland}\ and\ \citenamefont
  {Hughes}(2010)}]{Vigeland:2009pr}%
  \BibitemOpen
  \bibfield  {author} {\bibinfo {author} {\bibfnamefont {S.J.}\ \bibnamefont
  {Vigeland}}\ and\ \bibinfo {author} {\bibfnamefont {S.A.}\ \bibnamefont
  {Hughes}},\ }\bibfield  {title} {\enquote {\bibinfo {title} {{Spacetime and
  orbits of bumpy black holes}},}\ }\href {\doibase 10.1103/PhysRevD.81.024030}
  {\bibfield  {journal} {\bibinfo  {journal} {Phys.\ Rev.}\ }\textbf {\bibinfo
  {volume} {D81}},\ \bibinfo {pages} {024030} (\bibinfo {year} {2010})},\
  \Eprint {http://arxiv.org/abs/0911.1756} {arXiv:0911.1756 [gr-qc]}
  \BibitemShut {NoStop}%
\bibitem [{\citenamefont {Vigeland}\ \emph {et~al.}(2011)\citenamefont
  {Vigeland}, \citenamefont {Yunes},\ and\ \citenamefont
  {Stein}}]{Vigeland:2011ji}%
  \BibitemOpen
  \bibfield  {author} {\bibinfo {author} {\bibfnamefont {S.}~\bibnamefont
  {Vigeland}}, \bibinfo {author} {\bibfnamefont {N.}~\bibnamefont {Yunes}}, \
  and\ \bibinfo {author} {\bibfnamefont {L.}~\bibnamefont {Stein}},\ }\bibfield
   {title} {\enquote {\bibinfo {title} {{Bumpy Black Holes in Alternate
  Theories of Gravity}},}\ }\href {\doibase 10.1103/PhysRevD.83.104027}
  {\bibfield  {journal} {\bibinfo  {journal} {Phys. Rev.}\ }\textbf {\bibinfo
  {volume} {D83}},\ \bibinfo {pages} {104027} (\bibinfo {year} {2011})},\
  \Eprint {http://arxiv.org/abs/1102.3706} {arXiv:1102.3706 [gr-qc]}
  \BibitemShut {NoStop}%
\end{thebibliography}%

\end{document}